\documentclass[aps, twocolumn,superscriptaddress,nofootinbib,longbibliography]{revtex4-2}
\usepackage[utf8]{inputenc}
\usepackage[colorlinks=true,linkcolor=blue,citecolor=blue,urlcolor=blue]{hyperref}
\hypersetup{pdfborder={0 0 0}}

\usepackage{graphicx}  % needed for figures
\usepackage{dcolumn}   % needed for some tables
\usepackage{bm}        % for math
\usepackage{amssymb,amsmath}   % for math
\usepackage{uri}
\usepackage{float}
\usepackage{geometry}
\usepackage{comment}
\usepackage{fixmath}
\usepackage{makecell}
\begin{document}
\title{Phase Coexistence in Nonreciprocal Quorum-Sensing Active Matter}

\author{Yu Duan}
\affiliation{Max Planck Institute for Dynamics and Self-Organization (MPI-DS), 37077 G\"ottingen, Germany}

\author{Jaime Agudo-Canalejo}
\affiliation{Max Planck Institute for Dynamics and Self-Organization (MPI-DS), 37077 G\"ottingen, Germany}

\author{Ramin Golestanian} \email{ramin.golestanian@ds.mpg.de}
\affiliation{Max Planck Institute for Dynamics and Self-Organization (MPI-DS), 37077 G\"ottingen, Germany}
\affiliation{Rudolf Peierls Centre for Theoretical Physics, University of Oxford, Oxford OX1 3PU,
United Kingdom}

\author{Beno\^{\i}t Mahault} \email{benoit.mahault@ds.mpg.de}
\affiliation{Max Planck Institute for Dynamics and Self-Organization (MPI-DS), 37077 G\"ottingen, Germany}

\date{\today}

\begin{abstract}
Motility and nonreciprocity are two primary mechanisms for self-organization in active matter. In a recent study [\href{https://journals.aps.org/prl/abstract/10.1103/PhysRevLett.131.148301}{Phys. Rev. Lett. \textbf{131}, 148301 (2023)}], we explored their joint influence in a minimal model of two-species quorum-sensing active particles interacting via mutual motility regulation. Our results notably revealed a highly dynamic phase of chaotic chasing bands that is absent when either nonreciprocity or self-propulsion is missing. Here, we examine further the phase behavior of nonreciprocal quorum-sensing active particles, distinguishing between the regimes of weak and strong nonreciprocity. In the weakly nonreciprocal regime, this system exhibits multi-component motility-induced phase separation. We establish an analytical criterion for the associated phase coexistence, enabling a quantitative prediction of the phase diagram. For strong nonreciprocity, where the dynamics is chase-and-run-like, we determine the phase behavior and show that it strongly depends on the scale of observation. In small systems, our numerical simulations reveal a phenomenology consistent with phenomenological models, comprising traveling phase-separated domains and spiral-like defect patterns. However, we show that these structures are generically unstable in large systems, where they are superseded by bulk phase coexistence between domains that are either homogeneous or populated by mesoscopic chasing bands. Crucially, this implies that collective motion totally vanishes at large scales, while the breakdown of our analytical criterion for this phase coexistence with multi-scale structures prevents us from predicting the corresponding phase diagram.

\end{abstract}

\maketitle
\tableofcontents

\section{Introduction}

In recent years, systems in which the local breaking of detailed balance results in effective interactions that violate the action-reaction symmetry have emerged as a paradigm for active matter~\cite{bowick2022symmetry,golestanian2024non-reciprocal}. 
Such nonreciprocal couplings mediate a variety of collective behaviors including the emergence of dynamical oscillations~\cite{saha2020scalar,you2020nonreciprocity,fruchart2021non,guislain2024discontinuous}, 
odd elasticity~\cite{fruchart2023odd}, and long-range orientational order~\cite{dadhichi2020nonmutual,pisegna2024emergent}.
Nonreciprocity is commonly found in condensed matter, as it generally arises when agents interact via a nonequilibrium medium.
This situation encompasses examples as diverse as complex plasmas~\cite{morfill2009complex}, non-Hermitian quantum systems~\cite{chiacchio2023nonreciprocal}, moving topological defects~\cite{romano2024dynamical}, 
and active particles coupled via chemotaxis~\cite{soto2014self,meredith2020predator,liu2024self}, hydrodynamics~\cite{uchida2010synchronization,tan2022odd,maity2023spontaneous} or through an elastic medium~\cite{banerjee2022unjamming,maity2023spontaneous}. 
Nonreciprocal interactions are also a hallmark of living matter, 
where they emerge from asymmetric interactions between heterogeneous populations~\cite{yamanaka2014invitro,haluts2024active,blumenthal2024phase}.

Beyond minimal agent-based models including explicit nonreciprocal pairwise interactions~\cite{fruchart2021non,ivlev2015statistical,chiu2023phase,mandal2024robustness,hanai2024nonreciprocal,kreienkamp2024non},
chemically active matter offers a physically grounded, bottom-up framework for investigating
self-organization in nonreciprocally coupled mixtures~\cite{golestanian2024non-reciprocal,agudo2019active,duan2023dynamical,dinelli2023non,ouazan2023self,tucci2024nonreciprocal}.
Communication mediated by chemical signalling, also known as quorum sensing, 
is natural vector for nonreciprocal interactions in microbial populations.
Quorum sensing indeed regulates a wide range of collective behaviors of bacteria~\cite{Papenfort2016review},
including the regulation of motility in response of local variations in population density~\cite{Weiss2008}.
Such quorum-sensing interactions are involved in mechanisms driving collective aggregation of various bacteria~\cite{Liu2011Science,liu2019self,curatolo2020cooperative,fragkopoulos2021self,courcoubetis2022formation},
while they have also been evidenced in fire ants~\cite{anderson2022social}, and can be mimicked in synthetic systems such as light-activated colloids~\cite{bauerle2018self,lavergne2019Science} or motile robots~\cite{chen2024emergent}.
In turn, nonreciprocity naturally emerges in multi-species systems when agents from different populations respond asymmetrically to each other's presence.

For a single species, self-propelled particles with density-dependent motility serve as a prototypical model
for motility induced phase separation (MIPS)~\cite{cates2015motility,gnan2022critical,bertin2024biased,zhou2024clustering}.
The large scale dynamics of this system is captured by a conserved density field which obeys a generalized Cahn-Hilliard equation,
where activity appears through interface terms in the chemical potential that do not derive from a free energy functional~\cite{cates2015motility,wittkowski2014scalar}.
When extended to multiple species, the resulting nonreciprocal quorum-sensing (NRQS) 
model then provides a minimal microscopic description for active phase separation with nonreciprocal couplings~\cite{duan2023dynamical,dinelli2023non}. 

On scales much larger than the typical persistence length of the active motion, 
the dynamics of the NRQS model is described by the nonreciprocal Cahn-Hilliard (NRCH) model~\cite{duan2023dynamical,dinelli2023non,dinelli2024fluctuating}.
The NRCH equations generalize the multi-component Cahn-Hilliard model to include asymmetric
bulk contributions to the species chemical potentials~\cite{saha2020scalar,you2020nonreciprocity}, 
which enable the description of oscillatory behavior in conserved mixtures~\cite{frohoff2023nonreciprocal}. 
Studies of the NRCH model have revealed a rich phenomenology, comprising the emergence of stationary and traveling phase-separated domains~\cite{saha2020scalar,you2020nonreciprocity,frohoff2021suppression,brauns2024nonreciprocal} and dynamical defect patterns~\cite{rana2023defect,rana2024defect}, which are features shared with the NRQS dynamics~\cite{duan2023dynamical}.

However, we have also shown in Ref.~\cite{duan2023dynamical} that the interplay of motility and non-reciprocity in the NRQS model leads to mescoscale dynamic patterns not captured by the NRCH description.
Specifically, binary mixtures of NRQS active particles, when their self-propelled motion is sufficiently persistent, 
can form regular or chaotic patterns of thin chasing bands, with no counterpart in the NRCH model.
In addition, these bands significantly impact the phase behavior of the NRQS model,
leading to a new type of multi-scale phase separation where dense domains of chaotic chasing bands (CCB) coexist with a homogeneous gas.

In our previous work, Ref.~\cite{duan2023dynamical}, we have only explored a limited portion of the phase diagram of the two-species NRQS model. 
In this work, we present a more comprehensive investigation of its phase behavior, focusing on two key aspects. 
While the binodal densities of quorum-sensing MIPS can be predicted with good accuracy~\cite{solon2018generalized1,solon2018generalized2}, no general framework is available for nonreciprocal mixtures, despite recent progress in addressing specific scenarios~\cite{dinelli2023non,brauns2024nonreciprocal,greve2024maxwell,saha2024phase}.
Here, we thus propose a generalized common tangent construction that predicts the coexisting densities of stationary phase separation for a broader class of nonreciprocal dynamics.
This framework encompasses previously considered cases, and enables the quantitative prediction of the full phase diagram of weakly nonreciprocal mixtures.

Our second focus is on the regime of strong nonreciprocity, 
where our large-scale numerical simulations reveal a rich phase behavior.
Interestingly, the structure of the corresponding phase diagram resembles that of equilibrium mixtures with mutual attraction, 
but where the dense mixed phase is replaced by CCB in phase coexistence, as a result of a generic finite wavelength instability induced by nonreciprocity. 
This feature leads to the emergence of nonequilibrium phase-separated configurations characterized by the presence of multi-scale structures, and which exhibit an anomalous coarsening dynamics.
Despite the presence of finite-size traveling patterns, our analysis reveals the complete absence of large-scale collective motion in the NRQS dynamics, which we attribute to the presence of instabilities of interface profiles of traveling phase-separated domains.
These observations are are in stark contrast with predictions from the NRCH model~\cite{saha2020scalar,you2020nonreciprocity,brauns2024nonreciprocal,rana2023defect,pisegna2024emergent}, which highlights the need for a more advanced theoretical framework to understand the NRQS physics.

The remaining of the paper is organized as follows.
We introduce the microscopic NRQS model in Sec.~\ref{sec:micro-model} 
and its coarse-grained description in Sec.~\ref{sec:hydro_NRQS}.
After briefly reviewing results from the linear stability analysis in Sec.~\ref{sec:instability_homo},
we present in Sec.~\ref{sec:phase-equilibria} the generalized common tangent construction
that we use to predict the phase behavior of the NRQS model in the weakly nonreciprocal regime in Sec.~\ref{sec:wo-chasing}.
Section~\ref{sec:w-chasing} is in turn devoted to the exploration of the regime of strong nonreciprocity, 
characterized by the emergence of dynamical patterns.
We describe in Sec.~\ref{sec:wo-inhibition} chasing band patterns that emerge solely from nonreciprocal inter-species motility regulation,
while in Sec.~\ref{sec:w-inhibition} we examine the phase behavior of NRQS active particles when self-inhibition of their motilities additionally drives bulk phase separation.
We conclude this part with Sec.~\ref{sec:break-Maxwell}, where we illustrate the breakdown of the generalized common tangent construction for the description of phase coexistence involving CCB at large nonreciprocity.
We finally provide concluding remarks and perspectives on our results in Sec.~\ref{sec:conclusion}.

\section{The NRQS model}\label{sec:micro-model}
We consider a binary mixture of a total of $N$ active Brownian particles separated into two populations $S\in\{A, B\}$.
Particle $i$ of species $S$ moves on the two-dimensional plane at speed $v_S$ along its intrinsic polarity direction $\hat{\mathbf{u}}_{i,S}=(\cos \theta_{i,S}, \sin \theta_{i,S})$ that undergoes rotational diffusion with diffusivity $D_r$ assumed equal for both species.
The particle position $\mathbf{r}_{i,S}$ and orientation $\hat{\mathbf{u}}_{i,S}$ evolve at time $t$ according to
\begin{subequations} \label{eq:micro-model}
\begin{align}
    \dot{\mathbf{r}}_{i,S} &= v_S\left[\tilde{{\rho}}_A(\mathbf{r}_{i,S}), \tilde{\rho}_B(\mathbf{r}_{i,S})\right] \hat{\mathbf{u}}_{i,S}, \label{eq:micro-model-a}\\
    \dot{\theta}_{i,S} &= \sqrt{2D_r}\xi_i(t),
    \label{eq:micro-model-b}
\end{align}
\end{subequations}
where $\xi_i$ is a Gaussian white noise satisfying $\langle \xi_i(t)\rangle=0$ and $\langle \xi_i(t)\xi_j(t')\rangle=\delta_{ij}\delta(t-t')$.
To calculate the quorum-sensing interactions, we define $\tilde{\rho}_S(\mathbf{r},t)\equiv\sum_i \hat{w}(|\mathbf{r}-\mathbf{r}_{i,S}(t)|)$ as the coarse-grained density fields for species $S$ evaluated at position $\mathbf{r}$.
Here, $\hat{w}(r)$ is an isotropic short-range weight function that vanishes beyond a finite interaction scale $R$, and satisfies
$\int_{\mathbb{R}^2} {\rm d}\mathbf{r}\, \hat{w}(|\mathbf{r}|)=1$.
For simplicity, in this study we use a linear weight function, $\hat{w}(r)=\frac{3}{\pi R^2}(1-\frac{r}{R})$ if $r\le R$, and $0$ otherwise.

In order to keep the microscopic model minimal, we have neglected in Eq.~\eqref{eq:micro-model} the effect of steric repulsion between the particles, and assumed a instantaneous dependence of self-propulsion speed $v_S$ on the local density fields $\{\tilde{\rho}_S\}$.
These simplifications are discussed in Appendix~\ref{sec:repulsion}, where 
we moreover show that the presence of short-ranged soft repulsion does not prevent the emergence of the CCB pattern discussed below.

The functional form of the term $v_S(\tilde{\rho}_A,\tilde{\rho}_B)$  
appearing in Eq.~\eqref{eq:micro-model-a} 
determines the nature of the quorum-sensing interactions
and the ensuing collective dynamics.
For a single species, for example,  
motility-induced phase separation (MIPS) takes place 
when the speed $v(\tilde\rho)$ decays sufficiently fast as the local particle density increases.
The emergence of this collective behavior does not depend much on the exact form of $v(\tilde\rho)$~\cite{cates2015motility}, such that a broad class of models share the same qualitative features.
A similar picture holds for the two-species case, where the collective dynamics is also not very sensitive to the specific form of $v_{S}(\rho_A, \rho_B)$.
This is well illustrated by the linear instability analysis detailed in Sec.~\ref{sec:lin-stability-D}, which shows that under the local approximation
\begin{equation} \label{eq:local-approx}
    \tilde{\rho}_S(\mathbf{r},t) \simeq \rho_S(\mathbf{r},t)\equiv
    \sum_i \delta (\mathbf{r}-\mathbf{r}_{i,S}(t)),
\end{equation}
the linear stability of homogeneous configurations with global densities $(\bar{\rho}_A, \bar{\rho}_B)$ is controlled by the dimensionless couplings $\{\eta_{S U}(\bar{\rho}_A,\bar{\rho}_B)\}$, where
\begin{equation} \label{def:eta}
    \eta_{S U}(\rho_A, \rho_B) \equiv \rho_S \frac{\partial \ln v_S (\rho_A,\rho_B)}{\partial \rho_{U}},
\end{equation}
for $S,U \in \{A,B\}$.
Similarly to the single-species case, $\eta_{SU}$ characterizes how quickly the speed of species $S$ responds to local variations of the density of species $U$, 
such that the stability of the homogeneous phase does not depend on the specific form of $v_S(\rho_A, \rho_B)$. 
We thus assume the factorized expression
\begin{equation}\label{eq:v_S}
    v_{S}(\rho_A, \rho_B)=v_0 L_{SA}(\rho_A) L_{SB}(\rho_B),
\end{equation}
where $v_0$ is the nominal speed assumed equal for the two species, while $L_{SU}$ is a logistic function
\begin{equation}\label{def:L_SS'}
    L_{SU}(x)\equiv 1 + \kappa\tanh\left(
    \frac{\eta^0_{SU}}{\kappa}\frac{x-\rho_0}{\rho_0}
    \right),
\end{equation}
such that $v_S$ varies monotonously with both $\rho_A$ and $\rho_B$.
$\rho_0$ in~\eqref{def:L_SS'} is used as a reference density, yielding $v_S(\rho_0, \rho_0)=v_0$ and $\eta_{SU}(\rho_0, \rho_0)=\eta^0_{SU}$. % at the reference density value $\rho_0$.
Finally, the parameter $\kappa = 0.7$ ensures that $v_S$ remains strictly positive, 
thereby preventing the dynamics from freezing.

To understand how the couplings $\{\eta_{SU}\}$ defined in Eq.~\eqref{def:eta} induce effective interactions between the particles,
we note that active particles tend to accumulate in regions where they move slower~\cite{cates2015motility}.
Hence, 
for $\eta_{SU}<0$, particles from species $S$ shall accumulate in regions where the density of species $U$ is higher,
as if $S$ particles were attracted by $U$ particles.
On the other hand, ${\eta}_{SU}>0$ amounts for $S$ particles to experience an effective repulsion from $U$ particles.
For the same species $S$, a negative (positive) $\eta_{SS}$ thus leads to effective self-attraction (-repulsion), as a result of the self-inhibition (-activation) of their motility.
For two distinct species $A$ and $B$, the interactions can be non-reciprocal when $v_A$ varies differently with $\rho_B$ than $v_B$ with $\rho_A$.
In particular, effectively chasing interactions between species $A$ and $B$ are achieved when ${\eta}_{AB}{\eta}_{BA} < 0$~\cite{duan2023dynamical,dinelli2023non}.

In the following, we rescale space and time by setting both $R$ and $v_0$ to one, such that the remaining control parameters of the model are $\{\bar{\rho}_S/\rho_0\}$,
the ratio between the mean particle densities $\bar{\rho}_S$ and the reference density $\rho_0$
\footnote{In the mean-field approximation, the dynamics only depends on the ratio $\{\bar{\rho}_S/\rho_0\}$. %, irrespective of $\rho_0$. 
This remains true in simulations of the microscopic model when the particle densities are sufficiently large, thus ensuring limited density fluctuations in the evaluation of the coarse-grained densities $\tilde\rho_S$.},
the nominal couplings $\{\eta^0_{SU}\}$ and rotational diffusivity $D_r$.
For given $\{\bar{\rho}_S\}$ and $\rho_0$, we moreover define the P\'eclet number $\mathrm{Pe}\equiv v_B(\bar{\rho}_A,\bar{\rho}_B)/D_r$ to quantify self-propulsion strength,
while the parameter $\chi \equiv \eta_{AB}(\bar{\rho}_A,\bar{\rho}_B)\eta_{BA}(\bar{\rho}_A,\bar{\rho}_B)$ is used as a measure of the strength of nonreciprocity.

\section{The mean-field theory} \label{sec:continuous-theory}
\subsection{The continuous description of NRQS active particles} \label{sec:hydro_NRQS}
A continuum theory describing for the NRQS model can be derived from standard coarse-graining techniques in the mean field approximation.
Here, we briefly outline the derivation, and refer to Ref.~\cite{duan2023dynamical} for additional details.
The approach consists in writing the dynamical equation obeyed by the single-body distribution $P_S(\mathbf{r}, \theta, t)$ 
describing the probability that a particle of species $S$ has a position $\mathbf{r}$ and an orientation $\theta$ at time $t$.
In the local mean field approximation, this equation is closed by replacing the empirical densities $\tilde{\rho}_S(\mathbf{r},t)$ 
appearing in the quorum-sensing interactions with $\phi_S(\mathbf{r},t) \equiv \int_0^{2\pi}\mathrm{d}\theta P_S(\mathbf{r},\theta, t)$.
To further simplify the continuum description, we expand $P_S$ in its angular moments, 
and use the fact that the relevant long-time dynamics is well-captured by the low order moments.
We thus truncate the expansion by retaining only the first three angular moments, 
eliminate the third one using a quasi-stationary approximation, leading to the following pair of coupled equations for the  
density $\phi_S(\mathbf{r},t)$ and polarity $\mathbf{p}_S(\mathbf{r},t)\equiv \int_0^{2\pi}\mathrm{d}\theta\, \hat{\mathbf{u}}(\theta) P_S(\mathbf{r},\theta, t)$ fields:
\begin{subequations}\label{eq:rho-p}
    \begin{align}
\partial_t \phi_S = & -\nabla \cdot (v_S \mathbf{p}_S)\label{eq:rho-p-a},\\
		\partial_t \mathbf{p}_S = & 
		- \frac{1}{2} \nabla (v_S \phi_S) - D_r \mathbf{p}_S 
		+ \frac{v_S}{16D_r}\nabla^2 \left(v_S\mathbf{p}_S\right) \notag \\
		&+ \frac{1}{8D_r}\left[\nabla(v_S \mathbf{p}_S)\right]_{\rm ST} \cdot \nabla v_S,
		\label{eq:rho-p-b}
    \end{align}
\end{subequations}
where $\left[{\bf A}\right]_{\rm ST} \equiv \frac{1}{2}\left[{\bf A} + {\bf A}^{T} - {\bf I}{\rm Tr}({\bf A})\right]$ is the symmetric and traceless part of the tensor $\bf A$,
and for conciseness we kept the dependencies of $v_S$ in the densities $\phi_A$ and $\phi_B$ implicit. 

When the particle orientations dynamics are purely diffusive, as in Eq.~\eqref{eq:micro-model-b},
one generally uses the timescale separation between the density and polarity large-scale dynamics to 
enslave $\mathbf{p}_S$ to $\phi_S$, thus obtaining a field theory expressed solely in terms of the densities.
The resulting equations reasonably describe, e.g., MIPS for a single-species~\cite{cates2015motility}
and will be used in Sec.~\ref{sec:phase-equilibria} to predict the phase behavior of NRQS active particles with weak non-reciprocity.
However, we show in the following section that 
they fail to capture a short wavelength instability of the homogeneous solutions of Eqs.~\eqref{eq:rho-p} for the two-species case, 
which we find responsible for the emergence of finite-size traveling patterns observed in simulations of the microscopic model.

\subsection{Stability of homogeneous solutions}
\label{sec:instability_homo}

\subsubsection{Linear stability analysis}\label{sec:lin-stability-D}
\begin{figure*}
\centering
\includegraphics[width =\linewidth]{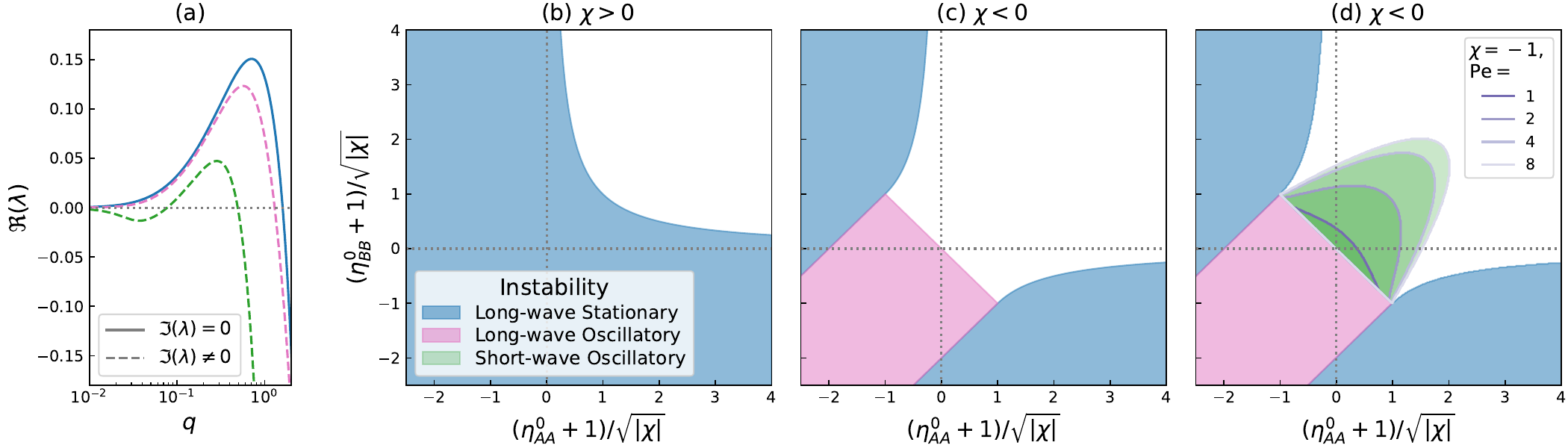}
\caption{Linear stability at equal composition $\bar{\rho}_A=\bar{\rho}_B=\rho_0$. 
(a) Typical dispersion relations $\lambda(q)$ for the three types of instabilities, 
solid lines correspond to $\Im(\lambda)=0$ and dashed lines to $\Im(\lambda)\neq 0$, the colour code is given in (b).
(b-d) Phase diagrams in the plane spanned by the normalized self-interaction couplings.
Regions where homogeneous configurations are stable appear in white.
(b) For $\chi>0$, only LSI can arise.
(c,d) When $\chi < 0$, both Eqs.~\eqref{eq:phi_mu}(c) and~\eqref{eq:rho-p}(d) predict the emergence of LOI.
Eqs.~\eqref{eq:rho-p} additionally exhibit SOI for sufficiently large $\chi$ and ${\rm Pe}$. 
Note that the region marking the existence of SOI, marked by the green areas with different transparency in (d),
enlarges when increasing $\mathrm{Pe}$.
}\label{fig:linear_stability_diagram}
\end{figure*}

We linearize Eqs.~\eqref{eq:rho-p} around the solution with homogeneous densities $(\bar{\rho}_A,\bar{\rho}_B)$ and vanishing polarities. 
We also assume equal composition, $\bar{\rho}_A=\bar{\rho}_B=\rho_0$, which allows to deal with simpler expressions,
while the general case leading to similar results was addressed in Ref.~\cite{duan2023dynamical}.
The stability of the homogeneous solution is determined by four control parameters:
$\chi=\eta_{AB}^0\eta_{BA}^0$ quantifying nonreciprocity, $\mathrm{Pe}=v_0/D_r$ measuring the self-propulsion strength,
and the normalized intra-species couplings $(\eta_{SS}^0+1)/\sqrt{|\chi|}$ with $S=A$, $B$.
The eigenvalue controlling the stability of the solution is~\cite{duan2023dynamical}
\begin{equation}
    \lambda (q) = -\frac{q}{2} \gamma(q) + \frac{q}{2}\sqrt{\gamma^2(q) -2\mu +2\sqrt{\alpha}},
\end{equation}
where $q$ is the wavenumber of the perturbation, while $\gamma(q)\equiv \frac{1}{\mathrm{Pe} q} + \frac{\mathrm{Pe}q}{16}$, 
$\mu\equiv 1+\frac{1}{2}(\eta_{AA}^0+\eta_{BB}^0)$,
 and $\alpha\equiv\chi +\frac{1}{4}(\eta_{AA}^0-\eta_{BB}^0)^2$.

As summarized in Fig.~\ref{fig:linear_stability_diagram}(a), $\lambda$ leads to three types of instabilities, which we refer to as
long-wave stationary (LSI) and oscillatory (LOI), or short-wave oscillatory (SOI).
When $\alpha>0$, $\lambda(q)$ is real as $q\to 0$, and LSI arises when $\mu<\sqrt{\alpha}$, i.e.\ $\mu <0$ or $(1 + {\eta}_{AA}^0)(1 + {\eta}_{BB}^0)<\chi$.
In turn, $\alpha = 0$ marks the presence of an exceptional point~\cite{saha2020scalar}, 
such that for $\alpha < 0$ the imaginary part of $\lambda(q)$ is always nonzero.
Its real part $\Re(\lambda)$ then becomes positive when~\cite{duan2023dynamical}
\begin{equation} \label{eq:pos_lambda_re}
    2\mu \gamma^2(q)+\alpha = 
    2\mu \left(
    \frac{1}{\mathrm{Pe} \, q} + \frac{\mathrm{Pe}\,q}{16}
    \right)^2+\alpha
    <0.
\end{equation}
For $\alpha<0$ and $\mu<0$, the condition \eqref{eq:pos_lambda_re} is always true down to $q=0$, giving rise to LOI.
The emergence of long-wave instabilities is thus primarily controlled by the self-interaction couplings $\eta_{SS}^0$,
and their nature (stationary or oscillatory) is set by the value of $\chi$, as shown in the phase diagrams of Figs.~\ref{fig:linear_stability_diagram}(b, c).
In particular, we note that both LSI and LOI can arise for arbitrary small $|\chi|$, and their existence is insensitive to the value of $\mathrm{Pe}$.

Equation~\eqref{eq:pos_lambda_re} further indicates that for $\alpha<0$ and $\mu>0$, although $\Re(\lambda)$ is always negative at small $q$, 
$2\mu\gamma^2(q) + \alpha$ may change sign at some finite wavenumber $q_c$ given by
\begin{equation*}
q_c^2 = -\frac{16}{{\rm Pe}^2}\left[ 1 + \frac{4 \alpha}{\mu} + 4\sqrt{\frac{\alpha}{\mu}\left(\frac{1}{2} + \frac{\alpha}{\mu}\right)} \right],
\end{equation*}
for $\alpha / \mu < -\tfrac{1}{2}$.
Since in this scenario $\lambda$ has a nonzero imaginary part ($\alpha < 0$), it corresponds to SOI.
In contrast to the conditions leading to LSI and LOI, 
$\alpha / \mu < -\tfrac{1}{2}$ is satisfied even when both species present weak or vanishing self-couplings,
while it requires sufficiently strong non-reciprocity.
Imposing further that $q_c < 2\pi/R$, where the interaction range $R$ serves as a short-scale cutoff for the continuous model, 
it is clear that SOI can only emerge when $\rm Pe$ is sufficiently large.
These features are summarized in the phase diagram of Fig.~\ref{fig:linear_stability_diagram}(d) plotted for $\chi = -1$.

\begin{figure*}
\centering
\includegraphics[width =\linewidth]{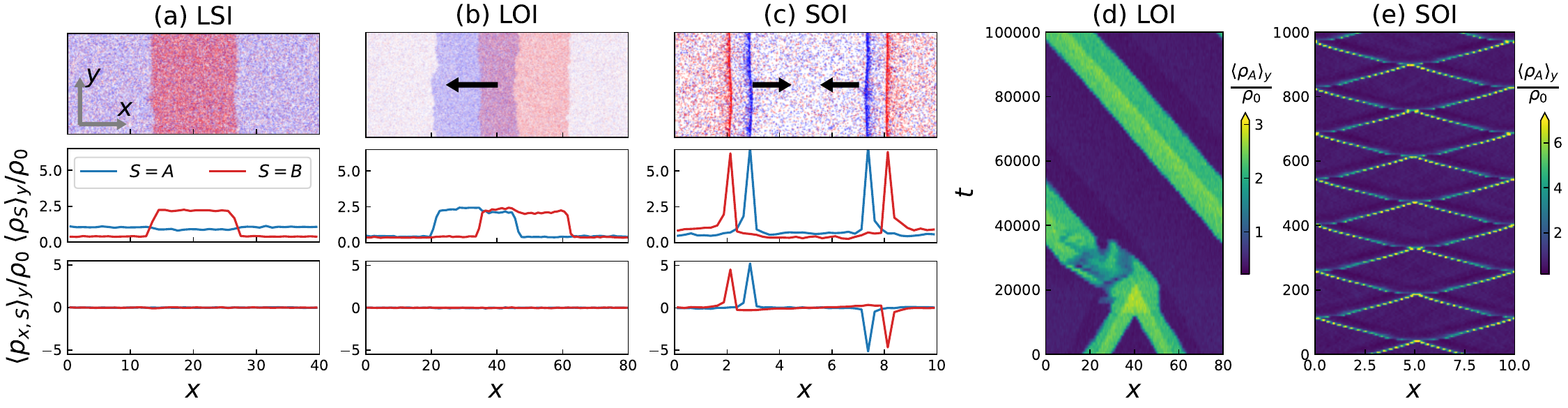}
\caption{The three types of patterns arsing in small systems from the instabilities reported in Fig.~\ref{fig:linear_stability_diagram}.
(a-c) From top to bottom: representative snapshot of the steady state,
the corresponding density and polarity profiles averaged over the vertical direction for stationary phase separation(a), 
wide(b) and thin(c) traveling bands.
On the top, the particles are coloured according to their species (blue for $A$ and red for $B$).
Arrows in (b, c) indicate the moving direction of the patterns.
(d, e) Kymographs showing the collision of two initially counter-propagating travelling bands induced by LOI(d) and SOI(e). 
Parameters: $D_r=0.1$ and $\bar{\rho}_A=\bar{\rho}_B=\rho_0=80$ in all cases, while 
(a) $\eta_{AA}^0=0$, $\eta_{BB}^0=-2$, $\eta_{AB}^0=-\eta_{BA}^0=0.1$, $L_x=2.5L_y=40$;
(b, d) $\eta_{AA}^0=\eta_{BB}^0=-2$, $\eta_{AB}^0=-\eta_{BA}^0=0.1$,  $L_x=80$, $L_y=32$ for (b) and $L_y=16$ for (d);
(c, e) $\eta_{AA}^0=\eta_{BB}^0=0$, $\eta_{AB}^0=-\eta_{BA}^0=1.8$, $L_x=2.5L_y=10$.
}
\label{fig:quasi-1D}
\end{figure*}

\subsubsection{Pattern classification at equal composition}\label{sec:quasi-1D}

To showcase the basic phenomenology that results from the instabilities identified above, 
we conducted numerical simulations of the NRQS model~\eqref{eq:micro-model} in small systems with equal composition for the two species.
The results presented here are in particular useful to apprehend the rich large-scale physics of the NRQS model explored in Secs.~\ref{sec:wo-chasing} and~\ref{sec:w-chasing}.
As shown in Fig.~\ref{fig:quasi-1D}(a-c), while LSI leads to static motility-induced phase separated clusters, 
both LOI and SOI induce partial demixing into dynamical states where macroscopic domains (LOI) or thin bands (SOI)
rich in either of the two species chase each other.
These last two types of traveling patterns, in particular, exhibit striking differences which we highlight below.

The cohesion of the phase-separated patterns emerging from LSI and LOI is ensured by sufficiently strong motility self-inhibition,
in line with the typical phenomenology of MIPS~\cite{cates2015motility}. 
For LOI, both species phase-separate and partially demix, forming a well defined interface between domains rich in $A$ and $B$ particles (Fig.~\ref{fig:quasi-1D}(b)).
In this case, the strong nonreciprocal couplings further drive the whole configuration to travel along the direction transverse to the interface,
the moving direction being determined by the relative signs of $\eta_{AB}^0$ and $\eta_{BA}^0$.
The width of the LOI traveling domains can be varied by changing the system size or the mean particle density, indicating the absence of size selection.
We illustrate this in Fig.~\ref{fig:quasi-1D}(d), which shows how the collision of two phase-separated domains results in the formation of a unique, wider, traveling domain. 
This behavior is reminiscent of results recently reported from studies of continuum descriptions such as the NRCH model~\cite{brauns2024nonreciprocal}.
We will make the connection between Eqs.~\eqref{eq:rho-p} and the NRCH model explicit in the following section. 

In contrast with the above, the SOI-induced chasing bands emerge even in the absence of intra-species couplings.
In this case, the recruitment of particles from the surrounding gas needed to balance the effect of rotational noise 
is thus operated via cross-species nonreciprocal interactions~\cite{duan2023dynamical}.
These structures moreover have a well-defined selected width and are separated by a finite spacing,
both of the order of the interaction range $R = 1$ (Fig.~\ref{fig:quasi-1D}(c)). 
Besides, the SOI-induced bands are characterized by the presence of strong local polarity, as pictured in the bottom panel of Fig.~\ref{fig:quasi-1D}(c).
The presence of this local polarity, in particular, constrains the spatial organization of these patterns.
Since the orientational dynamics of the particles is only determined by diffusion (Eq.~\eqref{eq:micro-model-b}), 
the spatially and noise-averaged polarities $\langle \mathbf{p}_{S}(\mathbf{r}, t)\rangle_{\mathbf{r}}$ must vanish in the steady state.
Hence, in small rectangular geometries such as in Fig.~\ref{fig:quasi-1D}(c), the bands settle in configurations involving an even number of pairs moving along opposite directions, such that the global polarities associated with both species are zero on average.
As we show in Sec.~\ref{sec:wo-inhibition}, larger systems allow for different configurations ensuring zero mean polarity, such as square or triangular band networks 
(see Fig.~\ref{fig:pure-NR} below).
The stability of these configurations then imposes that the thin chasing bands do not perturb each other upon crossings, contrary to LOI-induced domains.
This feature is illustrated in Fig.~\ref{fig:quasi-1D}(e), which shows that the SOI-induced bands indeed display a soliton-like behavior~\cite{tsyganov2003quasisoliton,yamanaka2014collision}, as they only weakly affect each others.

Although the above classification builds upon the results from the linear stability analysis, 
a complete characterization of the patterns arising in the NRQS model requires going beyond the linear level.
While this agenda is beyond the scope of this work, in the following section, we make a step forward by presenting an approximate 
derivation of the phase boundaries of static phase-separated configurations for a broad class of models of non-reciprocal mixtures
comprising the NRQS model~\eqref{eq:micro-model}.

\subsection{Generalized thermodynamics for the phase equilibria of non-reciprocal mixtures}\label{sec:phase-equilibria}
Predicting the phase boundaries of non-equilibrium phase-separating mixtures is a notoriously difficult task.
While, at equilibrium, one can rely on the common tangent or Maxwell equal area constructions,
the absence of a free energy minimization principle out-of-equilibrium prevents the use of these powerful tools.
In recent years, alternatives have been proposed in pursuit of a theory predicting the phase equilibria of MIPS in single-component active systems~\cite{takatori2015towards,solon2018generalized1,solon2018generalized2,paliwal2018chemical,hermann2021phase,speck2021coexistence,omar2023mechanical}. 
For multi-component systems, on the other hand, a general theory of phase coexistence is still lacking.
Despite recent efforts offering enlightening insights into the phase equilibria of nonreciprocal mixtures~\cite{brauns2024nonreciprocal,greve2024maxwell,saha2024phase},
the range of application of these works is restrained to simple continuum models unable to capture the physics of NRQS active particles.  

\subsubsection{The NRCH-like continuum description}

As shown in the bottom row of Fig.~\ref{fig:quasi-1D}(a,b), the phase-separated configurations originating from LSI and LOI do not exhibit any local polar order.
Therefore, we seek to describe them from a simplified version of the continuous description~\eqref{eq:rho-p}.
Namely, treating $\mathbf{p}_S$ in the quasi-static regime, we get $\mathbf{p}_S=-\frac{1}{2D_r}\nabla(v_S \phi_S)$ from Eq.~\eqref{eq:rho-p-b}.
Substituting this expression into Eq.~\eqref{eq:rho-p-a} then yields
\begin{equation} \label{eq:phi_mu}
    \partial_t \phi_S = \nabla\cdot (M_S\nabla \mu_S),
\end{equation}
where the collective mobility $M_S=\phi_Sv_S^2/{2D_r}$, while the effective chemical potential $\mu_S$ is given by
\begin{equation} \label{eq:mu_S}
    \mu_S(\phi_A, \phi_B) = \ln[ \phi_S v_S(\phi_A, \phi_B)].
\end{equation}
Note that, within the local approximation~\eqref{eq:local-approx}, $\mu_S$ is a function of the local densities only.

Equation~\eqref{eq:phi_mu} formally corresponds to the NRCH model proposed on phenomenological grounds in Refs.~\cite{saha2020scalar,you2020nonreciprocity}.
Here, however, the effective chemical potential~\eqref{eq:mu_S} is obtained by coarse-gaining the microscopic model~\eqref{eq:micro-model}, 
thereby linking the microscopic and macroscopic physics.
In fact, the r.h.s.\ of Eq.~\eqref{eq:phi_mu} coincides with the deterministic part of the fluctuating hydrodynamic description derived in Refs.~\cite{dinelli2023non,dinelli2024fluctuating}
from a similar particle-based model, such that it consistently corresponds to the mean field large-scale description of the dynamics.

The derivation of Eq.~\eqref{eq:phi_mu} indeed relies on a time scale separation between the dynamics of $\phi_S$ and $\mathbf{p}_S$, 
which formally holds only over macroscopic scales\footnote{
Formally, the relaxation time $\tau_\phi$ of a density perturbation of wavelength $\ell$ scales as $\ell^2$, while perturbations of the polarity
are all damped on a finite timescale $\propto D_r^{-1}$. In large systems, the density dynamics is then typically much slower than that of the polarity.}.
Because LSI and LOI take place at arbitrary large scales, Eq.~\eqref{eq:phi_mu} provides predictions for their emergence identical to that of Eqs.~\eqref{eq:rho-p}.
However, the reduced NRCH-like description~\eqref{eq:phi_mu} fails to capture the onset of SOI at large $\rm Pe$ and $\chi$ (compare panels (c) and (d) of Fig.~\ref{fig:linear_stability_diagram}).
Since SOI arises only over finite scales, the quasi-static approximation for $\mathbf{p}_S$ is indeed generally not verified in this case.

Nevertheless, the reduced model~\eqref{eq:phi_mu} constitute a simpler description of the phase-separated configurations emerging from long-wave instabilities. 
In the remaining of this section, we further show how it enables analytical progress for the prediction of the corresponding binodal densities.

\subsubsection{Departure from equilibrium at the coarse-grained scales}\label{sec:rho_eq_decompo}
The model~\eqref{eq:phi_mu} breaks detailed balance by allowing for asymmetric inter-species couplings.
To highlight the departure from equilibrium, we thus express the effective chemical potentials as
\begin{equation} \label{eq:mu_decomp}
    \mu_A = \partial_{\phi_A} (f^\mathrm{eq}+f^\mathrm{NR}), \quad
    \mu_B = \partial_{\phi_B} (f^\mathrm{eq}-f^\mathrm{NR}),
\end{equation}
where $f^\mathrm{eq}(\phi_A,\phi_B)$ acts as an equilibrium-like free energy density, 
while the anti-symmetric part $f^\mathrm{NR}(\phi_A,\phi_B)$ accounts for the presence of nonreciprocity.
We show in Appendix~\ref{sec:decomp-f} that for general $\mu_A$ and $\mu_B$ these functions take the form
\begin{subequations}\label{eq:feq_fNR}
    \begin{align}
        f^\mathrm{eq}(\phi_A,\phi_B) &=\int^{\phi_A}\mathrm{d}x \, \mu_A(x,\phi_B)+\int^{\phi_B}\mathrm{d}y \, \mu_B(\phi_A,y) \notag\\
        &-\int^{\phi_A}\mathrm{d}x\int^{\phi_B}\mathrm{d}y\, g_+(x,y), \label{eq:f_eq} \\
        f^\mathrm{NR}(\phi_A,\phi_B)&=\int^{\phi_A}\mathrm{d}x\int^{\phi_B}\mathrm{d}y \, g_-(x, y),\label{eq:fNR}
    \end{align}
\end{subequations}
where $g_\pm (\phi_A,\phi_B)\equiv \frac{1}{2}\left(\partial_{\phi_B} \mu_A \pm \partial_{\phi_A} \mu_B\right)$.
The macroscopic dynamics is therefore equilibrium-like whenever $g_-(\phi_A,\phi_B)=0$ for all $\phi_A$ and $\phi_B$.
This is for example the case when the inter-species couplings arise through identical dependencies of the chemical potentials in the total density,
such that $\mu_S(\phi_A,\phi_B) = \mu_{SS}(\phi_S) + \mu_{\rm tot}(\phi_A + \phi_B)$ for $S = A, B$~\cite{dinelli2023non}.
As we discuss in the following section, the phase behavior can then be predicted from the familiar common tangent construction on $f^{\rm eq}$.

Using the chemical potential~\eqref{eq:mu_S}, for the choice of motility-density coupling considered in this work (Eq.~\eqref{eq:v_S}),
the generalized free energies read
\begin{subequations} \label{eq:fS_fR_fNR-v_LSA_LSB}
    \begin{align}
    f^\mathrm{eq} &= f_{AA}(\phi_A) + f_{BB}(\phi_B) \notag,\\
            &+\frac{1}{2}\left[
            \phi_A \ln L_{AB}(\phi_B) + \phi_B\ln L_{BA}(\phi_A)
            \right], \label{eq:f_eq_v_LSA_LSB}\\
        f^\mathrm{NR} &= \frac{1}{2}\left[
            \phi_A \ln L_{AB}(\phi_B) - \phi_B\ln L_{BA}(\phi_A)
            \right], \label{eq:f_NR_v_LSA_LSB}
    \end{align}
\end{subequations}
where $f_{SS}(\phi_S)=\phi_S (\ln \phi_S -1) + \int^{\phi_S} \mathrm{d}x \ln L_{SS}(x)$, 
and $L_{SU}(x)$ is the logistic function defined in Eq.~\eqref{def:L_SS'}.
For the choice of inter-species couplings defined by Eqs.~\eqref{eq:v_S} and~\eqref{def:L_SS'}, 
the anti-symmetric part $f^\mathrm{NR}$ therefore only vanishes when $\eta_{AB}^0=\eta_{BA}^0=0$, i.e., 
when the dynamics of the two species are decoupled.
Note that, even when $\eta_{AB}^0=\eta_{BA}^0\neq 0$, 
$f^\mathrm{NR}$ cancels only when $\phi_A = \phi_B$, such that, even in this case, nonreciprocity generally prevails at macroscopic scales.

\subsubsection{The generalized common tangent construction} \label{sec:maxwell}
In this section, we consider a configuration involving the coexistence of $m$ phases with densities $(\phi_A^\alpha, \phi_B^\alpha)$ for $\alpha = 1, \ldots, m$.
In the limit of large systems, the higher order gradient terms that may enter the expression of $\mu_S$ will be subdominant, 
such that we neglect them in this discussion.
The characterization of the phase-coexistence then consists in determining the $2m$ densities and $m$ volumes of each phase under the constraints of total mass and volume conservation, leading to $3(m - 1)$ independent unknowns. 
We obtain $2(m-1)$ relations from the 
steady state condition on Eq.~\eqref{eq:phi_mu}, which imposes that both chemical potentials are constant across the phases:
    \begin{align}
        \mu_S(\phi_A^\alpha, \phi_B^\alpha)&=\bar{\mu}_S \quad \forall \alpha \in \{1,\ldots,m\}. \label{eq:equal-mu}
    \end{align}
At equilibrium, $f^\mathrm{NR} = 0$ and the remaining $m-1$ relations follow from the equality of the pressure among phases.
Namely, writing $\mu_S = \partial_{\phi_S} f^{\rm eq}$,
the thermodynamic pressure reads $p(\phi_A, \phi_B)=\phi_A \mu_A +\phi_B \mu_B - f^\mathrm{eq}$, and we obtain
\begin{equation}\label{cd:equal-p}
    p(\phi_A^\alpha, \phi_B^\alpha) = \bar p \quad \forall \alpha \in \{1,\ldots,m\}.
\end{equation}
Taken together with the total volume and mass conservation, Eqs.~\eqref{eq:equal-mu} and~\eqref{cd:equal-p} then fully determine the phase behavior.
In geometrical terms, solving these equations amounts to construct the plane tangent to the surface defined by $f^{\rm eq}(\phi_A,\phi_B)$ with an orientation determined by the vectors ${\bf t}_A = (1,0,\bar\mu_A)$ and ${\bf t}_B = (0,1,\bar\mu_B)$, and with intercept at the origin given by $-\bar{p}$.
Although~\eqref{eq:equal-mu} remains valid to describe stationary configurations out of equilibrium, the definition of the pressure relies on the fact that both chemical potentials derive from the same free energy,
such that it breaks down when $f^{\rm NR} \ne 0$.

To derive an alternative to~\eqref{cd:equal-p}, we take inspiration from Refs.~\cite{solon2018generalized1,solon2018generalized2}
and seek a mapping to pseudo-variables $\{\psi_S(\phi_A, \phi_B)\}_{S=A,B}$ allowing to express the stationary condition on the phases 
in terms of the minimization of a generalized free energy density $V$ with respect to $\psi_S$.
While the existence of such mapping is not guaranteed in general, below we give explicit solutions for $\psi_S$ for two broad classes of problems.

To simplify the discussion, we now consider two coexisting phases $\alpha$ and $\beta$ separated by a thin flat interface along the $y$ direction,
such that the densities only vary along $x$.
In particular, we look for pseudo-variables $\psi_S$ and a potential $V$ satisfying
\begin{equation}\label{def:psi}
    \frac{\partial V(\psi_A, \psi_B)}{\partial \psi_S} = \mu_S(\phi_A, \phi_B),
\end{equation}
for both species, and where for conciseness we omit to specify that both $\phi_{A,B}$ are functions of $(\psi_A,\psi_B)$.
Equation~\eqref{def:psi} directly implies that $\partial_x V (\psi_A, \psi_B)= \mu_A\partial_x \psi_A + \mu_B \partial_x \psi_B$
which, when integrated across the interface separating the two phases, yields
\begin{equation}\label{eq:equal-h0}
        h(\psi_A^\alpha, \psi_B^\alpha) = h(\psi_A^\beta, \psi_B^\beta) = \bar{h},
\end{equation}
where we have defined the pseudo-pressure
\begin{equation}\label{def:h0}
    h(\psi_A, \psi_B) = \psi_A \frac{\partial V}{\partial \psi_A} + \psi_B\frac{\partial V}{\partial \psi_B} -V(\psi_A, \psi_B).
\end{equation}
Generalizing the argument to $m$ phases, Eq.~\eqref{eq:equal-h0} gives $m-1$ relations and thus provides an alternative to the pressure balance~\eqref{cd:equal-p}
for nonequilibrium dynamics. 
Note that this is only true when the relations~\eqref{eq:equal-h0} and~\eqref{eq:equal-mu} are independent, which further constrains the mapping.
For example, although $V = \psi_A\psi_B$ together with $\psi_A=\mu_B$ and $\psi_B=\mu_A$ always verify Eq.~\eqref{def:psi},
the corresponding relation~\eqref{eq:equal-h0} reads $\bar{h} = \bar{\mu}_A\bar\mu_B$, and is not independent of~\eqref{eq:equal-mu}.

As observed previously~\cite{solon2018generalized1,solon2018generalized2}, although $V$ and $h$ play analogue roles to the free energy and the pressure in the generalized thermodynamics mapping,
they do not bear a straightforward physical interpretation.
Furthermore, we also note that the mapping to pseudo-variables does not correspond to a reformulation in terms of an effective equilibrium dynamics,
since it leads to a force in Eq.~\eqref{eq:phi_mu} equal to $-\nabla \mu_S = -\nabla \partial_{\psi_S}V$,
different from the expression $-\nabla \partial_{\phi_S}V$ necessary to ensure that detailed balance is verified~\cite{cates2022active}.
  
\subsubsection{Application to the NRQS model}
As we have mentioned above, for general expressions of $\mu_S(\phi_A, \phi_B)$ the existence of pseudo-variables satisfying~\eqref{def:psi} is not guaranteed.
Indeed, Schwarz's theorem imposes the consistency condition
\begin{equation} \label{eq:find-RA-RB}
    \frac{\partial \mu_A(\phi_A, \phi_B)}{\partial \psi_B} =  \frac{\partial\mu_B(\phi_A, \phi_B)}{\partial \psi_A},
\end{equation}
which can be used to determine the pseudo-variables $\psi_S$.
Equation~\eqref{eq:find-RA-RB} is not straightforward to solve in general,
but it can be substantially simplified by assuming that the pseudo-variables $\psi_S$ are only function of $\phi_S$, 
such that $\partial_{\phi_{U}}\psi_S = 0$ for $S\neq U$.
In this case, it reduces to
\begin{equation}\label{eq:find-RA-RB-2}
    \psi'_A(\phi_A) \frac{\partial\mu_A(\phi_A, \phi_B)}{\partial \phi_B} = %\frac{\partial \phi_B}{\partial \psi_B} =
    \psi'_B(\phi_B)\frac{\partial\mu_B(\phi_A, \phi_B)}{\partial \phi_A}, %\frac{\partial \phi_A}{\partial \psi_A}.
\end{equation}
where primes denote derivatives w.r.t.\ $\phi_S$.
Decomposing the chemical potentials in terms of self- and cross-species couplings as $\mu_S= f_{SS}'(\phi_S) + \partial_{\phi_S} f_{SU}(\phi_A, \phi_B)$ (with $S\neq U$),
it is clear from Eq.~\eqref{eq:find-RA-RB-2}, that the intra-species terms $f_{SS}(\phi_S)$ play no role in determining $\psi_S$.

Equation~\eqref{eq:find-RA-RB-2} admits solutions for two classes of problems.
In the first one, the inter-species couplings $f_{SU}$ both derive from the same function:
\begin{equation} \label{eq:solvable-NRCH-1}
    f_{SU}(\phi_A, \phi_B)=\alpha_{SU}\, g(\phi_A, \phi_B)
\end{equation}
up to some coefficients $\{\alpha_{SU}\}$ which can be different.
When the function $g(\phi_A, \phi_B)$ satisfies $\partial^2_{\phi_A\phi_B} g = 0$, the dynamics is equilibrium-like and the mapping is trivial.
In the nontrivial case, a solution of Eq.~\eqref{eq:find-RA-RB-2} reads
\begin{equation} \label{eq:R-solution-1}
\psi_A (\phi_A) = \alpha_{BA}\phi_A \quad \mathrm{and}\quad \psi_B (\phi_B) = \alpha_{AB}\phi_B.
\end{equation}

The second class of problems for which Eq.~\eqref{eq:find-RA-RB-2} can be solved corresponds to situations where $f_{SU}$ take the factorized form:
\begin{equation}\label{eq:solvable-NRCH-2}
    f_{SU}(\phi_A, \phi_B) = g_{SA}(\phi_A) g_{SB}(\phi_B).
\end{equation}
Substituting Eq.~\eqref{eq:solvable-NRCH-2} into Eq.~\eqref{eq:find-RA-RB-2}, we find that the pseudo-variables must satisfy
\begin{equation} \label{eq:RA_RB_gSA_gSB}
    \frac{\psi_A'(\phi_A)g'_{AA}(\phi_A)}{g'_{BA}(\phi_A)}
    = \frac{\psi_B'(\phi_B)g'_{BB}(\phi_B)}{g'_{AB}(\phi_B)}.
\end{equation}
Given that the two sides of Eq.~\eqref{eq:RA_RB_gSA_gSB} depend only on $\phi_A$ and $\phi_B$, respectively, 
the equality imposes that both are equal to the same constant,
which leads to a solution
\begin{equation}\label{eq:R-solution-2}
\psi_S(\phi_S) = \int^{\phi_S} \mathrm{d}x\, \frac{g'_{US}(x)}{g'_{SS}(x)},
\end{equation}
for $U \ne S$.

The particular case of linear cross-species couplings $f_{SU}=\alpha_{SU}\phi_S\phi_{U}$ satisfies both Eqs.~\eqref{eq:solvable-NRCH-1} and~\eqref{eq:solvable-NRCH-2}, 
such that in this case the solution~\eqref{eq:R-solution-2} reduces to~\eqref{eq:R-solution-1}.
This scenario corresponds to the linearly coupled NRCH model recently studied in~\cite{brauns2024nonreciprocal,greve2024maxwell,saha2024phase}.
Equations~\eqref{eq:R-solution-1} and~\eqref{eq:R-solution-2}, in turn, provide systematic mappings for a wider class of nonlinear problems.
In particular, the generalized free energies~\eqref{eq:fS_fR_fNR-v_LSA_LSB} obtained by coarse-graining the NRQS active particle model
correspond to inter-species couplings $f_{S U}=\phi_S \ln [L_{S U}(\phi_{U})]$, which fall into the class defined by Eq.~\eqref{eq:solvable-NRCH-2}. 
We then immediately obtain from Eq.~\eqref{eq:R-solution-2}
\begin{equation}\label{eq:solution-R-NRQS}
    \psi_A(\phi_A)= \ln L_{BA}(\phi_A),\quad
    \psi_B(\phi_B)= \ln L_{AB}(\phi_B),
\end{equation}
while the pseudo-potential defined by Eq.~\eqref{def:psi} is given by
\begin{align}\label{eq:psi_NRQS}
    V(\psi_A, \psi_B) &= \psi_A \psi_B +\int^{\psi_A} \mathrm{d}x\, \ln[\phi_A(x)L_{AA}(\phi_A(x))] \nonumber \\
    &+ \int^{\psi_B} \mathrm{d}x\, \ln[\phi_B(x) L_{BB}(\phi_B(x))].
\end{align}

Equations~\eqref{eq:solution-R-NRQS} and~\eqref{eq:psi_NRQS} can be directly used to predict the coexisting densities of stationary phase-separated configurations such as those arising from LSI in numerical simulations of the NRQS model (Fig.~\ref{fig:quasi-1D}(a)).
Note that the extension of the approach to the description of the traveling domains emerging from LOI is not straightforward, since in this case the equalities~\eqref{eq:equal-mu}
obtained from the sationnary condition on Eq.~\eqref{eq:phi_mu} need not be verified.
Furthermore, as we will discuss in Sec.~\ref{sec:w-chasing}, these traveling configurations become generically unstable in sufficiently large systems, leading to a more intricate dynamics.
In the following section, we therefore confront the predictions for the coexisting densities obtained from the generalized common tangent construction to direct numerical simulations of the microscopic model in the regime of weak non-reciprocity, i.e. for which only stationary phase-separated configurations exist.

\section{Phase behavior at weak non-reciprocity}\label{sec:wo-chasing}
From now on, we systematically explore the phase behavior of the NRQS model~\eqref{eq:micro-model} upon varying the system's composition and interaction couplings.
In this section, we start by investigating the regime of weak-nonreciprocity for which only LSI ---and therefore stationary patterns--- emerge.
As we have discussed in Sec.~\ref{sec:instability_homo} for the case of equal composition, and can be generalized to the general case of unequal compositions~\cite{duan2023dynamical}, this regime amounts to imposing $\eta_{AB}^0\eta_{BA}^0\geq 0$, such that effective chasing interactions between the two species are prohibited.

\begin{figure}
\centering
\includegraphics[width =\linewidth]{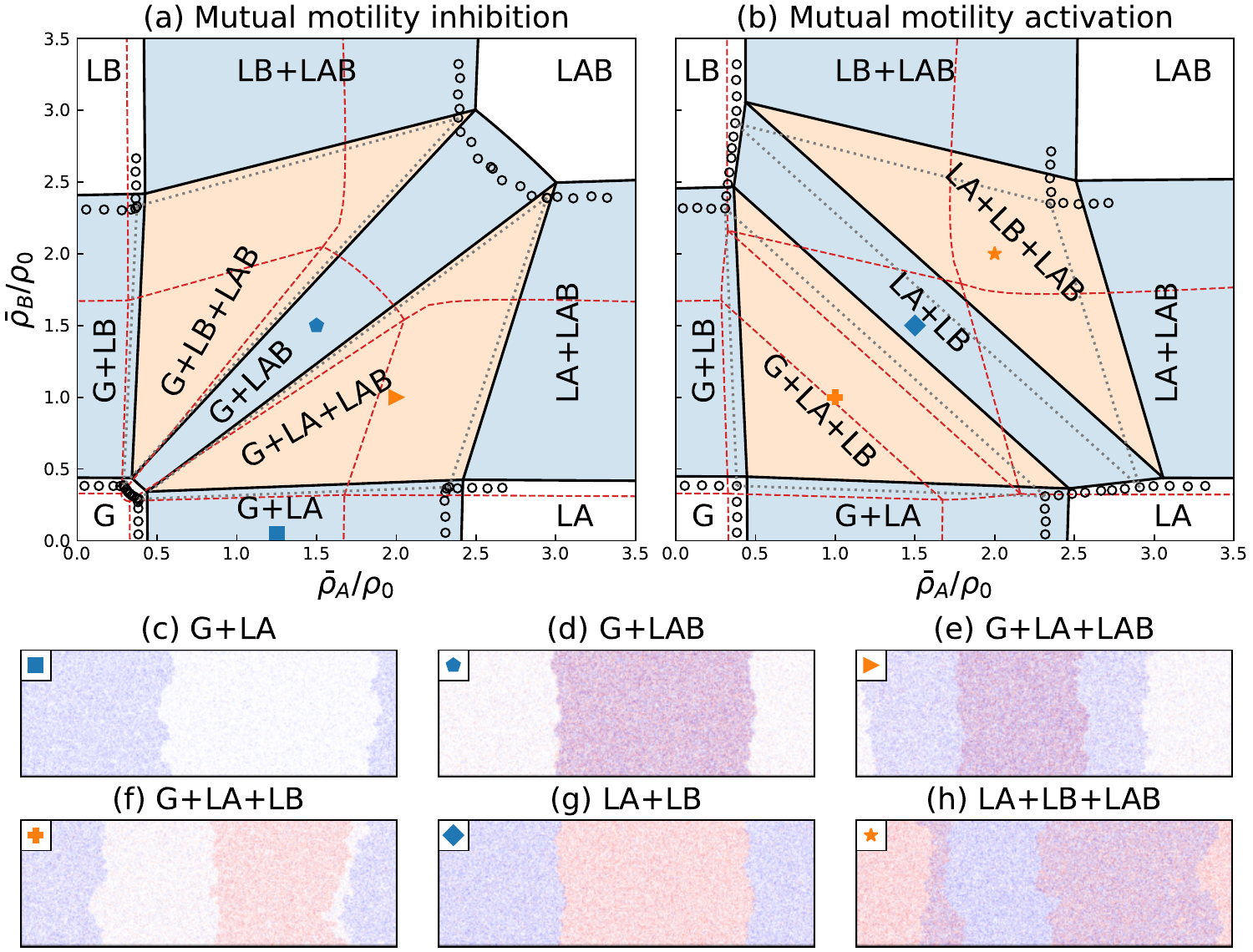}
\caption{Phase diagrams for the NRQS model with mutual motility inhibition $\eta_{AB}^0=\eta_{BA}^0=-0.1$(a) and activation $\eta_{AB}^0=\eta_{BA}^0=0.1$(b), respectively.
In both cases, $\eta_{AA}^0=\eta_{BB}^0=-2$, $D_r = 1$ and $\rho_0 = 10$.
Homogeneous, two-phase and three-phase coexistence phases are coloured in white, blue and orange, respectively.
The black solid lines indicate the binodal densities predicted by the generalized common tangent construction using the criteria~\eqref{eq:equal-mu} and~\eqref{eq:equal-h0}, 
while the red dashed lines correspond to the equilibrium construction using~\eqref{eq:equal-mu} and~\eqref{cd:equal-p}. 
The hollow circles mark the coexisting densities obtained from simulations of the microscopic model~\eqref{eq:micro-model}.
(c-h) Representative snapshots obtained for the cases with mutual inhibition(c-e) and activation(f-h), with system sizes $L_x=3L_y=120$.}  
\label{fig:mutual-attr-rep}
\end{figure}

To induce the spontaneous aggregation of both species, we fix $\eta_{AA}^0=\eta_{BB}^0=-2$, while we set for simplicity $\eta_{AB}^0=\eta_{BA}^0$. 
Figures~\ref{fig:mutual-attr-rep}(a) and (b) show the resulting phase diagrams in the composition plane when the effective interactions between the species are mutually attractive ($\eta_{AB}^0 < 0$) and repulsive ($\eta_{AB}^0 > 0$), respectively.
In both cases, the phase diagrams feature four stable homogeneous phases that are found at the corners of the composition plane, and denoted as G (gas), LA (liquid rich in $A$ particles), LB (liquid rich in $B$ particles) and LAB (mixed liquid rich in both $A$ and $B$ particles).
As a result of LSI, the system can phase separate into configurations involving the coexistence of two or three of these phases, 
as pictured in the simulation snapshots of Figs.~\ref{fig:mutual-attr-rep}(c-h).

Although this phase behavior closely resembles that of multi-component phase separation at equilibrium, the numerically determined binodal densities do not match with those predicted by the common tangent construction on the symmetrized free energy $f^{\rm eq}$ defined by Eqs.~\eqref{eq:equal-mu} and~\eqref{cd:equal-p} (compare the black open symbols and the dashed red lines in Figs.~\ref{fig:mutual-attr-rep}(a,b)).
As discussed above, this discrepancy is due to the fact that the asymmetric contribution $f^{\rm NR}$ is generally not vanishing (Eq.~\eqref{eq:f_NR_v_LSA_LSB}),
leading to the breakdown of the equilibrium construction.

Conversely, the binodal curves depicted by solid black lines in Figs.~\ref{fig:mutual-attr-rep}(a,b) are predicted by the generalized common tangent construction outlined in Sec.~\ref{sec:maxwell}, therefore determining the binodal densities from Eqs.~\eqref{eq:equal-mu} and~\eqref{eq:equal-h0}, together with the pseudo-variables defined in~\eqref{eq:solution-R-NRQS} and~\eqref{eq:psi_NRQS}.
Comparing these theoretical predictions with the binodal densities determined in numerical simulations of the microscopic dynamics (marked by the open symbols in Figs.~\ref{fig:mutual-attr-rep}(a, b)) reveals very good agreement.
Hence, combining the mean field coarse-graining approach with the generalized common tangent construction,
we can quantitatively predict the phase behavior of the NRQS model in the weakly nonreciprocal regime, without requiring any fitting parameter.
Turning on chasing interactions, our approach stills allows to predict the coexisting densities of stationary phase separated configurations (Fig.~\ref{fig:rho_v_profile}(a)). 
However, as we discuss in the following section, those are restricted to limited regions of the phase diagram, as most of the phase behavior is dominated by the emergence of dynamical patterns.

\section{Phase behavior at strong nonreciprocity} \label{sec:w-chasing}
As summarized by Figs.~\ref{fig:linear_stability_diagram} and~\ref{fig:quasi-1D}, turning on the effective chasing interactions ($\eta_{AB}^0\eta_{BA}^0<0$) enables the emergence of oscillatory behavior~\cite{saha2020scalar,you2020nonreciprocity,fruchart2021non}.
Hence, we refer to this regime as being strongly nonreciprocal.
We have shown in Sec.~\ref{sec:instability_homo} that nonreciprocity in this context can lead to two types of instabilities producing different dynamical patterns.
While LOI relies on sufficiently strong intra-species motility inhibition and leads to traveling phase-separated domains,
SOI emerges solely from inter-species nonreciprocity and drives the formation of finite-size chasing bands.
To distinguish the separate effects of nonreciprocity and aggregation via self-inhibition of motilities,
we first explore in Sec.~\ref{sec:wo-inhibition} the large-scale behavior of the SOI-induced patterns in the absence of intra-species couplings.
We then turn on intra-species motility inhibition in Sec.~\ref{sec:w-inhibition}, and characterize how the interplay of self-propulsion, cross-species nonreciprocity and self-induced aggregation structure the collective behavior of NRQS active particles. 

\subsection{Absence of intra-species motility inhibition}\label{sec:wo-inhibition}

\begin{figure}
\centering
\includegraphics[width =\linewidth]{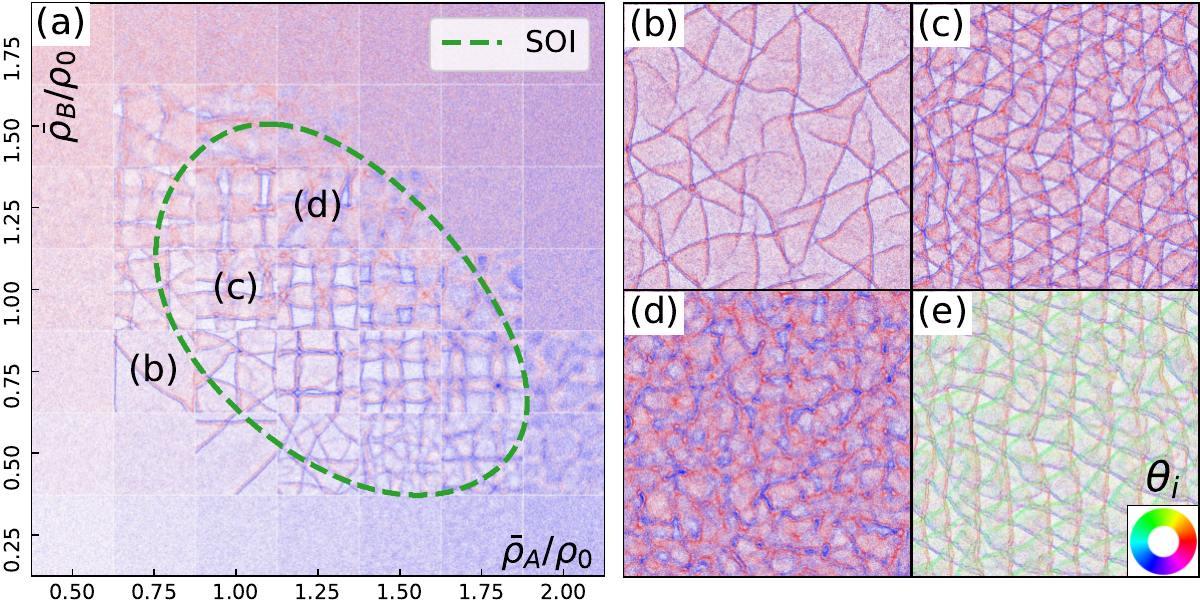}
\caption{Phase behavior in the composition plane for $\eta_{AA}^0=\eta_{BB}^0=0$, $\eta_{AB}^0=-\eta_{BA}^0=1$, $D_r=0.1$, and $\rho_0=80$. (a) Snapshots obtained from microscopic simulations in square domain of size $L_x=20$. 
The green dashed line encloses region where the linear stability analysis of %homogeneous solutions of 
Eqs.~\eqref{eq:rho-p} predicts SOI.
% the boundary of SOI predicted by the linear stability analysis for Eq.~\eqref{eq:rho-p}. 
(b-d) Simulations in square domains of size $L_x=80$, with composition $(\bar{\phi}_A/\rho_0,\bar{\phi}_B/\rho_0)=(0.75, 0.75)$, $(1, 1)$ and $(1.25, 1.25)$, respectively. 
(e) Same as (c), except that the particles are colored according to their moving direction $\theta_i$.
}
\label{fig:pure-NR}
\end{figure}

In this section, we therefore set $\eta_{AA}^0=\eta_{BB}^0=0$, such that quorum-sensing interactions only take place between particles belonging to different species.
As we have shown in Sec.~\ref{sec:instability_homo}, when the effective cross-species interactions are chase-and-run-like, this scenario allows for the emergence of band patterns at strong enough self-propulsion and nonreciprocity (see Fig.~\ref{fig:quasi-1D}(c))~\cite{duan2023dynamical}. 
These patterns, in particular, arise from SOI and are thus not captured by the mapping~\eqref{eq:phi_mu} to the NRCH model.

We now explore the phase behavior in the composition plane at fixed $D_r = 0.1$ and $\eta_{AB}^0=-\eta_{AB}^0=1$ in small square domains of size $L_x=20$.
As shown in Fig.~\ref{fig:pure-NR}(a), homogeneous configurations are unstable when both $\bar\rho_{A,B} \approx \rho_0$, leading to a diversity of patterns of thin traveling bands.
The emergence of these patterns moreover closely overlaps with the region encircled by the 
green dashed line in Fig.~\ref{fig:pure-NR}(a), marking the apparition of SOI.
We recall that the strength of nonreciprocal interactions is measured by the parameter $\chi\equiv \eta_{AB}(\bar{\rho}_A,\bar{\rho}_B)\eta_{BA}(\bar{\rho}_A,\bar{\rho}_B)$.
From the definition of $\eta_{SU}$ given in Eq.~\eqref{def:eta},
it is clear that for fixed nominal couplings $\{\eta_{S U}^0\}$ the value of $|\chi|$ is maximal when both densities $\bar\rho_{A,B} \approx \rho_0$.
Hence,  
the traveling band patterns typically emerge in regions of strong effective chasing interactions.

Beyond the linear level, Fig.~\ref{fig:pure-NR}(a) reveals that for moderate densities the chasing bands generally self-organize into large-scale triangular or square networks.
These networks are formed by the superposition of regular smectic arrangements of traveling bands, reminiscent of patterns recently observed in simulations of flocking models~\cite{kursten2020dry,chate2020dry} in a different context.
Similarly to the results obtained in a small quasi-one-dimensional domain in Fig.~\ref{fig:quasi-1D}(c), both triangular and square band networks preserve a vanishing spatially-averaged polar order, as illustrated in Fig.~\ref{fig:pure-NR}(e).
However, the origin of the mechanism leading to the selection of the symmetries of these macro structures remains so far elusive.

Focusing on the diagonal line in the composition plane, we carry out simulations in larger square domains of size $L_x=80$.
As shown in Figs.~\ref{fig:pure-NR}(b-d), the typical number of chasing bands increases with the species densities.
This feature is expected when the density of the gas in which the bands cruise does not vary much with the composition.  
In particular, we find that the network pattern is most regular at moderate densities $\approx \rho_0$ (Fig.~\ref{fig:pure-NR}(c)).
As lower densities, in contrast, the system does not contain enough particles to form system-spanning bands, leading them to break easily (Fig.~\ref{fig:pure-NR}(b)).
On the other hand, configurations at high densities are too crowded to accommodate regular arrangements of bands, leading to the emergence of chaotic chasing bands (CCB) (Fig.~\ref{fig:pure-NR}(d)).

Be they regular of chaotic, the dynamical patterns induced by SOI remain globally homogeneous at large scales, so that they do not correspond to a regime of phase separation. We will confirm this in Sec.~\ref{sec:coarsening}, where we show that the coarsening dynamics of the CCB band pattern is arrested.

\subsection{With intra-species motility inhibition}\label{sec:w-inhibition}
As discussed in Sec.~\ref{sec:wo-chasing}, sufficiently strong intra-species motility inhibition induces bulk phase separation, with a phase diagram resembling that of equilibrium mixtures in the weakly nonreciprocal regime.
Here, we explore the phase behavior of the model for strong nonreciprocity ($\eta_{AB}^0\eta_{BA}^0< 0$), leading to effective chasing interactions. 
In particular, we illustrate how the interplay of bulk-phase separation and the SOI-induced chasing bands leads to a rich phenomenology, which can nevertheless be understood as bulk-phase separation involving the phase of chaotic chasing bands.
Hence, we set $\eta_{AA}^0=\eta_{BB}^0=-2$ and focus on the case with antisymmetric inter-species couplings: $\eta_{AB}^0=-\eta_{BA}^0=0.5$, while we keep $D_r = 0.1$. 

\subsubsection{Finite-size phase behavior}

\begin{figure*}
\centering
\includegraphics[width =\linewidth]{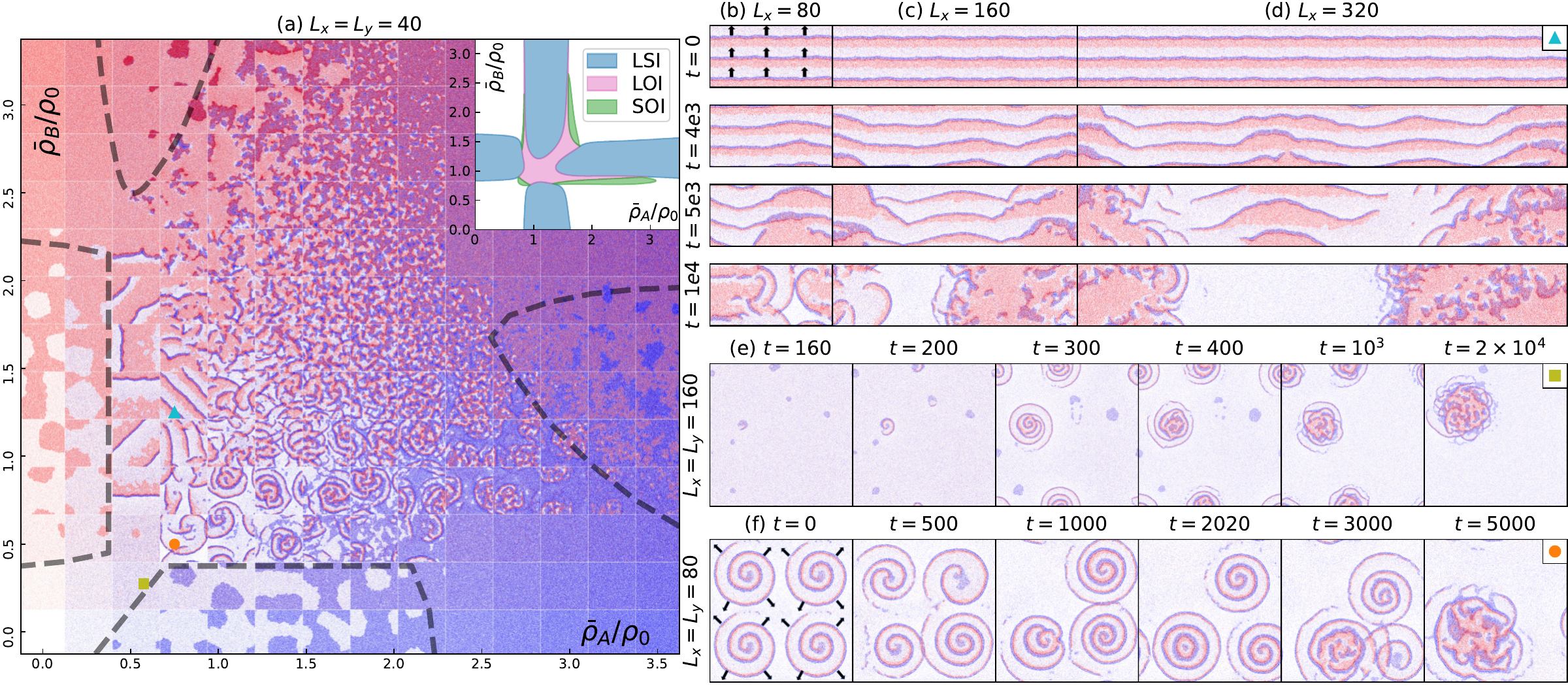}
\caption{Phase behavior in the composition plane for $\eta^0_{AA}=\eta^0_{BB}=-2$, $\eta^0_{AB}=-\eta^0_{BA}=0.5$, $D_r=0.1$, and $\rho_0=10$.
(a) Snapshots of representative configurations obtained from simulations in square boxes of linear size $L_x=40$. 
The regions of stationary phase separation are indicated by the black dashed lines as guides to the eye.
Inset: the corresponding linear stability diagram.
(b-d) Evolution of an initial condition built by duplicating a regular band configuration along the horizontal ($x$) direction for different system sizes with $L_y = 40$, showing the corresponding transverse instability.
(e, f) Instability of the rotating spirals revealed by the evolution of a homogeneous(e) and a duplicated spirals(f) initial conditions. 
Compositions: $(\bar{\rho}_A/\rho_0,\bar{\rho}_B/\rho_0)=(0.75, 1.25)$ for (b-d), $(0.575, 0.275)$ for (e) and $(0.75, 0.5)$ for (f), as marked by the corresponding colored symbols in (a).
}
\label{fig:PD-strong_NR-small_L}
\end{figure*}

To identify the different phases and patterns emerging in this regime, we first explore the phase diagram of the model in the composition plane, shown in Fig.~\ref{fig:PD-strong_NR-small_L}(a) for simulations performed in small square domains of size $L_x=40$.
As for weak nonreciprocity, the phase diagram features the homogeneous G, LA, LB and LAB phases at its corners, in agreement with the prediction from the linear stability analysis of Eqs.~\eqref{eq:rho-p} shown in the inset.
In addition, we identify four lobes, encircled by the black dashed lines in Fig.~\ref{fig:PD-strong_NR-small_L}(a), where stationary phase-separation takes place.
These regions qualitatively correspond to the predictions for the location of LSI (shown in blue in the inset).

In turn, the region near the diagonal $\bar\rho_A = \bar\rho_B$ displays a variety of dynamical patterns, consistently where LOI and SOI take place in the composition plane (pink and green regions in the inset, respectively).
At high densities ($\bar\rho_A \approx \bar\rho_B \approx 2\rho_0$), the homogeneous LAB liquid undergoes SOI, and is thus superseded by a chaotic phase of chasing clusters.  
Regular system-spanning patterns only become discernible at lower densities. 
In particular, we observe travelling bands analogue to the traveling domains of Fig.~\ref{fig:quasi-1D}(b) when $\bar{\rho}_A\lesssim \bar{\rho}_B$, while for $\bar{\rho}_A\gtrsim \bar{\rho}_B$ these bands self-organize in rotating spirals. 
Note that, due to our choice of inter-species couplings $\eta_{AB}^0 > 0$ and $\eta_{BA}^0 < 0$, the traveling bands are made of a domain rich in $B$ particles chasing a domain rich in $A$ particles. 
The positive curvature of the outward-going band in spirals therefore requires the $A$-rich domain to be longer than the $B$-rich domain, such that this configuration is indeed favoured when $\bar{\rho}_A > \bar{\rho}_B$.
On the other hand, $\bar{\rho}_A < \bar{\rho}_B$ leads to a relative excess of $B$ particles, such that $B$ domains become wider and the curvature of the interface is reduced.
For sufficiently large density difference between the two species, the domains then adopt a straight shape and form regular trains of bands such as those shown in the top panel of Fig.~\ref{fig:PD-strong_NR-small_L}(b).

Similar band and spiral patterns were reported in simulations of the NRCH model~\cite{saha2020scalar,rana2023defect} where, as discussed above, they originate from LOI.
In the dynamical systems nomenclature, this instability corresponds to a conserved-Hopf bifurcation~\cite{frohoff2023nonreciprocal}, which is the counterpart of the Hopf bifurcation for dynamics with conserved densities. 
The dynamical patterns reported here are also commonly encountered in reaction-diffusion dynamics~\cite{kondo2010reaction,reichenbach2007noise,vanag2009cross,luo2023influence}, which can often be mapped to the complex Ginzburg-Landau equation~\cite{aranson2002world} in the vicinity of a (non-conserved) Hopf bifurcation~\cite{cross1993pattern}.
These similarities therefore highlight connections between the NRQS and NRCH models, despite the continuous description of the NRQS model being typically more complex.
However, we show in the following section that the NRQS model exhibits a qualitatively distinct phase behavior in large systems, where the band and spiral patterns are generally unstable.

\subsubsection{Instability of travelling bands and rotating spirals} \label{sec:instability-SB-spiral}

To probe the stability of the regular traveling band patterns, we create trains of long bands by duplicating a stable configuration obtained in a system of size $40\times40$ (top row of Figs.~\ref{fig:PD-strong_NR-small_L}(b-d)).
As pictured in the bottom rows of Figs.~\ref{fig:PD-strong_NR-small_L}(b-d), sufficiently long band profiles undergo an instability, which eventually leads to their destruction.
Comparing the short-time oscillations of the band profiles in the second row of Figs.~\ref{fig:PD-strong_NR-small_L}(b-d), we estimate the characteristic wavelength of the instability to be about $80$ (in units of $R$), which is consistent with the fact that shorter bands are found stable.
Bands whose length exceeds this wavelength, however, are replaced by a macroscopic CCB domain coexisting with a dilute homogeneous gas, as shown in the last row of Figs.~\ref{fig:PD-strong_NR-small_L}(c,d). 
The front instability is reminiscent of that found in the contexts of driven granulars~\cite{parteli2011transverse}, chemotaxis~\cite{alert2022cellular}, as well as bacterial growth~\cite{allen2019bacterial},
but to the best of our knowledge were not discussed for conserved non-reciprocal mixtures\footnote{Undulations of travelling bands were reported in a simplified version of the NRCH model~\cite{brauns2024nonreciprocal}. In this case, however, their amplitude is bounded and they do not destabilize the fronts.}.
We leave the theoretical understanding of this instability for future works.

The rotating spirals made of curved trains of bands are also unstable in sufficiently large systems, as illustrated in Figs.~\ref{fig:PD-strong_NR-small_L}(e,f).
The snapshots of Fig.~\ref{fig:PD-strong_NR-small_L}(e)
depict the evolution of an unstable homogeneous configuration in a system of size $160\times160$ at a composition such that stable spirals are observed in a smaller system of size $40\times40$.
Following the instability of the homogeneous phase, small clusters of $A$ particles firstly nucleate ($t = 160$) and serve as nucleation sites for spirals ($t = 200$) which develop quickly ($t=300$).
As the spirals grow beyond a certain size, their core becomes chaotic ($t=400$), leading to the formation of a CCB cluster ($t=10^3$).
Finally, these clusters merge by coalescence, giving after a long coarsening time a single CCB domain coexisting with a disordered gas ($t=2\times10^4$).
Figure~\ref{fig:PD-strong_NR-small_L}(f), in turn, shows how a configuration made of four duplicated spirals becomes quickly unstable and results in a similar coexistence phase at long times.
Interestingly, the snapshot of Fig.~\ref{fig:PD-strong_NR-small_L}(f) at $t=2020$, features a target (bottom left) consisting of concentric circular bands that move radially outward.
Similar structures were reported in simulations of reaction-diffusion systems~\cite{hendrey2000target,jiang2009emergence}, as well as in the NRCH model~\cite{rana2023defect}, 
but appear to be only transient here. 
As noted in our previous work~\cite{duan2023dynamical}, both traveling bands and rotating spirals are generically unstable to long-wavelength perturbations, such that these dynamical patterns are replaced by phase-coexisting configurations involving macroscopic domains of CCB.
In the following section, we characterize further the large-scale phase behavior of the NRQS model.

\subsubsection{Construction of the large-scale phase diagram}
\begin{figure*}
\centering
\includegraphics[width =\linewidth]{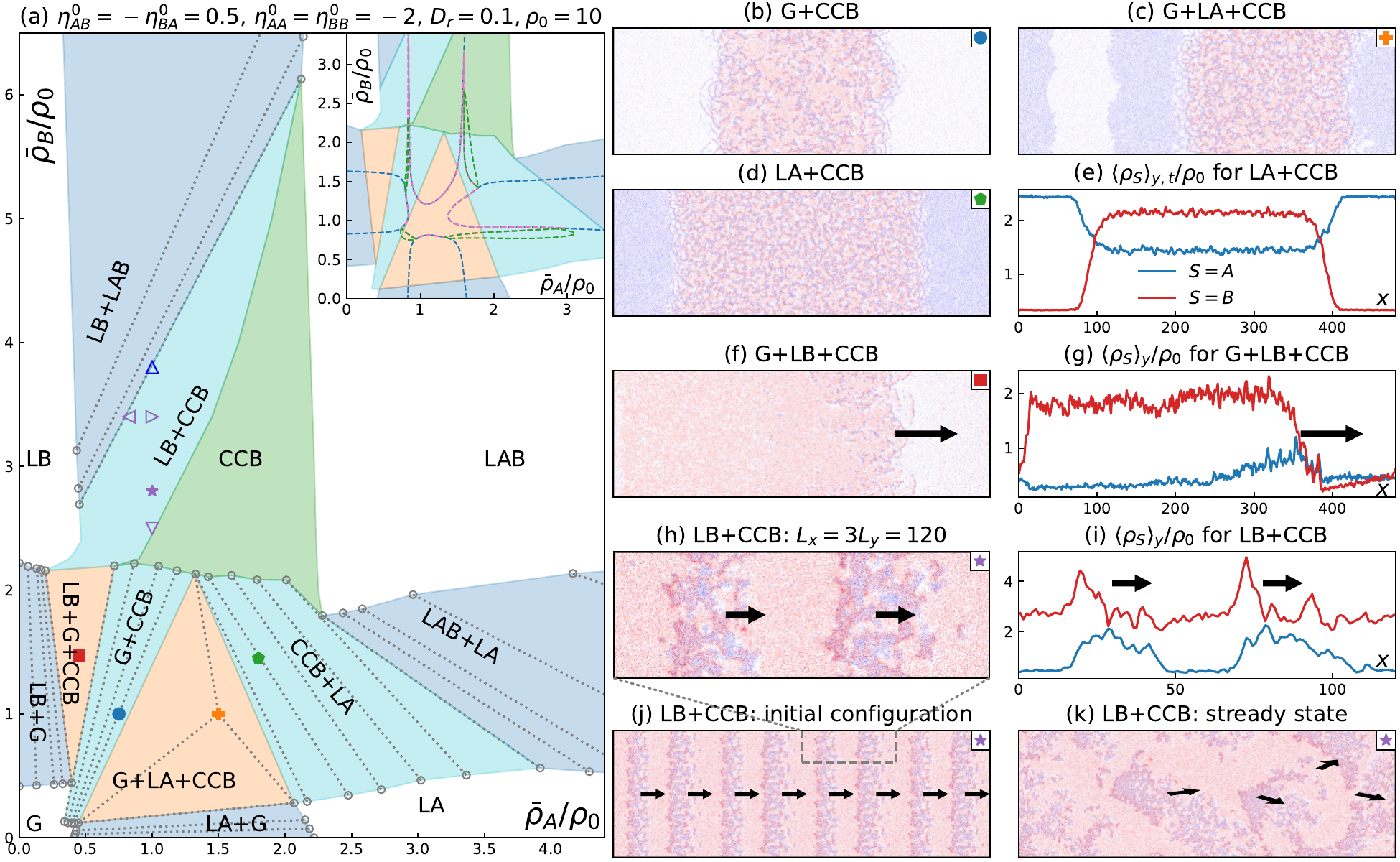}
\caption{(a) Phase diagram summarizing the large-scale phase behavior of NRQS active particles in the composition plane. 
The white regions correspond to the homogeneous phases: G, LA, LB, LAB, while the CCB phase appears in green. 
The blue regions indicate stationary phase coexistence between two homogeneous phases (similarly to Fig.~\ref{fig:mutual-attr-rep}), while cyan and orange regions respectively correspond to two- and three-phase coexistence involving CCB.
Hollow circles connected by dotted lines mark the coexisting densities measured in simulations (see text for details).
Hollow triangles mark the composition associated with the simulations results presented in Fig.~\ref{fig:osi_travelling}.
Inset: spinodal densities (dashed lines, color-code identical as in the inset of Fig.~\ref{fig:PD-strong_NR-small_L}(a)) predicted by linear stability analysis vs. the numerically determined phase boundaries.
(b-k) Typical snapshots and $y$-averaged density profiles at the compositions marked by the corresponding solid symbols in (a). 
The profiles in (e,g,i) correspond to the phases shown in (d,f,h), respectively. The system sizes for all snapshots are $L_x=3L_y=480$, except for (h) where $L_x=3L_y=120$.
Compositions: $(\bar{\rho}_A/\rho_0,\bar{\rho}_B/\rho_0)=(0.75, 1)$(b), $(1.5, 1)$(c), $(1.8, 1.45)$(d), $(0.45, 1.47)$(f), $(1, 2.8)$(h,j,k).
}
\label{fig:PD-strong_NR-large_L}
\end{figure*}

The pattern instabilities reported previously imply that the actual phase diagram of the NRQS model is quite different from the one pictured in Fig.~\ref{fig:PD-strong_NR-small_L}(a).
Performing numerical simulations of the particle-based model in large systems indeed reveals a rich phase behavior featuring various two- and three-phase coexistence, as summarized in Fig.~\ref{fig:PD-strong_NR-large_L}.

\begin{figure}
\centering
\includegraphics[width =\linewidth]{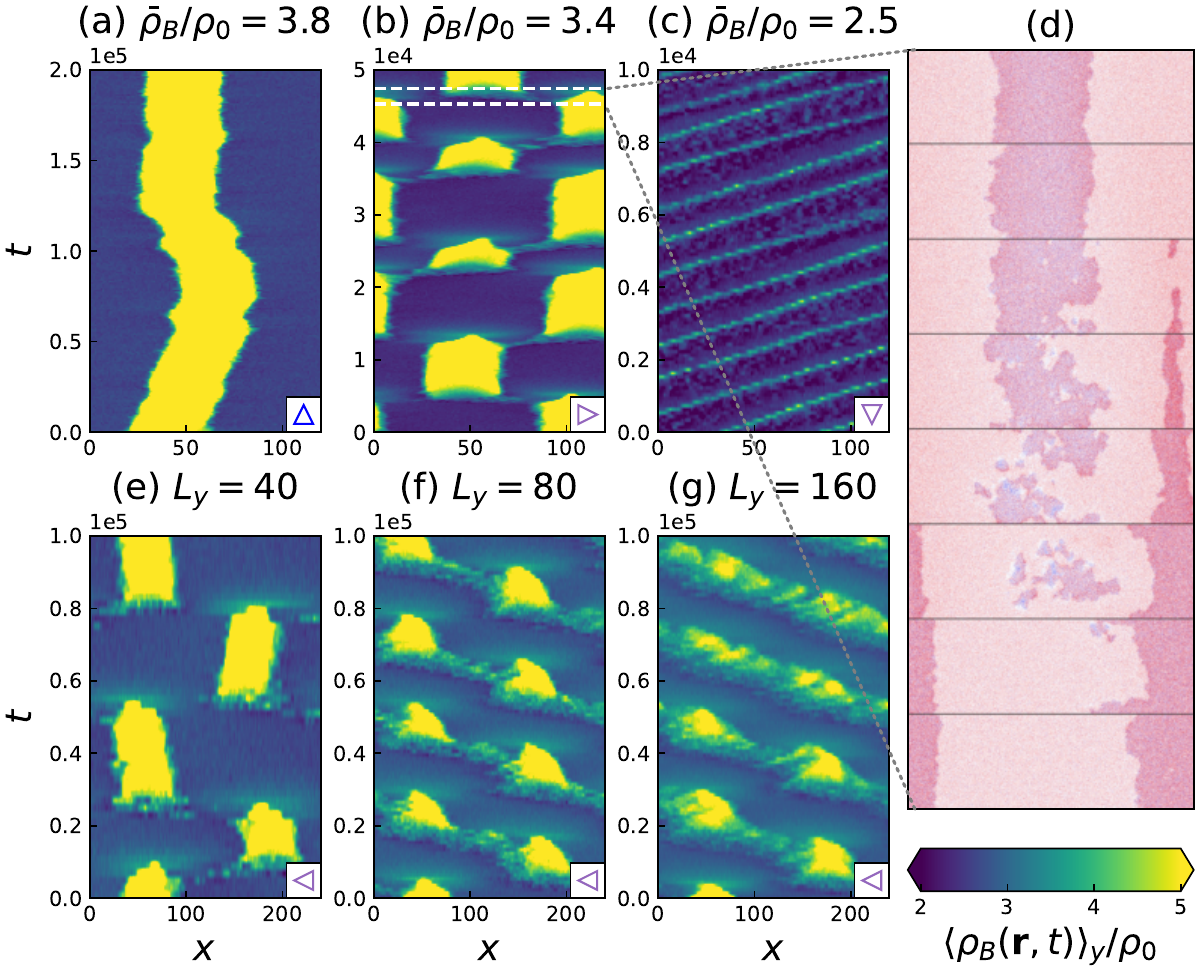}
\caption{Transition from the LB+LAB to the LB+CCB phases. (a-c) Kymographs showing the spatiotemporal dynamics of $\langle \rho_B(\mathbf{r},t)\rangle_y/\rho_0$, the $y$-averaged density profiles for species $B$. Setting $\bar{\rho}_A/\rho_0=1$ we observe, as $\bar{\rho}_B/\rho_0$ is decreased, stationary phase separation(a), oscillations(b) and steady travelling bands(c) in a system of size $L_x = 120$ and $L_y = 40$. 
(d) Time sequence of snapshots corresponding to(b), with times ranging from $4.53\times 10^4$ to $4.74\times 10^4$, from bottom to top.
(e-g) Same kymographs as (a-c) for  $\bar{\rho}_A/\rho_0=0.8$, $\bar{\rho}_B/\rho_0=3.4$, and $L_x=240$. Increasing $L_y$, oscillations become more frequent(e,f) and are eventually superseded by travelling patterns(g).
The compositions are marked by the corresponding hollow triangles in Fig.~\ref{fig:PD-strong_NR-large_L}(a), while other parameters identical to those of Fig.~\ref{fig:PD-strong_NR-large_L}.
}
\label{fig:osi_travelling}
\end{figure}

The structure of the phase diagram shown in Fig.~\ref{fig:PD-strong_NR-large_L}(a) shares striking similarities with that of Fig.~\ref{fig:mutual-attr-rep}(a) for the case of mutual motility inhibition leading to effective attractive couplings between the two species. 
The main difference between the two cases is that non-reciprocity induces, in addition to G, LA, LB and LAB, a CCB phase resulting from the instability of the dense $AB$ liquid (marked by the green region in Fig.~\ref{fig:PD-strong_NR-large_L}(a)).
The CCB phase then replaces LAB when it coexists with the gas, leading to the new G+CCB, G+LA+CCB, and G+LB+CCB coexistence phases pictured in Figs.~\ref{fig:PD-strong_NR-large_L}(b,c,f).
The CCB phase is, in addition, involved in two-phase coexistence with both LA and LB, resulting in the LA+CCB and LB+CCB phases shown in Figs.~\ref{fig:PD-strong_NR-large_L}(d,h).

As pointed out by Figs.~\ref{fig:PD-strong_NR-large_L}(b-d), although the internal dynamics of CCB domains is highly active, the G+CCB, G+LA+CCB and LA+CCB coexistence phases are not associated with any directed motion at macroscopic scales.
This is because these configurations are globally symmetric, leading to a balance of the mass flux at interfaces, similarly to stationary homogeneous phase-separated domains found at weak nonreciprocity.
Since in rectangular domains these phase-coexisting configurations produce flat interfaces spanning the shortest dimension, say $y$, we use time- and $y$-averaged density profiles to determine the corresponding coexisting densities.
As exemplified by Fig.~\ref{fig:PD-strong_NR-large_L}(e) for the case of LA+CCB, the phase-separated domains indeed exhibit nearly constant average bulk densities, and are separated by well-defined interfaces.
The bulk densities of both species hence define a pair of points in the composition plane joined by a tie line (hollow circles and grey dotted lines in Fig.~\ref{fig:PD-strong_NR-large_L}(a)).
Generalizing this approach to G+CCB and G+LA+CCB then enables the evaluation of the corresponding binodal densities, which can be used to determine the boundaries of these phases, similarly to the case of stationary phase separation addressed in Sec.~\ref{sec:wo-chasing}.
As detailed in Appendix.~\ref{sec:lever-ruel} (see also Ref.~\cite{duan2023dynamical}), shifting the composition of the system along the tie line simply affects the relative fractions of each phase, while leaving their characteristics intact, in agreement with the lever rule expected in this context.

Determining the boundaries of the G+LB+CCB and LB+CCB phases is, however, more complex.
Indeed, as shown in Figs.~\ref{fig:PD-strong_NR-large_L}(f,h) these phase-separated configurations are typically asymmetric, resulting in large-scale directed motion. 
The G+LB+CCB phase consists of a wide LB domain which constantly nucleates CCB at its front and cruises amidst a dilute homogeneous gas.
The asymmetry of this configuration is further highlighted by the instantaneous vertically averaged density profiles shown in Fig.~\ref{fig:PD-strong_NR-large_L}(g), 
which do not exhibit well-defined coexisting densities.
Hence, we define the boundaries of the G+LB+CCB phase from the rightmost tie line of the LB+G phase and leftmost tie line of the G+CCB phase, respectively. 
These two tie lines meet for low $\bar\rho_B$ at a quadruple point for the phases G, G+LB, G+LB+CCB, and G+CCB, while their two upper endpoints define a third boundary at high $\bar\rho_B$ with the LB+CCB phase (see the leftmost orange triangle in Fig.~\ref{fig:PD-strong_NR-large_L}(a)).

The LB+CCB phase consists of chaotic chasing band clusters cruising in a homogeneous $B$ liquid. 
In sufficiently small rectangular domains, the CCB clusters take the form of bands organized in periodic one-dimensional arrays, as shown in Fig.~\ref{fig:PD-strong_NR-large_L}(h).
Here as well, the corresponding instantaneous density profiles shown in Fig.~\ref{fig:PD-strong_NR-large_L}(i) reveal an asymmetric structure ensuring globally directed motion.
In larger rectangular domains, the clusters loose positional order but maintain some global motion (Fig.~\ref{fig:PD-strong_NR-large_L}(j,k), see Sec.~\ref{sec:fragility_CM} below for a detailed discussion on collective motion).
The LB+CCB coexistence phase (upper cyan region in Fig.~\ref{fig:PD-strong_NR-large_L}(a)) is bordered by LB and CCB at lower and larger $\bar\rho_A$, respectively; and by G+LB+CCB and LB+LAB respectively at lower and larger $\bar\rho_B$.  
The boundary with LB is determined in moderate systems of size $120 \times 40$ by visually identifying the compositions corresponding to the nucleation of a single traveling CCB cluster.
The boundary separating LB+CCB from CCB is obtained from a similar protocol, and corresponds to the absence of unidirectional global motion induced by the vanishing of the LB phase. 

Finally, the boundary between LB+CCB and LB+LAB is delineated by the lowest tie line of the latter, below which stationary phase separation becomes unstable. 
Numerical simulations in the vicinity of this transition however reveal the existence of an intermediate regime, characterized by spatio-temporal oscillations of LAB patches in a LB background, as illustrated by the kymographs of Figs.~\ref{fig:osi_travelling}(a-c) (with compositions marked by the corresponding upward-, right- and downward-pointing hollow triangles in Fig.~\ref{fig:PD-strong_NR-large_L}(a)).
As shown by the sequence of snapshots in Fig.~\ref{fig:osi_travelling}(d),
oscillations of LAB domains are initiated by the local nucleation of traveling clusters at their interface.
After nucleation, these clusters, which contain an excess of $A$ particles at their front, move into LB where they progressively grow and form a new LAB domain at the expense of the initial one.
Increasing the interface length, e.g., when considering a larger system, increases the probability of such nucleation events, resulting in more frequent oscillations (Figs.~\ref{fig:osi_travelling}(e,f)).
In large enough systems with sufficiently fast oscillations, the symmetry of the detaching clusters cannot be fully restored within a single period, leading to the steady motion of the denser domain (Fig.~\ref{fig:osi_travelling}(g)).
The spatio-temporal LAB oscillations therefore constitute a finite-size sub-regime of the LB+CCB phase, 
such that the LB+LAB/LB+CCB transition is direct and marked by the limit of stability of stationary phase separation. 

The remaining phase boundary separating LAB from CCB is fixed by visual inspection for the emergence of finite-size traveling clusters. Combining the results of this section then leads to the phase diagram of Fig.~\ref{fig:PD-strong_NR-large_L}(a), which highlights the rich phenomenology of NRQS active particles and the far-reaching influence of the CCB phase on their phase behavior. This phase diagram is moreover markedly different from that obtained from simulations of small systems in Fig.~\ref{fig:PD-strong_NR-small_L}(a), from which it could hardly be inferred. In what follows, we characterise the properties of the new coexistence phases involving CCB, focusing on two attributes points regarding their coarsening dynamics and whether they present large-scale collective motion.

\subsubsection{The coarsening dynamics}\label{sec:coarsening}

\begin{figure}
\centering
\includegraphics[width =\linewidth]{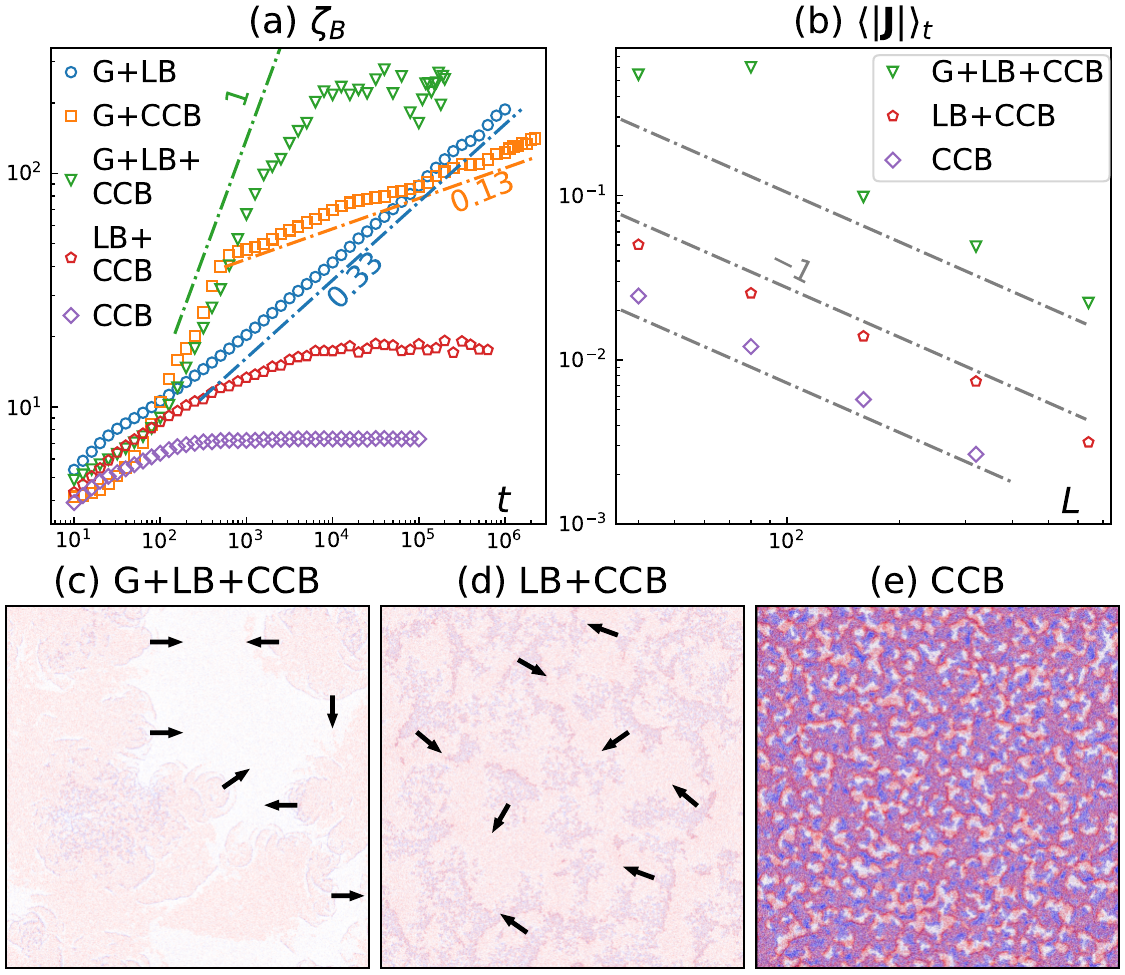}
\caption{(a) Measure of the typical domain size $\zeta_B$ as a function of time for randomly initialized systems in various phases.
System sizes are $1280^2$ for G+LB and G+CCB, and $640^2$ for the other cases. 
(b) Scaling of the order parameter $\langle|\mathbf{J}|\rangle_t$ vs. system size $L_x=L_y=L$ in the G+LB+CCB, LB+CCB and pure CCB phases.
(c-e) Typical snapshots of the C+LB+CCB, LB+CCB and pure CCB phases. System sizes are $640^2$ for (c,d) and $160^2$ for (e).
Compositions: $(\bar{\rho}_A/\rho_0,\bar{\rho}_B/\rho_0)=(0.3, 0.6)$(G+LB), $(0.55, 0.385)$(G+CCB), $(1.7, 2.25)$(pure CCB), and are identical to those of Fig.~\ref{fig:PD-strong_NR-large_L} for the other cases.
}
\label{fig:R_J_snaps}
\end{figure}

To investigate the coarsening dynamics of the coexistence phases, we monitor the typical length scale of phase-separated domains via 
\begin{equation}\label{eq:R_X_t}
    \zeta_X(t) =  \frac{2\pi\int \mathrm{d}q\, \tilde{S}_X(q,t)}{\int \mathrm{d}q\, q\tilde{S}_X(q,t)},
    % \zeta_X (t)=2\pi \int \mathrm{d} q\, \tilde{S}_X(q,t)\bigg/\int \mathrm{d}q\, q \tilde{S}_X(q,t),
\end{equation}
where $X\in \{A,B\}$ and 
$S_X(\mathbf{q},t)=\langle\rho_X(\mathbf{q},t)\rho_X(-\mathbf{q},t) \rangle$
is the structure factor associated with the density of species $X$,
while $\tilde{S}_X(q,t)\equiv \langle S_X(\mathbf{q}, t)\rangle_{|\mathbf{q}|=q}$ denotes the corresponding average over all orientations of $\bf q$. 
For systems initialized in the homogeneous phase, the time evolution of $\zeta_A(t)$ and $\zeta_B(t)$ are similar, such that Fig.~\ref{fig:R_J_snaps}(a) only reports data for $\zeta_B(t)$.

Despite the presence of strong nonreciprocity, the stationary phase-separated configurations G+LA, G+LB, LA+LAB and LB+LAB exhibit conventional coarsening, where the relevant $\zeta_X(t) \simeq t^{1/3}$, consistently with the Lifshitz–Slyozov law~\cite{Lifshitz1961JPCS,BrayAdvPhys2002}, as shown for G+LB by the blue symbols in Fig.~\ref{fig:R_J_snaps}(a).
This behavior is consistent with results obtained for the coarsening of dense MIPS domains in suspensions of active particles interacting via quorum-sensing~\cite{Stenhammar2014SoftMatt} and pairwise forces~\cite{shi2020self,caporusso2023dynamics}.

The presence of the CCB phase in the other coexistence phases, on the other hand, deeply modifies their coarsening behavior.
As expected from the analysis of Sec.~\ref{sec:wo-inhibition}, the coarsening of the pure CCB phase is arrested, as indicated by the saturation of $\zeta_B(t)$ at long times (purple symbols in Fig.~\ref{fig:R_J_snaps}(a)).
This reflects the fact that the typical chaotic chasing cluster size is selected, resulting in the absence of macroscopic structures (Fig.~\ref{fig:R_J_snaps}(e)).
In addition, the red symbols in Fig.~\ref{fig:R_J_snaps}(a) reveal that the coarsening of the LB+CCB phase is arrested as well, with a typical cluster size much larger than the one associated with CCB.
This feature also clearly appears in the snapshot of Fig.~\ref{fig:R_J_snaps}(d), showing that the large scale LB+CCB phase consists of large domains of CCB traveling in a $B$-rich liquid.

Moving on to G+LB+CCB, the green symbols in Fig.~\ref{fig:R_J_snaps}(a) indicate that the corresponding typical domain size first exhibit a fast, ballistic, growth and saturates at longer times.   
The ballistic scaling $\zeta_B(t) \simeq t$ starting at $t \simeq 10^2$ follows an initial nucleation-dominated regime, and results from the fast merging of traveling clusters similar to the one pictured in Fig.~\ref{fig:PD-strong_NR-large_L}(f).
This regime ends at $t \simeq 10^4$ where, as shown in Fig.~\ref{fig:R_J_snaps}(c), the typical domain size is comparable to the system size $L_x=L_y=640$. 
We nevertheless note that the cohesion of the LB+CCB patches is rather weak, such that CCB clusters occasionally detach from these domains, and move into the gas where they nucleate new dense aggregates. 
These events result in strong fluctuations of $\zeta_B(t)$ at long times (green symbols in Fig.~\ref{fig:R_J_snaps}(a)).
It is then reasonable to suppose that the coarsening of is asymptotically arrested, with a selected aggregate size much larger than the scales currently accessible in simulations.   

In contrast with the above, G+CCB, LA+CCB and G+LA+CCB appear to fully coarsen, resulting in domains whose sizes scale with the simulation box.
As shown by the orange symbols in Fig.~\ref{fig:R_J_snaps}(a) the initial stage of G+CCB coarsening exhibits a similar trend to that of G+LB+CCB. 
In particular, the ballistic coarsening regime for $10^2 \lesssim t \lesssim 10^3$ follows the nucleation of traveling bands, which progressively merge and collapse into CCB domains, as pictured in Fig.~\ref{fig:PD-strong_NR-small_L}(e).
The further coarsening of these domains at $t \gtrsim 10^3$ marks a crossover to a long-time scaling regime for $\zeta_B(t)$, which is associated with an exponent $\approx 0.13$ significantly smaller than $1/3$ (note that the power law behavior holds for more than three orders of magnitude in time).
We find that the late-time coarsening of the LA+CCB and G+LA+CCB phases is associated with a similar anomalous exponent. 
A detailed study of this abnormal coarsening behavior will be presented elsewhere~\cite{TBP}.

\subsubsection{Absence of macroscopic collective motion} \label{sec:fragility_CM}
We now turn to the question of macroscopic collective motion.
While large-scale collective motion is absent from the CCB, G+CCB, G+LA+CCB and LA+CCB phases, we cannot \textit{a priori} exclude that it originates from couplings between traveling domains in LB+CCB and G+LB+CCB. 

To quantify macroscopic collective motion, we use the time-averaged global order parameter $\langle |\mathbf{J}(t)| \rangle_t$, where $\mathbf{J}$ is defined as~\cite{saha2020scalar}
\begin{equation}
    \mathbf{J}(t) \equiv \frac{\langle \rho_A(\mathbf{r}, t) \nabla\rho_B(\mathbf{r}, t)
    -  \rho_B(\mathbf{r}, t) \nabla\rho_A(\mathbf{r}, t)
    \rangle_\mathbf{r}}{\bar{\rho}_A\bar{\rho}_B},
\end{equation}
from the spatially-averaged density fields and their gradients.
The order parameter $\langle |\mathbf{J}(t)| \rangle_t$ vanishes for stationary states or randomly moving clusters, but remains nonzero for gloablly travelling configurations involving a finite overlap of $A$ and $B$-rich domains such as pictured in Fig.~\ref{fig:quasi-1D}(b)\footnote{
Intuitively, this can be understood by noting that $\mathbf{J}$ changes sign under reversal of space coordinates $\mathbf{r}\leftrightarrow -\mathbf{r}$, 
such that $\langle |\mathbf{J}(t)| \rangle_t$ formally measures the lack of global parity symmetry.
Since collective motion in nonreciprocal mixtures is usually associated with the breaking of parity symmetry, as suggested by Figs.~\ref{fig:PD-strong_NR-large_L}(e,g,i), $\langle |\mathbf{J}(t)| \rangle_t$ als serves as an order parameter for the emergence of collective motion.}.

\begin{table*}
    \centering
    \caption{Large-scale properties of the coexisting phases appearing in Fig.~\ref{fig:PD-strong_NR-large_L}(a). 
    The phase-separated interfaces can either be stationary or propagating when they are associated to symmetric or asymmetric configurations, respectively.
    For fully coarsening domains, the typical domain size grows like $t^\nu$ at long times.
    The coarsening is deemed normal when $\nu=\tfrac{1}{3}$, while it is anomalously slow when $\nu<\tfrac{1}{3}$. 
    Global collective motion is absent in all phases, although various dynamical patterns can arise in sufficiently small systems as summarized in the bottom row and Figs.~\ref{fig:PD-strong_NR-small_L} to~\ref{fig:osi_travelling}. 
    }
    \label{tab:phase_behavior}
    \begin{tabular}{l c c c c c}
    \hline
    \hline
         Phases & \makecell[c]{LA(LB)+G\\ 
         LA(LB)+LAB
         } & CCB & LB+CCB  & \makecell[c]{G+CCB\\LA+CCB\\G+LA+CCB} & G+LB+CCB \\
        \hline
        \makecell[l]{Interfaces}& stationary & \makecell[c]{not \\applicable} & propagating & stationary  &propagating 
        \vspace{2mm}
        \\
        %%%
        Coarsening  & normal & arrested & arrested  & anomalously slow  & \makecell[c]{arrested?} 
        \vspace{2mm}
        \\
        %%%
        \makecell[l]{Finite-size \\collective motion}& none & \makecell[c]{polar chasing \\bands}  & \makecell[c]{travelling CCB domains, \\ spatio-temporal oscillations}& \makecell[c]{travelling bands,\\rotating spirals} & \makecell[c]{travelling phase-\\separated domains} \\
        \hline
        \hline
    \end{tabular}
\end{table*}

Note that $\mathbf{J}(t)$ is different from the mean particle orientation $\langle\hat{\mathbf{u}}_i \rangle_i$, which is conventionally used as an order parameter in flocking models~\cite{chate2020dry}, 
but trivially vanishes here as a result of the rotational noise.  
Hence, we evaluate the presence of large-scale collective motion by measuring the steady state average $\langle |\mathbf{J}(t)| \rangle_t$ in long-time simulations performed in square domains of increasing sizes $L$.
As shown by the purple symbols in Fig.~\ref{fig:R_J_snaps}(b), $\langle |\mathbf{J}(t)| \rangle_t$ decays as $L^{-1}$ in the pure CCB phase, highlighting the absence of collective motion in this case.
In fact, a similar scaling is observed for the LB+CCB coexistence (red symbols in Fig.~\ref{fig:R_J_snaps}(b)), indicating that large-scale motion is absent there too.
For G+LB+CCB, $\langle |\mathbf{J}(t)| \rangle_t$ is nearly constant at small sizes ($L \lesssim 10^2$, green symbols in Fig.~\ref{fig:R_J_snaps}(b)), but then quickly drops in larger systems.
We attribute this behavior to the intermittent breakings of large LB+CCB clusters, which only arise in sufficiently large systems.
The decay of $\langle |\mathbf{J}(t)| \rangle_t$ as the inverse of system size rules out the possibility of global directed motion, while it also exclude the presence of scale-free correlations associated with the presence of Goldstone modes in phases with spontaneously-broken rotational symmetry.

\subsubsection{Summary}
The large-scale properties of the various phases identified in the phase diagram of Fig.~\ref{fig:PD-strong_NR-large_L}(a) are summarized in Table~\ref{tab:phase_behavior}. 
In particular, our analysis reveals that large-scale traveling states emerge only for configurations involving CCB where the chasing species (here, $B$) is in majority.
A surprising finding reported in Table~\ref{tab:phase_behavior} is the absence of global collective motion in all phases.
This is in contrast with recent results 
suggesting that the order parameter $\langle |\mathbf{J}(t)| \rangle_t$ exhibits polar long-range order in the minimal NRCH model~\cite{pisegna2024emergent}, implying that large-scale collective motion survives the introduction of conservative noise in Eq.~\eqref{eq:phi_mu}.
This feature is another indication that the collective dynamics of NRQS active particles is not fully captured, even qualitatively, by the minimal NRCH model.
We note that an analogous situation occurs in the physics of flocks, 
where large-scale collective motion is found in minimal settings~\cite{toner1995long,mahault2019quantitative}, 
but is highly susceptible and can easily be destroyed in rather generic situations~\cite{duan2021breakdown,besse2022metastability,benvegnen2023metastability,codina2022small}.

Although we have focused our numerical efforts on the case with anti-symmetric inter-species couplings ($\eta^0_{AB} = -\eta^0_{BA}$), we expect the layout of the phase diagram of Fig.~\ref{fig:PD-strong_NR-large_L}(a), and the large-scale properties of associated phases summarized in Table~\ref{tab:phase_behavior}, to be representative of more general situations. 
For completeness, such cases are briefly discussed in Appendix~\ref{sec:weak-chasing}.

We finally note that the linear stabilty analysis of Sec.~\ref{sec:instability_homo} suggests that CCB may arise from SOI only for sufficiently large self-propulsion, here measured by the Péclet number ${\rm Pe} = v_0/D_r = 10$.
Hence, one cannot exclude \textit{a priori} that large traveling domains such as pictured in Fig.~\ref{fig:quasi-1D}(b) would be stable at lower values of $\rm Pe$, thus leaving a possibility for large-scale collective motion in this parameter regime. 
We present preliminary simulation results in Appendix~\ref{sec:instability-SB-Dr1}, which indicate that for $\rm Pe = 1$ sufficiently long traveling bands remain unstable.
Whether or not collective motion can be restored at lower $\rm Pe$ remains an open question which should be addressed in future works.

\subsection{Breakdown of the generalized common tangent construction}\label{sec:break-Maxwell}

Constructing the phase diagram of Fig.~\ref{fig:PD-strong_NR-large_L}(a) requires a substantial numerical effort.
The main reason for this is that bulk phase separation involving CCB only occurs in large enough systems (see Fig.~\ref{fig:PD-strong_NR-small_L}(a)). 
We further note that, unfortunately, the linear stability analysis is of little help in predicting the large-scale phase behavior of NRQS active particles, even qualitatively, as can be seen from the inset of Fig.~\ref{fig:PD-strong_NR-large_L}(a).
Indeed, the homogeneous phases are unstable in large areas of the composition plane where the linear stability analysis predicts that they are stable. 
In addition, dynamical patterns (both at finite and large scales) arise in regions where the homogeneous state is stable or undergoes a stationary instability.
On the other hand, globally static domains of CCB are found in regions overlapping with the presence of an oscillatory instability, 
such that using the presence of the latter as a criterion for the emergence of large-scale traveling states is unsatisfactory.

\begin{figure}
\centering
\includegraphics[width =\linewidth]{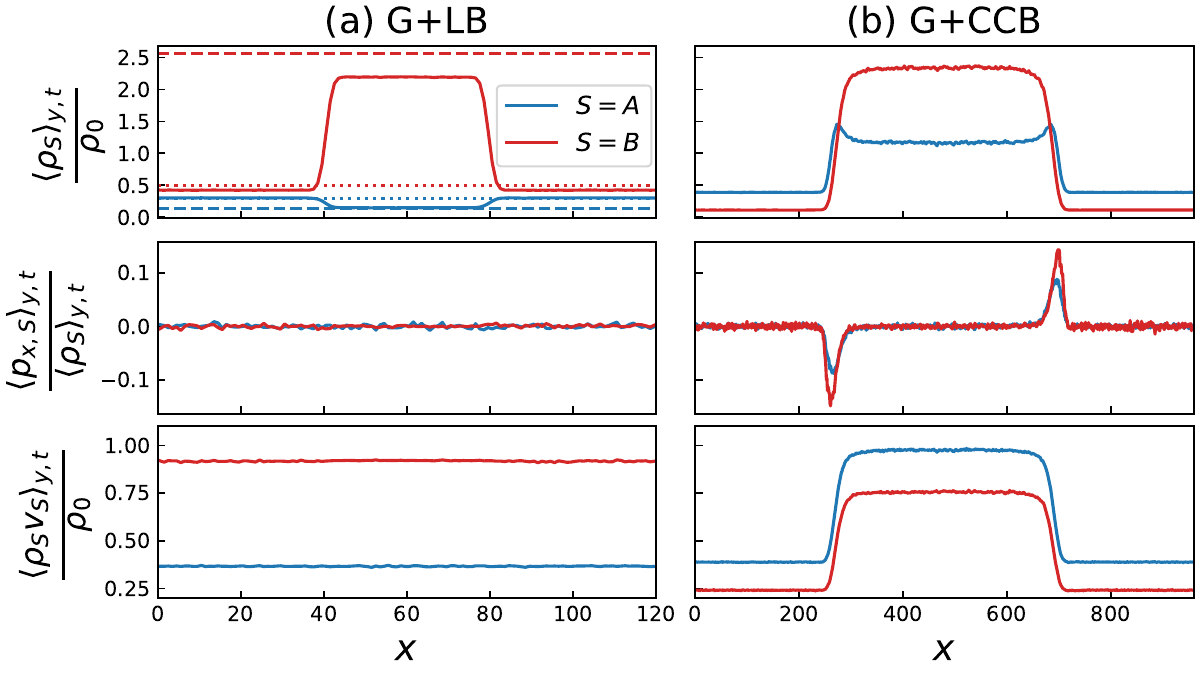}
\caption{Phase coexistence of G+LB(a) and G+CCB(b) in rectangular domains of sizes
$120 \times 40$ and $960 \times 160$, respectively.
The three rows show the $y-$ and time-averaged profiles of, from top to bottom, the particle densities,
the $x$ component of their polarities,
and the product of their local speed and density.
For G+LB, predictions of the binodal densities from the generalized common tangent construction 
are marked with dotted(G) and dashed(LB) lines.
Compositions: $(\bar{\rho}_A/\rho_0,\bar{\rho}_B/\rho_0) = (0.25,1)$(a) and $(0.75,1)$(b), while the other parameters are identical to those of Fig.~\ref{fig:PD-strong_NR-large_L}.
}
\label{fig:rho_v_profile}
\end{figure}

Given these limitations, we naturally explore the validity of the generalized common tangent construction proposed in Sec.~\ref{sec:maxwell} in the regime of strong nonreciprocity.
As shown in the top panel of Fig.~\ref{fig:rho_v_profile}(a) for G+LB, the binodal densities predicted by Eqs.~\eqref{eq:equal-mu} and~\eqref{eq:equal-h0} agree well with those measured from numerical simulations (compare the dotted and dashed lines with the density profiles shown with continuous lines).
Hence, the generalized common tangent construction remains accurate in predicting the phase boundaries of the coexistence between homogeneous phases, despite the presence of chasing interactions between the two species.

As discussed in the previous section, in large systems the coexistence phases appearing in the fifth column of Table~\ref{tab:phase_behavior} also exhibit stationary phase separation, albeit involving the inhomogeneous CCB phase.
Since the generalized common tangent construction is based on the reduced description~\eqref{eq:phi_mu} that cannot account for CCB, it however breaks down when applied to these phases.
This is exemplified in Fig.~\ref{fig:rho_v_profile}(b) for the G+CCB configuration. 
In this case, the common tangent construction notably gives unphysical predictions, such as a volume fraction for the gas phase greater than one.

This result can be rationalized from the observation that the presence of CCB significantly affects the properties of the interfaces of phase-separated domains.
The middle row of Fig.~\ref{fig:rho_v_profile} indeed shows that the constant ejection of polar chasing bands from the CCB domain to the gas results in polarized interfaces of the G+CCB phase, in contrast with G+LB where polarity vanishes everywhere. 
Moreover, the quantity $\rho_S v_S$ ---which is constant across G+LB interfaces due to the constraint of uniform chemical potential $\mu_S=\ln (\rho_S v_S)$---
significantly varies at G+CCB interfaces, as revealed by the bottom row of Fig.~\ref{fig:rho_v_profile}.
The G+CCB configuration therefore violates the condition~\eqref{eq:equal-mu} of equality of the effective chemical potentials in the two phases, which forms the basis of the generalized common tangent construction.

\section{Conclusion and outlook}
\label{sec:conclusion}

In this work, we have investigated the phase behavior of NRQS active particles, which are among the simplest microscopic models of chemically active mixtures.
Despite the simplicity of the microscopic dynamics, we have uncovered, both via numerical simulations and the study of continuous models derived from coarse-graining, a rich phase behavior featuring a variety of multi-scale patterns with intriguing properties.

In the regime of weak nonreciprocity where the effective interactions between the species are mutually attractive or repulsive (albeit possibly asymmetric), we have shown in Sec.~\ref{sec:wo-chasing} that the phase behavior of NRQS active particles qualitatively resembles that of equilibrium mixtures. 
Because of the absence free energy structure, the prediction of the associated phase boundaries however requires the use of the generalized common tangent construction introduced in Sec.~\ref{sec:phase-equilibria}, showcasing the influence of nonreciprocity at large scales.
Combined with the derivation of the continuous NRCH-like model~\eqref{eq:phi_mu} via coarse-graining, this approach notably enables an analytical, parameter-free, prediction of the phase diagram.

Increasing non-reciprocity so as to induce effective chasing interactions between the two species, we have demonstrated that the dynamics of NRQS active particles is no more captured by the NRCH model, such that its description requires a new theoretical framework.
This feature was first revealed by the characterisation of the thin chasing band phase in Sec.~\ref{sec:wo-inhibition} which arises without inter-species interactions, a parameter regime for which pattern formation is prohibited in the NRCH model.
As predicted from the linear stability of homogeneous solutions of the coarse-grained description~\eqref{eq:rho-p} which retain the dynamics of the polarity fields, 
the chasing bands emerge from a short-wave oscillatory instability that originates from the interplay of non-reciprocity and self-propulsion, and thereby has no counterpart in the physics of the NRCH model, MIPS and reaction-diffusion systems~\cite{duan2023dynamical}.

Turning-on intra-species motility inhibition to induce bulk-phase separation, we have shown that the NRQS model supports a multitude of dynamical patterns including traveling bands, rotating spirals, and oscillations of phase-separated domains similar to those reported in related models~\cite{saha2020scalar,you2020nonreciprocity,rana2023defect}.
All these patterns, however, turn out to be unstable in large systems where they are superseded by various coexistences phases displaying multi-scale structures, and where chaotic chasing bands play a central role, as summarized by Fig.~\ref{fig:PD-strong_NR-large_L}.

Also noteworthy is the complete absence of large-scale collective motion, at odds with results obtained previously from the NRCH model~\cite{saha2020scalar,you2020nonreciprocity,brauns2024nonreciprocal,rana2023defect,pisegna2024emergent}.
We further note that this situation is distinct from the case studied in Ref.~\cite{saha2022effervescent}, where it is shown that the presence of nonlinearities suppressing nonreciprocity with increasing particle density can also lead to a destruction of global collective motion.
In the present work, since the motility responses of both species are monotonous in the densities, the couplings in the coarse-grained NRCH-like description~\eqref{eq:phi_mu} are highly nonlinear but remain monotonous, as reflected by Eq.~\eqref{eq:f_NR_v_LSA_LSB}.
In addition, we observed that the restoration of global parity symmetry ---and the associated loss of global motion--- only arise at large enough scales,
while its origin seems rooted in the destabilisation of interfaces between traveling phase-separated domains (Figs.~\ref{fig:PD-strong_NR-small_L}(b,c)).
A more complete comprehension of the mechanisms underlying the destruction of collective motion in NRQS mixtures will therefore require a deeper understanding of the physics of active interfaces~\cite{besse2023interface,osat2024escaping}.

Furthermore, the mesoscopic chasing bands populating the phase-separated domains in Fig.~\ref{fig:PD-strong_NR-large_L} deeply affect the properties of their interfaces, resulting, e.g., in oscillations as well as arrested or anomalous coarsening.
Importantly, the presence of chasing bands also leads to a breakdown of the generalized common tangent construction, 
while extending this approach to predict the coexisting densities of configurations involving CCB domains poses a significant challenge.
Another uncertainty lies in the validity of the mean-field approximation that underlies the derivation of the continuous model.
Connecting microscopic and macroscopic descriptions beyond mean-field is a long standing issue in the field of active matter~\cite{kourbane2018exact,kursten2021quantitative,martin2023exact,mason2024dynamical}.
Whether or not the physics of NRQS active particles in the strongly nonreciprocal regime is qualitatively captured by the mean-field theory remains to be explored.

Despite the theoretical challenges, the interaction rules of the NRQS model can be straightforwardly engineered in light-controlled self-propelled colloids~\cite{lavergne2019Science,bauerle2018self} and assemblies of robots~\cite{chen2024emergent}.
Besides, the possibilities offered by synthetic biology place the implementation of nonreciprocal motility regulation in generically engineered {\it E. coli} potentially within reach~\cite{Liu2011Science,curatolo2020cooperative}.
Hence, the wide range of self-organized behaviors of NRQS active particles, as well as additional features not captured by our minimal description, could be observed in both synthetic and living systems.
This especially applies to the pure CCB phase, regular travelling bands, rotating spirals, targets and spatio-temporal oscillations, which all emerge in systems of moderate sizes.

\appendix

\section{Influence of the presence of pairwise repulsion} \label{sec:repulsion}
To keep the microscopic model~\eqref{eq:micro-model} minimal, we have neglected steric interactions between the particles,
which is well-justified so long as the range of quorum-sensing interactions $R$ remains much larger than the typical particles size $\sigma$.
In synthetic systems, $R$ can be tuned arbitrarily, while $R = 10\sigma$ was used in Ref.~\cite{bauerle2018self}.
In biological systems where quorum sensing is regulated by exchange of chemicals, 
the scale $R$ is set by the typical distance a signaling molecule can diffuse before being degraded. 
For the acyl-homoserine lactone molecules that mediate the quorum-sensing interactions of \textit{E. coli} in Refs.~\cite{Liu2011Science,curatolo2020cooperative}, 
$R$ can reach a few millimeters~\cite{marenda2016modeling} thus several orders of magnitude larger than the typical size of a bacterium.

\begin{figure}
\centering
\includegraphics[width =\linewidth]{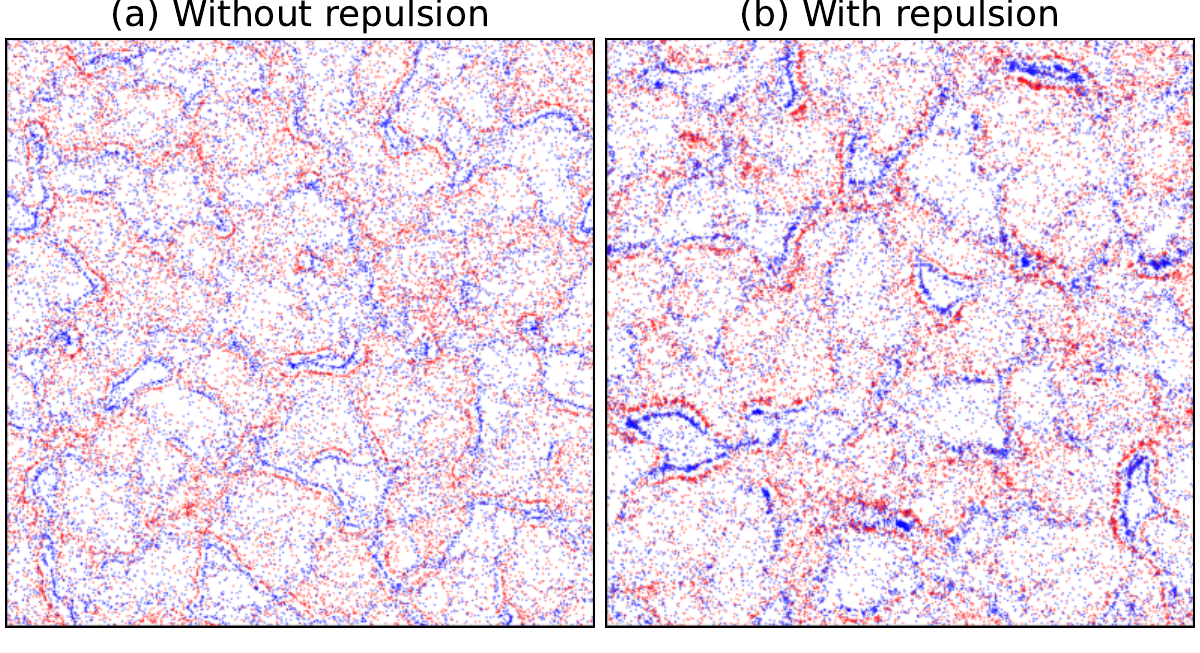}
\caption{Steady states of the CCB phase in the absence ($k=0$,a) or with ($k=100$,b) short-ranged repulsion. 
Other simulation parameters: $\eta_{AA}^0=\eta_{BB}^0=0$, $\eta_{AB}^0=-\eta_{BA}^0=2$, 
$D_r=0.1$, $\bar{\rho}_A=\bar{\rho}_B=\rho_0=10$, $L_x=L_y=40$, $\sigma=0.1$.
}
\label{fig:snaps_wo_w_rep}
\end{figure}

To check the influence of short-ranged pairwise repulsion on the dynamics of NRQS active particles,  
we denote $\sigma$ as their diameter and assume that they interact with each other via soft harmonic forces with sitfness $k$. 
Equation~\eqref{eq:micro-model-a} then becomes
\begin{equation*}
     \dot{\mathbf{r}}_{i,S} = v_S\left[\tilde{{\rho}}_A(\mathbf{r}_{i,S}), \tilde{\rho}_B(\mathbf{r}_{i,S})\right]
     \hat{\mathbf{u}}_{i, S} + k\sum_{j\sim i} (r_{ij}-\sigma) \hat{\mathbf{r}}_{ij},
\end{equation*}
where $r_{ij}\equiv |\mathbf{r}_j-\mathbf{r}_i|$ denotes the distance between particles $i$ and $j$, $\hat{\mathbf{r}}_{ij}=(\mathbf{r}_j-\mathbf{r}_i)/r_{ij}$ is the unit vector pointing from $i$ to $j$, and the sum is taken over
all $j$ particles within distance $\sigma$ of $i$, irrespective of their species $S$.
Here, we set $\sigma=R/10=0.1$, and $k=100$ to mimic the effect of short-ranged steric repulsion between the particles.
As shown in Fig.~\ref{fig:snaps_wo_w_rep}, the presence of such repulsion does not prevent the emergence of CCB.
Hence, we expect that the results presented in the main text won't be qualitatively affected by the presence of repulsion.
Similarly, we have checked that our results remain qualitatively unchanged when using other forms of quorum-sensing interaction kernel $\hat{w}(r)$ (see the definition below Eqs.~\eqref{eq:micro-model}), such as square-shaped and bell-shaped functions.

\section{Decomposition of the effective chemical potential}\label{sec:decomp-f}
In this section, we outline the derivation of Eqs.~\eqref{eq:feq_fNR}. 
We first rewritte Eqs.~\eqref{eq:mu_decomp} as
\begin{subequations}\label{def:mu_tilde}
    \begin{align}
        \tilde{\mu}_A &= \mu_A - \partial_{\phi_A} f^\mathrm{NR} = \partial_{\phi_A} f^\mathrm{eq},\\
        \tilde{\mu}_B &= \mu_B + \partial_{\phi_B} f^\mathrm{NR} = \partial_{\phi_B} f^\mathrm{eq},
    \end{align}
\end{subequations}
where $\tilde{\mu}_S$ is the chemical potential associated with the symmetrized effective free energy $f^\mathrm{eq}$.
Given that $\partial_{\phi_B}\tilde{\mu}_A=\partial_{\phi_A}\tilde{\mu}_B=\partial_{\phi_A}\partial_{\phi_B} f^\mathrm{eq}$,
we obtain from~\eqref{def:mu_tilde}
\begin{equation} \label{eq:double_partial_f_NR}
    \partial_{\phi_A}\partial_{\phi_B}f^\mathrm{NR} = \frac{1}{2} \left(\frac{\partial \mu_A}{\partial \phi_B}- \frac{\partial \mu_B}{\partial \phi_A}\right).   
\end{equation}
For convenience, we thus introduce
\begin{equation}
    g_\pm(\phi_A,\phi_B) = \frac{1}{2} \left(\frac{\partial \mu_A}{\partial \phi_B}\pm \frac{\partial \mu_B}{\partial \phi_A}\right).
\end{equation}
Integrating Eq.~\eqref{eq:double_partial_f_NR}, we then obtain
\begin{equation} \label{eq:fNR_appB}
    f^\mathrm{NR} =\int^{\phi_A}\mathrm{d}x\int^{\phi_B}\mathrm{d}y\, g_-(x,y) + c_A(\phi_A) + c_B(\phi_B),
\end{equation}
where constants of integration $c_S(\phi_S)$ can be absorbed into $f^\mathrm{eq}$ since they satisfy $\partial_{\phi_A}\partial_{\phi_B}c_S(\phi_S)=0$. 
Hence, we discard these terms and recover Eq.~\eqref{eq:fNR} of the main text.

Meanwhile, Eq.~\eqref{def:mu_tilde} yields
\begin{align}
    f^\mathrm{eq} = & \int^{\phi_A} \mathrm{d}x \, \tilde{\mu}_A (x,\phi_B)+ \int^{\phi_B} \mathrm{d}y \, \tilde{\mu}_B(\phi_A,y) \nonumber \\
    & - \int^{\phi_A} \mathrm{d} x \int^{\phi_B} \mathrm{d} y \, \frac{\partial \tilde{\mu}_A(x,y)}{\partial y}.
\end{align}
Substituting $\tilde{\mu}_S$ and $f^{\rm NR}$ by their expressions given in~\eqref{def:mu_tilde} and~\eqref{eq:fNR_appB}, 
we finally recover the expression of $f^\mathrm{eq}$ given by Eq.~\eqref{eq:f_eq} of the main text.

Note that one could naively write from Eq.~\eqref{eq:mu_decomp}
\begin{subequations}\label{eq:wrong_decomp_f}
    \begin{align}
        f^\mathrm{eq} &=\frac{1}{2} \left(\int^{\phi_A} \mathrm{d}x \mu_A(x,\phi_B) + \int^{\phi_B} \mathrm{d}y\mu_B(\phi_A,y)\right), \\
        f^\mathrm{NR} &=\frac{1}{2} \left(\int^{\phi_A} \mathrm{d}x \mu_A(x,\phi_B) - \int^{\phi_B} \mathrm{d}y\mu_B(\phi_A,y)\right). \label{eq:wrong_decomp_f-b}
    \end{align}
\end{subequations}
These expressions are, however, misleading.
This can be seen, for example, considering the equilibrium-like scenario $\mu_A = \mu_B = \mu(\phi_A+\phi_B)$,
for which Eq.~\eqref{eq:wrong_decomp_f-b} leads to a non-vanishing $f^\mathrm{NR}$, 
in contrast with the expression Eq.~\eqref{eq:fNR} which correctly gives $f^{\rm NR} = 0$. 
In fact, the non-vanishing $f^\mathrm{NR}$ given by Eq.~\eqref{eq:wrong_decomp_f-b} satisfies $\partial_{\phi_A}\partial_{\phi_B}f^\mathrm{NR}=0$, 
such that it should be absorbed into the definition of $f^\mathrm{eq}$, in agreement with the absence of nonreciprocity.

\section{From mutual attraction/repulsion to chasing interactions}\label{sec:weak-chasing}
\begin{figure}
\centering
\includegraphics[width =\linewidth]{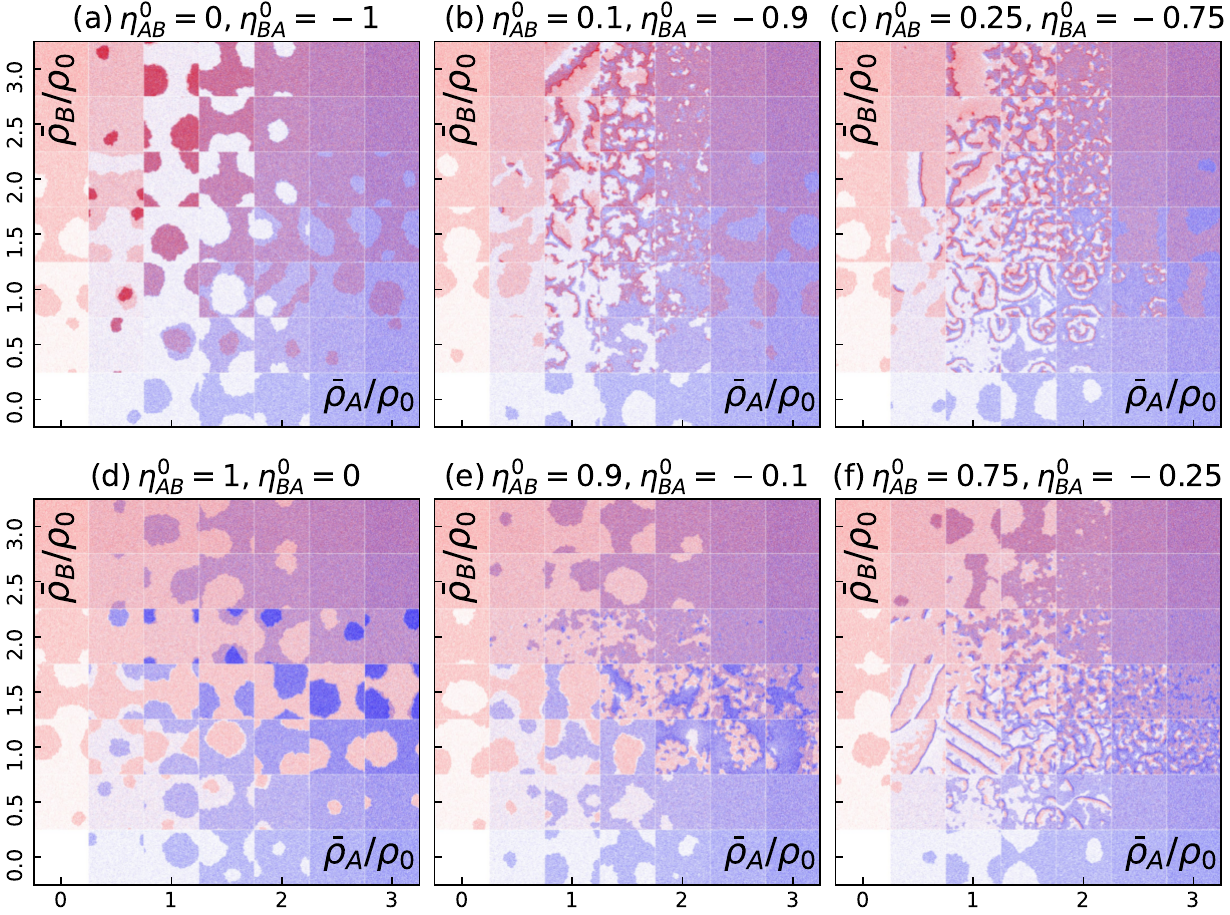}
\caption{Phase behavior in composition plane as the strength of chasing interaction increases left to right, 
while $D_r=0.1$ and $\rho_0=20$.
Snapshots are obtained from particle simulations in square domains of size $L_x=40$.
}
\label{fig:PD_varied_J}
\end{figure}

The analysis performed in Sec.~\ref{sec:w-inhibition} focuses on the purely asymmetric case with $\eta_{AB}^0=-\eta_{BA}^0=0.5$.
Here, we explore the phase behavior of the NRQS model for a more general family of couplings which interpolate from equilibrium-like
phase behavior to chasing interactions.
We thus set $\eta^0_{AA}=\eta^0_{BB}=-2$ and introduce a parameter $\tilde\eta \in [0;1]$ such that the inter-species couplings are given by $\eta_{AB}^0=\tilde\eta$ and $\eta_{BA}^0=\tilde\eta-1$.

As shown in Figs.~\ref{fig:PD_varied_J}(a,d), setting $\tilde\eta = 0$(1) leads to a phase diagram in composition plane with a similar structure as when the species exhibit mutual motility inhibition(activation) (c.f.\ Fig.~\ref{fig:mutual-attr-rep}).
For weak chasing interactions in the vicinity of the mutual motility activation regime, e.g. with $\tilde\eta = 0.1$ in Fig.~\ref{fig:PD_varied_J}(b), 
small dynamical clusters form in the parameter region corresponding to the G+LAB coexistence phase in Fig.~\ref{fig:PD_varied_J}(a), 
while the other coexisting phases remain mostly stable.
On the other hand, setting $\tilde\eta = 0.9$ as in Fig.~\ref{fig:PD_varied_J}(e) mainly destabilizes of the LA+LB phase.
In both cases, the G+LAB phases for $\tilde\eta \to 0$ and LA+LB for $\tilde\eta \to 1$ become unstable via the nucleation of 
small traveling clusters at the interfaces of phase-separated domains.
Because of the structure of the quorum-sensing interactions defined by Eqs.~(\ref{eq:v_S},\ref{def:L_SS'}), 
chasing interactions are strongest when the densities of both species are close to the reference value $\rho_0$.
For both G+LAB and LA+LB when $\tilde\eta \to 0$ and $\tilde\eta \to 1$, respectively, 
the bulk densities of both species are far from $\rho_0$, but become close to $\rho_0$ around the interfaces.
Hence, these phases are more susceptible to the presence of chasing interactions.

On the other hand, phases for which the density of one of the species is almost uniform are more robust.
For instance, the G+LA phase exhibits strong spatial variations of $\rho_A$, while $\rho_B$ remains small everywhere.
Similarly, in the three-phase coexistence G+LA+LB the interfaces of $A$-rich and $B$ rich domains do not overlap, 
such that they only become unstable for strong chasing interactions.
This feature is illustrated in Figs.~\ref{fig:PD_varied_J}(c, f), which respectively correspond to $\tilde\eta = 0.25$ and $\tilde\eta = 0.75$.
At these values of $\tilde\eta$, one also perceives regular dynamical patterns such as bands and spirals, 
which become more prominent in the purely antisymmetric case $\tilde\eta=0.5$, as shown in Fig.~\ref{fig:PD-strong_NR-small_L}(a). 
Hence, we expect that the large scale phase behavior of NRQS active particles summarized in Fig.~\ref{fig:PD-strong_NR-large_L} and Table~\ref{tab:phase_behavior} for $\eta_{AB}^0=-\eta_{BA}^0=0.5$ is representative of most choices of chasing interactions.

\section{Instability of travelling domains at small $\mathrm{Pe}$}\label{sec:instability-SB-Dr1}

\begin{figure}
\centering
\includegraphics[width =\linewidth]{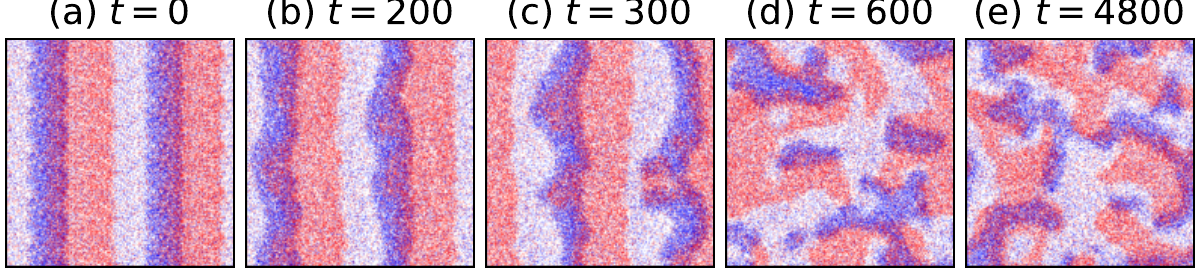}
\caption{Instability of travelling bands for $D_r=1$, 
system size $L_x=L_y=20$, 
$\eta_{AA}^0=\eta_{BB}^0=-2$, 
$\eta_{AB}^0=-\eta_{BA}^0=0.5$, 
$\bar{\rho}_A/\rho_0=1$, $\bar{\rho}_B/\rho_0=1.375$, and $\rho_0=80$.
}
\label{fig:SB_instability_Dr1}
\end{figure}

As illustrated in Fig.~\ref{fig:SB_instability_Dr1}, 
we find that sufficiently long traveling bands emerging in simulations of the NRQS model remain unstable 
even when the Péclet number is decreased.
Namely, the initial condition corresponding to the simulation results shown in Fig.~\ref{fig:SB_instability_Dr1}
was built by duplicating stable bands obtained in a system of size $L_x = 20$ and $L_y = 4$,
and with the same values of couplings $\{\eta_{SU}^0\}$ as those used for Fig.~\ref{fig:PD-strong_NR-small_L}, 
but with a larger rotational diffusivity $D_r=1$.
Similarly to the simulation results at larger $\rm Pe$ shown in Fig.~\ref{fig:PD-strong_NR-small_L}(b), 
the bands quickly develop an instability at their front (Figs.~\ref{fig:SB_instability_Dr1}(b,c)), 
break up, and are eventually replaced by smaller chasing clusters (Figs.~\ref{fig:SB_instability_Dr1}(d,e)).

\section{Lever rule for the gas-CCB coexistence phase} \label{sec:lever-ruel}

\begin{figure}
\centering
\includegraphics[width =\linewidth]{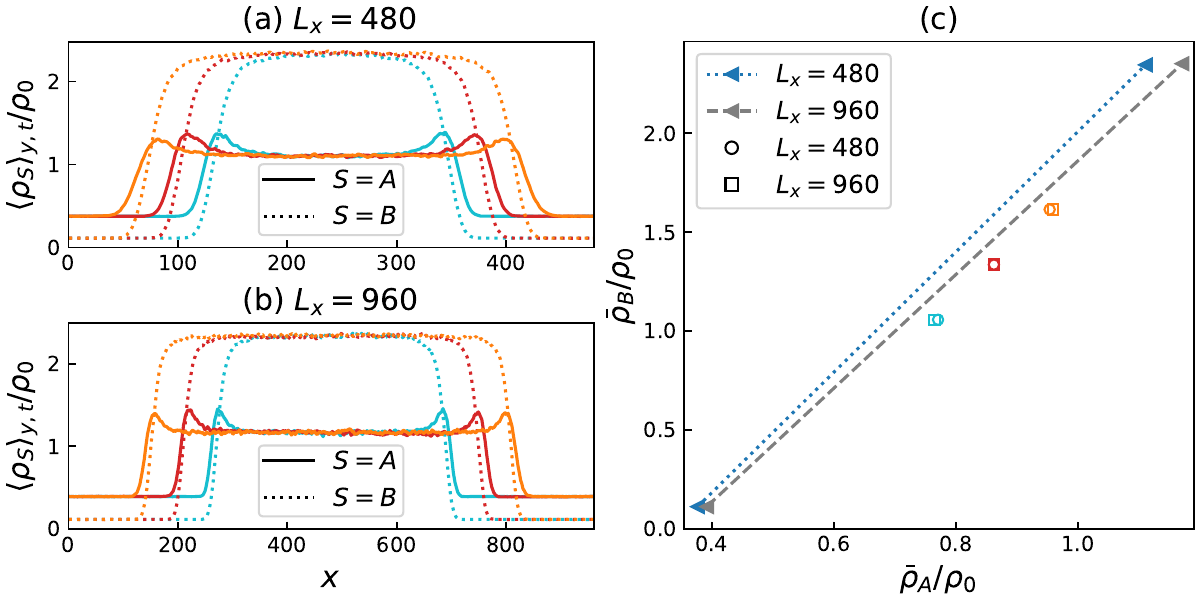}
\caption{Lever rule for the G+CCB coexistence phase. 
(a, b) $y-$ and time-averaged density profiles for both species at various compositions marked by the hollow symbols of the same color in (c). System sizes: $L_x=480$ (a) and $L_x=960$ (b), while $L_y=160$ for both. 
(c) The three compositions (hollow circles for $L_x=480$, hollow squares for $L_x=960$) at which the G+CCB phase coexistence was analyzed. The binodals (solid triangles) and tie lines were estimated from the coexistence densities in (a,b).
}
\label{fig:lever-rule}
\end{figure}

To demonstrate the lever rule in the G+CCB phase coexistence regime, 
we performed simulations in large rectangular domains following the procedure detailed in Sec.~\ref{sec:w-inhibition}.
This way, we can easily infer the bulk coexisting densities of both species from the time-averaged density profiles
shown in Figs.~\ref{fig:lever-rule}(a,b).

Namely, we define the coexistence densities from the values of $\langle\rho_{A,B}\rangle_{y,t}$ far from the interfaces, 
where the density profiles are nearly flat.
As shown by the solid triangles and dashed lines in Fig.~\ref{fig:lever-rule}(c), 
these values define two points joined by a tie line in the composition plane.
Varying the system composition along that tie line, 
the density profiles in Figs.~\ref{fig:lever-rule}(a,b) show that the relative volume fractions of the gas and CCB phase vary, 
while the coexisting densities are unchanged.
Because of the small density peaks in the vicinity of the interfaces, 
the global compositions marked by hollow symbols in Fig.~\ref{fig:lever-rule}(c) do not exactly fall on the tie line.
As the width of these peaks is not extensive, however, the tie line converges to these points when the system size is increased. 
For the LA+CCB and G+LA+CCB phase, no such peaks near the interface are present (Fig.~\ref{fig:PD-strong_NR-large_L}(e)),
such that the tie lines pass through the corresponding global composition points even in finite systems.

\acknowledgments
This work has received support from the Max Planck School Matter to Life and the MaxSynBio Consortium, which are jointly funded by the Federal Ministry of Education and Research (BMBF) of Germany, and the Max Planck Society.

\bibliography{ref.bib}

\begin{thebibliography}{99}%
\makeatletter
\providecommand \@ifxundefined [1]{%
 \@ifx{#1\undefined}
}%
\providecommand \@ifnum [1]{%
 \ifnum #1\expandafter \@firstoftwo
 \else \expandafter \@secondoftwo
 \fi
}%
\providecommand \@ifx [1]{%
 \ifx #1\expandafter \@firstoftwo
 \else \expandafter \@secondoftwo
 \fi
}%
\providecommand \natexlab [1]{#1}%
\providecommand \enquote  [1]{``#1''}%
\providecommand \bibnamefont  [1]{#1}%
\providecommand \bibfnamefont [1]{#1}%
\providecommand \citenamefont [1]{#1}%
\providecommand \href@noop [0]{\@secondoftwo}%
\providecommand \href [0]{\begingroup \@sanitize@url \@href}%
\providecommand \@href[1]{\@@startlink{#1}\@@href}%
\providecommand \@@href[1]{\endgroup#1\@@endlink}%
\providecommand \@sanitize@url [0]{\catcode `\\12\catcode `\$12\catcode
  `\&12\catcode `\#12\catcode `\^12\catcode `\_12\catcode `\%12\relax}%
\providecommand \@@startlink[1]{}%
\providecommand \@@endlink[0]{}%
\providecommand \url  [0]{\begingroup\@sanitize@url \@url }%
\providecommand \@url [1]{\endgroup\@href {#1}{\urlprefix }}%
\providecommand \urlprefix  [0]{URL }%
\providecommand \Eprint [0]{\href }%
\providecommand \doibase [0]{https://doi.org/}%
\providecommand \selectlanguage [0]{\@gobble}%
\providecommand \bibinfo  [0]{\@secondoftwo}%
\providecommand \bibfield  [0]{\@secondoftwo}%
\providecommand \translation [1]{[#1]}%
\providecommand \BibitemOpen [0]{}%
\providecommand \bibitemStop [0]{}%
\providecommand \bibitemNoStop [0]{.\EOS\space}%
\providecommand \EOS [0]{\spacefactor3000\relax}%
\providecommand \BibitemShut  [1]{\csname bibitem#1\endcsname}%
\let\auto@bib@innerbib\@empty
%</preamble>
\bibitem [{\citenamefont {Bowick}\ \emph {et~al.}(2022)\citenamefont {Bowick},
  \citenamefont {Fakhri}, \citenamefont {Marchetti},\ and\ \citenamefont
  {Ramaswamy}}]{bowick2022symmetry}%
  \BibitemOpen
  \bibfield  {author} {\bibinfo {author} {\bibfnamefont {M.~J.}\ \bibnamefont
  {Bowick}}, \bibinfo {author} {\bibfnamefont {N.}~\bibnamefont {Fakhri}},
  \bibinfo {author} {\bibfnamefont {M.~C.}\ \bibnamefont {Marchetti}},\ and\
  \bibinfo {author} {\bibfnamefont {S.}~\bibnamefont {Ramaswamy}},\ }\bibfield
  {title} {\bibinfo {title} {{Symmetry, Thermodynamics, and Topology in Active
  Matter}},\ }\href {https://doi.org/10.1103/PhysRevX.12.010501} {\bibfield
  {journal} {\bibinfo  {journal} {Phys. Rev. X}\ }\textbf {\bibinfo {volume}
  {12}},\ \bibinfo {pages} {010501} (\bibinfo {year} {2022})}\BibitemShut
  {NoStop}%
\bibitem [{\citenamefont {Golestanian}(2024)}]{golestanian2024non-reciprocal}%
  \BibitemOpen
  \bibfield  {author} {\bibinfo {author} {\bibfnamefont {R.}~\bibnamefont
  {Golestanian}},\ }\bibfield  {title} {\bibinfo {title} {Non-reciprocal
  active-matter: a tale of ``loving hate, brawling love'' across the scales},\
  }\href {https://doi.org/10.1051/epn/2024305} {\bibfield  {journal} {\bibinfo
  {journal} {Europhys. News}\ }\textbf {\bibinfo {volume} {55}},\ \bibinfo
  {pages} {12} (\bibinfo {year} {2024})}\BibitemShut {NoStop}%
\bibitem [{\citenamefont {Saha}\ \emph {et~al.}(2020)\citenamefont {Saha},
  \citenamefont {Agudo-Canalejo},\ and\ \citenamefont
  {Golestanian}}]{saha2020scalar}%
  \BibitemOpen
  \bibfield  {author} {\bibinfo {author} {\bibfnamefont {S.}~\bibnamefont
  {Saha}}, \bibinfo {author} {\bibfnamefont {J.}~\bibnamefont
  {Agudo-Canalejo}},\ and\ \bibinfo {author} {\bibfnamefont {R.}~\bibnamefont
  {Golestanian}},\ }\bibfield  {title} {\bibinfo {title} {{Scalar active
  mixtures: The nonreciprocal Cahn-Hilliard model}},\ }\href
  {https://doi.org/10.1103/PhysRevX.10.041009} {\bibfield  {journal} {\bibinfo
  {journal} {Phys. Rev. X}\ }\textbf {\bibinfo {volume} {10}},\ \bibinfo
  {pages} {041009} (\bibinfo {year} {2020})}\BibitemShut {NoStop}%
\bibitem [{\citenamefont {You}\ \emph {et~al.}(2020)\citenamefont {You},
  \citenamefont {Baskaran},\ and\ \citenamefont
  {Marchetti}}]{you2020nonreciprocity}%
  \BibitemOpen
  \bibfield  {author} {\bibinfo {author} {\bibfnamefont {Z.}~\bibnamefont
  {You}}, \bibinfo {author} {\bibfnamefont {A.}~\bibnamefont {Baskaran}},\ and\
  \bibinfo {author} {\bibfnamefont {M.~C.}\ \bibnamefont {Marchetti}},\
  }\bibfield  {title} {\bibinfo {title} {Nonreciprocity as a generic route to
  traveling states},\ }\href {https://doi.org/10.1073/pnas.2010318117}
  {\bibfield  {journal} {\bibinfo  {journal} {Proc. Natl. Acad. Sci. USA.}\
  }\textbf {\bibinfo {volume} {117}},\ \bibinfo {pages} {19767} (\bibinfo
  {year} {2020})}\BibitemShut {NoStop}%
\bibitem [{\citenamefont {Fruchart}\ \emph {et~al.}(2021)\citenamefont
  {Fruchart}, \citenamefont {Hanai}, \citenamefont {Littlewood},\ and\
  \citenamefont {Vitelli}}]{fruchart2021non}%
  \BibitemOpen
  \bibfield  {author} {\bibinfo {author} {\bibfnamefont {M.}~\bibnamefont
  {Fruchart}}, \bibinfo {author} {\bibfnamefont {R.}~\bibnamefont {Hanai}},
  \bibinfo {author} {\bibfnamefont {P.~B.}\ \bibnamefont {Littlewood}},\ and\
  \bibinfo {author} {\bibfnamefont {V.}~\bibnamefont {Vitelli}},\ }\bibfield
  {title} {\bibinfo {title} {Non-reciprocal phase transitions},\ }\href
  {https://doi.org/10.1038/s41586-021-03375-9} {\bibfield  {journal} {\bibinfo
  {journal} {Nature}\ }\textbf {\bibinfo {volume} {592}},\ \bibinfo {pages}
  {363} (\bibinfo {year} {2021})}\BibitemShut {NoStop}%
\bibitem [{\citenamefont {Guislain}\ and\ \citenamefont
  {Bertin}(2024)}]{guislain2024discontinuous}%
  \BibitemOpen
  \bibfield  {author} {\bibinfo {author} {\bibfnamefont {L.}~\bibnamefont
  {Guislain}}\ and\ \bibinfo {author} {\bibfnamefont {E.}~\bibnamefont
  {Bertin}},\ }\bibfield  {title} {\bibinfo {title} {Discontinuous phase
  transition from ferromagnetic to oscillating states in a nonequilibrium
  mean-field spin model},\ }\href
  {https://doi.org/https://doi.org/10.1103/PhysRevE.109.034131} {\bibfield
  {journal} {\bibinfo  {journal} {Phys. Rev. E}\ }\textbf {\bibinfo {volume}
  {109}},\ \bibinfo {pages} {034131} (\bibinfo {year} {2024})}\BibitemShut
  {NoStop}%
\bibitem [{\citenamefont {Fruchart}\ \emph {et~al.}(2023)\citenamefont
  {Fruchart}, \citenamefont {Scheibner},\ and\ \citenamefont
  {Vitelli}}]{fruchart2023odd}%
  \BibitemOpen
  \bibfield  {author} {\bibinfo {author} {\bibfnamefont {M.}~\bibnamefont
  {Fruchart}}, \bibinfo {author} {\bibfnamefont {C.}~\bibnamefont
  {Scheibner}},\ and\ \bibinfo {author} {\bibfnamefont {V.}~\bibnamefont
  {Vitelli}},\ }\bibfield  {title} {\bibinfo {title} {Odd viscosity and odd
  elasticity},\ }\href
  {https://doi.org/https://doi.org/10.1146/annurev-conmatphys-040821-125506}
  {\bibfield  {journal} {\bibinfo  {journal} {Annu. Rev. Condens. Matter
  Phys.}\ }\textbf {\bibinfo {volume} {14}},\ \bibinfo {pages} {471} (\bibinfo
  {year} {2023})}\BibitemShut {NoStop}%
\bibitem [{\citenamefont {Dadhichi}\ \emph {et~al.}(2020)\citenamefont
  {Dadhichi}, \citenamefont {Kethapelli}, \citenamefont {Chajwa}, \citenamefont
  {Ramaswamy},\ and\ \citenamefont {Maitra}}]{dadhichi2020nonmutual}%
  \BibitemOpen
  \bibfield  {author} {\bibinfo {author} {\bibfnamefont {L.~P.}\ \bibnamefont
  {Dadhichi}}, \bibinfo {author} {\bibfnamefont {J.}~\bibnamefont
  {Kethapelli}}, \bibinfo {author} {\bibfnamefont {R.}~\bibnamefont {Chajwa}},
  \bibinfo {author} {\bibfnamefont {S.}~\bibnamefont {Ramaswamy}},\ and\
  \bibinfo {author} {\bibfnamefont {A.}~\bibnamefont {Maitra}},\ }\bibfield
  {title} {\bibinfo {title} {Nonmutual torques and the unimportance of motility
  for long-range order in two-dimensional flocks},\ }\href
  {https://doi.org/10.1103/PhysRevE.101.052601} {\bibfield  {journal} {\bibinfo
   {journal} {Phys. Rev. E}\ }\textbf {\bibinfo {volume} {101}},\ \bibinfo
  {pages} {052601} (\bibinfo {year} {2020})}\BibitemShut {NoStop}%
\bibitem [{\citenamefont {Pisegna}\ \emph {et~al.}(2024)\citenamefont
  {Pisegna}, \citenamefont {Saha},\ and\ \citenamefont
  {Golestanian}}]{pisegna2024emergent}%
  \BibitemOpen
  \bibfield  {author} {\bibinfo {author} {\bibfnamefont {G.}~\bibnamefont
  {Pisegna}}, \bibinfo {author} {\bibfnamefont {S.}~\bibnamefont {Saha}},\ and\
  \bibinfo {author} {\bibfnamefont {R.}~\bibnamefont {Golestanian}},\
  }\bibfield  {title} {\bibinfo {title} {Emergent polar order in non-polar
  mixtures with non-reciprocal interactions},\ }\href
  {https://arxiv.org/abs/2404.05396} {\bibfield  {journal} {\bibinfo  {journal}
  {arXiv preprint arXiv:2404.05396}\ } (\bibinfo {year} {2024})}\BibitemShut
  {NoStop}%
\bibitem [{\citenamefont {Morfill}\ and\ \citenamefont
  {Ivlev}(2009)}]{morfill2009complex}%
  \BibitemOpen
  \bibfield  {author} {\bibinfo {author} {\bibfnamefont {G.~E.}\ \bibnamefont
  {Morfill}}\ and\ \bibinfo {author} {\bibfnamefont {A.~V.}\ \bibnamefont
  {Ivlev}},\ }\bibfield  {title} {\bibinfo {title} {Complex plasmas: An
  interdisciplinary research field},\ }\href
  {https://doi.org/10.1103/RevModPhys.81.1353} {\bibfield  {journal} {\bibinfo
  {journal} {Rev. Mod. Phys.}\ }\textbf {\bibinfo {volume} {81}},\ \bibinfo
  {pages} {1353} (\bibinfo {year} {2009})}\BibitemShut {NoStop}%
\bibitem [{\citenamefont {Chiacchio}\ \emph {et~al.}(2023)\citenamefont
  {Chiacchio}, \citenamefont {Nunnenkamp},\ and\ \citenamefont
  {Brunelli}}]{chiacchio2023nonreciprocal}%
  \BibitemOpen
  \bibfield  {author} {\bibinfo {author} {\bibfnamefont {E.~I.~R.}\
  \bibnamefont {Chiacchio}}, \bibinfo {author} {\bibfnamefont {A.}~\bibnamefont
  {Nunnenkamp}},\ and\ \bibinfo {author} {\bibfnamefont {M.}~\bibnamefont
  {Brunelli}},\ }\bibfield  {title} {\bibinfo {title} {{Nonreciprocal Dicke
  Model}},\ }\href {https://doi.org/10.1103/PhysRevLett.131.113602} {\bibfield
  {journal} {\bibinfo  {journal} {Phys. Rev. Lett.}\ }\textbf {\bibinfo
  {volume} {131}},\ \bibinfo {pages} {113602} (\bibinfo {year}
  {2023})}\BibitemShut {NoStop}%
\bibitem [{\citenamefont {Romano}\ \emph {et~al.}(2024)\citenamefont {Romano},
  \citenamefont {Mahault},\ and\ \citenamefont
  {Golestanian}}]{romano2024dynamical}%
  \BibitemOpen
  \bibfield  {author} {\bibinfo {author} {\bibfnamefont {J.}~\bibnamefont
  {Romano}}, \bibinfo {author} {\bibfnamefont {B.}~\bibnamefont {Mahault}},\
  and\ \bibinfo {author} {\bibfnamefont {R.}~\bibnamefont {Golestanian}},\
  }\bibfield  {title} {\bibinfo {title} {Dynamical theory of topological
  defects {II}: universal aspects of defect motion},\ }\href
  {https://doi.org/10.1088/1742-5468/ad2ddb} {\bibfield  {journal} {\bibinfo
  {journal} {J. Stat. Mech.: Theory Exp.}\ }\textbf {\bibinfo {volume}
  {2024}}\bibinfo  {number} { (3)},\ \bibinfo {pages} {033208}}\BibitemShut
  {NoStop}%
\bibitem [{\citenamefont {Soto}\ and\ \citenamefont
  {Golestanian}(2014)}]{soto2014self}%
  \BibitemOpen
\bibfield  {number} {  }\bibfield  {author} {\bibinfo {author} {\bibfnamefont
  {R.}~\bibnamefont {Soto}}\ and\ \bibinfo {author} {\bibfnamefont
  {R.}~\bibnamefont {Golestanian}},\ }\bibfield  {title} {\bibinfo {title}
  {Self-assembly of catalytically active colloidal molecules: Tailoring
  activity through surface chemistry},\ }\href
  {https://doi.org/10.1103/PhysRevLett.112.068301} {\bibfield  {journal}
  {\bibinfo  {journal} {Phys. Rev. Lett.}\ }\textbf {\bibinfo {volume} {112}},\
  \bibinfo {pages} {068301} (\bibinfo {year} {2014})}\BibitemShut {NoStop}%
\bibitem [{\citenamefont {Meredith}\ \emph {et~al.}(2020)\citenamefont
  {Meredith}, \citenamefont {Moerman}, \citenamefont {Groenewold},
  \citenamefont {Chiu}, \citenamefont {Kegel}, \citenamefont {van Blaaderen},\
  and\ \citenamefont {Zarzar}}]{meredith2020predator}%
  \BibitemOpen
  \bibfield  {author} {\bibinfo {author} {\bibfnamefont {C.~H.}\ \bibnamefont
  {Meredith}}, \bibinfo {author} {\bibfnamefont {P.~G.}\ \bibnamefont
  {Moerman}}, \bibinfo {author} {\bibfnamefont {J.}~\bibnamefont {Groenewold}},
  \bibinfo {author} {\bibfnamefont {Y.-J.}\ \bibnamefont {Chiu}}, \bibinfo
  {author} {\bibfnamefont {W.~K.}\ \bibnamefont {Kegel}}, \bibinfo {author}
  {\bibfnamefont {A.}~\bibnamefont {van Blaaderen}},\ and\ \bibinfo {author}
  {\bibfnamefont {L.~D.}\ \bibnamefont {Zarzar}},\ }\bibfield  {title}
  {\bibinfo {title} {Predator--prey interactions between droplets driven by
  non-reciprocal oil exchange},\ }\href
  {https://doi.org/https://doi.org/10.1038/s41557-020-00575-0} {\bibfield
  {journal} {\bibinfo  {journal} {Nat. Chem.}\ }\textbf {\bibinfo {volume}
  {12}},\ \bibinfo {pages} {1136} (\bibinfo {year} {2020})}\BibitemShut
  {NoStop}%
\bibitem [{\citenamefont {Liu}\ \emph {et~al.}()\citenamefont {Liu},
  \citenamefont {Kailasham}, \citenamefont {Moerman}, \citenamefont {Khair},\
  and\ \citenamefont {Zarzar}}]{liu2024self}%
  \BibitemOpen
  \bibfield  {author} {\bibinfo {author} {\bibfnamefont {Y.}~\bibnamefont
  {Liu}}, \bibinfo {author} {\bibfnamefont {R.}~\bibnamefont {Kailasham}},
  \bibinfo {author} {\bibfnamefont {P.~G.}\ \bibnamefont {Moerman}}, \bibinfo
  {author} {\bibfnamefont {A.~S.}\ \bibnamefont {Khair}},\ and\ \bibinfo
  {author} {\bibfnamefont {L.}~\bibnamefont {Zarzar}},\ }\bibfield  {title}
  {\bibinfo {title} {Self-organized patterns in non-reciprocal active droplet
  systems},\ }\href {https://doi.org/https://doi.org/10.1002/anie.202409382}
  {\bibfield  {journal} {\bibinfo  {journal} {Angew. Chem. Int. Ed.}\ }\textbf
  {\bibinfo {volume} {2024}},\ \bibinfo {pages} {e202409382}}\BibitemShut
  {NoStop}%
\bibitem [{\citenamefont {Uchida}\ and\ \citenamefont
  {Golestanian}(2010)}]{uchida2010synchronization}%
  \BibitemOpen
  \bibfield  {author} {\bibinfo {author} {\bibfnamefont {N.}~\bibnamefont
  {Uchida}}\ and\ \bibinfo {author} {\bibfnamefont {R.}~\bibnamefont
  {Golestanian}},\ }\bibfield  {title} {\bibinfo {title} {Synchronization and
  collective dynamics in a carpet of microfluidic rotors},\ }\href
  {https://doi.org/10.1103/PhysRevLett.104.178103} {\bibfield  {journal}
  {\bibinfo  {journal} {Phys. Rev. Lett.}\ }\textbf {\bibinfo {volume} {104}},\
  \bibinfo {pages} {178103} (\bibinfo {year} {2010})}\BibitemShut {NoStop}%
\bibitem [{\citenamefont {Tan}\ \emph {et~al.}(2022)\citenamefont {Tan},
  \citenamefont {Mietke}, \citenamefont {Li}, \citenamefont {Chen},
  \citenamefont {Higinbotham}, \citenamefont {Foster}, \citenamefont {Gokhale},
  \citenamefont {Dunkel},\ and\ \citenamefont {Fakhri}}]{tan2022odd}%
  \BibitemOpen
  \bibfield  {author} {\bibinfo {author} {\bibfnamefont {T.~H.}\ \bibnamefont
  {Tan}}, \bibinfo {author} {\bibfnamefont {A.}~\bibnamefont {Mietke}},
  \bibinfo {author} {\bibfnamefont {J.}~\bibnamefont {Li}}, \bibinfo {author}
  {\bibfnamefont {Y.}~\bibnamefont {Chen}}, \bibinfo {author} {\bibfnamefont
  {H.}~\bibnamefont {Higinbotham}}, \bibinfo {author} {\bibfnamefont {P.~J.}\
  \bibnamefont {Foster}}, \bibinfo {author} {\bibfnamefont {S.}~\bibnamefont
  {Gokhale}}, \bibinfo {author} {\bibfnamefont {J.}~\bibnamefont {Dunkel}},\
  and\ \bibinfo {author} {\bibfnamefont {N.}~\bibnamefont {Fakhri}},\
  }\bibfield  {title} {\bibinfo {title} {Odd dynamics of living chiral
  crystals},\ }\href {https://doi.org/10.1038/s41586-022-04889-6} {\bibfield
  {journal} {\bibinfo  {journal} {Nature}\ }\textbf {\bibinfo {volume} {607}},\
  \bibinfo {pages} {287} (\bibinfo {year} {2022})}\BibitemShut {NoStop}%
\bibitem [{\citenamefont {Maity}\ and\ \citenamefont
  {Morin}(2023)}]{maity2023spontaneous}%
  \BibitemOpen
  \bibfield  {author} {\bibinfo {author} {\bibfnamefont {S.}~\bibnamefont
  {Maity}}\ and\ \bibinfo {author} {\bibfnamefont {A.}~\bibnamefont {Morin}},\
  }\bibfield  {title} {\bibinfo {title} {Spontaneous demixing of binary
  colloidal flocks},\ }\href
  {https://doi.org/https://doi.org/10.1103/PhysRevLett.131.178304} {\bibfield
  {journal} {\bibinfo  {journal} {Phys. Rev. Lett.}\ }\textbf {\bibinfo
  {volume} {131}},\ \bibinfo {pages} {178304} (\bibinfo {year}
  {2023})}\BibitemShut {NoStop}%
\bibitem [{\citenamefont {Banerjee}\ \emph {et~al.}(2022)\citenamefont
  {Banerjee}, \citenamefont {Mandal}, \citenamefont {Banerjee}, \citenamefont
  {Thutupalli},\ and\ \citenamefont {Rao}}]{banerjee2022unjamming}%
  \BibitemOpen
  \bibfield  {author} {\bibinfo {author} {\bibfnamefont {J.~P.}\ \bibnamefont
  {Banerjee}}, \bibinfo {author} {\bibfnamefont {R.}~\bibnamefont {Mandal}},
  \bibinfo {author} {\bibfnamefont {D.~S.}\ \bibnamefont {Banerjee}}, \bibinfo
  {author} {\bibfnamefont {S.}~\bibnamefont {Thutupalli}},\ and\ \bibinfo
  {author} {\bibfnamefont {M.}~\bibnamefont {Rao}},\ }\bibfield  {title}
  {\bibinfo {title} {Unjamming and emergent nonreciprocity in active ploughing
  through a compressible viscoelastic fluid},\ }\href
  {https://doi.org/10.1038/s41467-022-31984-z} {\bibfield  {journal} {\bibinfo
  {journal} {Nat. Commun.}\ }\textbf {\bibinfo {volume} {13}},\ \bibinfo
  {pages} {1} (\bibinfo {year} {2022})}\BibitemShut {NoStop}%
\bibitem [{\citenamefont {Yamanaka}\ and\ \citenamefont
  {Kondo}(2014)}]{yamanaka2014invitro}%
  \BibitemOpen
  \bibfield  {author} {\bibinfo {author} {\bibfnamefont {H.}~\bibnamefont
  {Yamanaka}}\ and\ \bibinfo {author} {\bibfnamefont {S.}~\bibnamefont
  {Kondo}},\ }\bibfield  {title} {\bibinfo {title} {In vitro analysis suggests
  that difference in cell movement during direct interaction can generate
  various pigment patterns in vivo},\ }\href
  {https://doi.org/10.1073/pnas.1315416111} {\bibfield  {journal} {\bibinfo
  {journal} {Proc. Natl. Acad. Sci. U.S.A.}\ }\textbf {\bibinfo {volume}
  {111}},\ \bibinfo {pages} {1867} (\bibinfo {year} {2014})}\BibitemShut
  {NoStop}%
\bibitem [{\citenamefont {Haluts}\ \emph {et~al.}(2024)\citenamefont {Haluts},
  \citenamefont {Gorbonos},\ and\ \citenamefont {Gov}}]{haluts2024active}%
  \BibitemOpen
  \bibfield  {author} {\bibinfo {author} {\bibfnamefont {A.}~\bibnamefont
  {Haluts}}, \bibinfo {author} {\bibfnamefont {D.}~\bibnamefont {Gorbonos}},\
  and\ \bibinfo {author} {\bibfnamefont {N.~S.}\ \bibnamefont {Gov}},\
  }\bibfield  {title} {\bibinfo {title} {Active particle models for animal
  behavior based on effective nonreciprocal forces and potentials},\ }\href
  {https://arxiv.org/abs/2401.14850} {\bibfield  {journal} {\bibinfo  {journal}
  {arXiv preprint arXiv:2401.14850}\ } (\bibinfo {year} {2024})}\BibitemShut
  {NoStop}%
\bibitem [{\citenamefont {Blumenthal}\ \emph {et~al.}(2024)\citenamefont
  {Blumenthal}, \citenamefont {Rocks},\ and\ \citenamefont
  {Mehta}}]{blumenthal2024phase}%
  \BibitemOpen
  \bibfield  {author} {\bibinfo {author} {\bibfnamefont {E.}~\bibnamefont
  {Blumenthal}}, \bibinfo {author} {\bibfnamefont {J.~W.}\ \bibnamefont
  {Rocks}},\ and\ \bibinfo {author} {\bibfnamefont {P.}~\bibnamefont {Mehta}},\
  }\bibfield  {title} {\bibinfo {title} {Phase transition to chaos in complex
  ecosystems with nonreciprocal species-resource interactions},\ }\href
  {https://doi.org/10.1103/PhysRevLett.132.127401} {\bibfield  {journal}
  {\bibinfo  {journal} {Phys. Rev. Lett.}\ }\textbf {\bibinfo {volume} {132}},\
  \bibinfo {pages} {127401} (\bibinfo {year} {2024})}\BibitemShut {NoStop}%
\bibitem [{\citenamefont {Ivlev}\ \emph {et~al.}(2015)\citenamefont {Ivlev},
  \citenamefont {Bartnick}, \citenamefont {Heinen}, \citenamefont {Du},
  \citenamefont {Nosenko},\ and\ \citenamefont
  {L\"owen}}]{ivlev2015statistical}%
  \BibitemOpen
  \bibfield  {author} {\bibinfo {author} {\bibfnamefont {A.~V.}\ \bibnamefont
  {Ivlev}}, \bibinfo {author} {\bibfnamefont {J.}~\bibnamefont {Bartnick}},
  \bibinfo {author} {\bibfnamefont {M.}~\bibnamefont {Heinen}}, \bibinfo
  {author} {\bibfnamefont {C.-R.}\ \bibnamefont {Du}}, \bibinfo {author}
  {\bibfnamefont {V.}~\bibnamefont {Nosenko}},\ and\ \bibinfo {author}
  {\bibfnamefont {H.}~\bibnamefont {L\"owen}},\ }\bibfield  {title} {\bibinfo
  {title} {{Statistical Mechanics where Newton's Third Law is Broken}},\ }\href
  {https://doi.org/10.1103/PhysRevX.5.011035} {\bibfield  {journal} {\bibinfo
  {journal} {Phys. Rev. X}\ }\textbf {\bibinfo {volume} {5}},\ \bibinfo {pages}
  {011035} (\bibinfo {year} {2015})}\BibitemShut {NoStop}%
\bibitem [{\citenamefont {Chiu}\ and\ \citenamefont
  {Omar}(2023)}]{chiu2023phase}%
  \BibitemOpen
  \bibfield  {author} {\bibinfo {author} {\bibfnamefont {Y.-J.}\ \bibnamefont
  {Chiu}}\ and\ \bibinfo {author} {\bibfnamefont {A.~K.}\ \bibnamefont
  {Omar}},\ }\bibfield  {title} {\bibinfo {title} {{Phase coexistence
  implications of violating Newton's third law}},\ }\href
  {https://doi.org/10.1063/5.0146822} {\bibfield  {journal} {\bibinfo
  {journal} {J. Chem. Phys.}\ }\textbf {\bibinfo {volume} {158}},\ \bibinfo
  {pages} {164903} (\bibinfo {year} {2023})}\BibitemShut {NoStop}%
\bibitem [{\citenamefont {Mandal}\ \emph {et~al.}(2024)\citenamefont {Mandal},
  \citenamefont {Jaramillo},\ and\ \citenamefont
  {Sollich}}]{mandal2024robustness}%
  \BibitemOpen
  \bibfield  {author} {\bibinfo {author} {\bibfnamefont {R.}~\bibnamefont
  {Mandal}}, \bibinfo {author} {\bibfnamefont {S.~S.}\ \bibnamefont
  {Jaramillo}},\ and\ \bibinfo {author} {\bibfnamefont {P.}~\bibnamefont
  {Sollich}},\ }\bibfield  {title} {\bibinfo {title} {Robustness of traveling
  states in generic nonreciprocal mixtures},\ }\href
  {https://doi.org/https://doi.org/10.1103/PhysRevE.109.L062602} {\bibfield
  {journal} {\bibinfo  {journal} {Phys. Rev. E}\ }\textbf {\bibinfo {volume}
  {109}},\ \bibinfo {pages} {L062602} (\bibinfo {year} {2024})}\BibitemShut
  {NoStop}%
\bibitem [{\citenamefont {Hanai}(2024)}]{hanai2024nonreciprocal}%
  \BibitemOpen
  \bibfield  {author} {\bibinfo {author} {\bibfnamefont {R.}~\bibnamefont
  {Hanai}},\ }\bibfield  {title} {\bibinfo {title} {Nonreciprocal frustration:
  Time crystalline order-by-disorder phenomenon and a spin-glass-like state},\
  }\href {https://doi.org/https://doi.org/10.1103/PhysRevX.14.011029}
  {\bibfield  {journal} {\bibinfo  {journal} {Phys. Rev. X}\ }\textbf {\bibinfo
  {volume} {14}},\ \bibinfo {pages} {011029} (\bibinfo {year}
  {2024})}\BibitemShut {NoStop}%
\bibitem [{\citenamefont {Kreienkamp}\ and\ \citenamefont
  {Klapp}(2024)}]{kreienkamp2024non}%
  \BibitemOpen
  \bibfield  {author} {\bibinfo {author} {\bibfnamefont {K.~L.}\ \bibnamefont
  {Kreienkamp}}\ and\ \bibinfo {author} {\bibfnamefont {S.~H.}\ \bibnamefont
  {Klapp}},\ }\bibfield  {title} {\bibinfo {title} {Non-reciprocal alignment
  induces asymmetric clustering in active repulsive mixtures},\ }\href
  {https://doi.org/10.48550/arXiv.2403.19291} {\bibfield  {journal} {\bibinfo
  {journal} {arXiv preprint arXiv:2403.19291}\ } (\bibinfo {year}
  {2024})}\BibitemShut {NoStop}%
\bibitem [{\citenamefont {Agudo-Canalejo}\ and\ \citenamefont
  {Golestanian}(2019)}]{agudo2019active}%
  \BibitemOpen
  \bibfield  {author} {\bibinfo {author} {\bibfnamefont {J.}~\bibnamefont
  {Agudo-Canalejo}}\ and\ \bibinfo {author} {\bibfnamefont {R.}~\bibnamefont
  {Golestanian}},\ }\bibfield  {title} {\bibinfo {title} {{Active Phase
  Separation in Mixtures of Chemically Interacting Particles}},\ }\href
  {https://doi.org/10.1103/PhysRevLett.123.018101} {\bibfield  {journal}
  {\bibinfo  {journal} {Phys. Rev. Lett.}\ }\textbf {\bibinfo {volume} {123}},\
  \bibinfo {pages} {018101} (\bibinfo {year} {2019})}\BibitemShut {NoStop}%
\bibitem [{\citenamefont {Duan}\ \emph {et~al.}(2023)\citenamefont {Duan},
  \citenamefont {Agudo-Canalejo}, \citenamefont {Golestanian},\ and\
  \citenamefont {Mahault}}]{duan2023dynamical}%
  \BibitemOpen
  \bibfield  {author} {\bibinfo {author} {\bibfnamefont {Y.}~\bibnamefont
  {Duan}}, \bibinfo {author} {\bibfnamefont {J.}~\bibnamefont
  {Agudo-Canalejo}}, \bibinfo {author} {\bibfnamefont {R.}~\bibnamefont
  {Golestanian}},\ and\ \bibinfo {author} {\bibfnamefont {B.}~\bibnamefont
  {Mahault}},\ }\bibfield  {title} {\bibinfo {title} {{Dynamical Pattern
  Formation without Self-Attraction in Quorum-Sensing Active Matter: The
  Interplay between Nonreciprocity and Motility}},\ }\href
  {https://doi.org/https://doi.org/10.1103/PhysRevLett.131.148301} {\bibfield
  {journal} {\bibinfo  {journal} {Phys. Rev. Lett.}\ }\textbf {\bibinfo
  {volume} {131}},\ \bibinfo {pages} {148301} (\bibinfo {year}
  {2023})}\BibitemShut {NoStop}%
\bibitem [{\citenamefont {Dinelli}\ \emph {et~al.}(2023)\citenamefont
  {Dinelli}, \citenamefont {O'Byrne}, \citenamefont {Curatolo}, \citenamefont
  {Zhao}, \citenamefont {Sollich},\ and\ \citenamefont
  {Tailleur}}]{dinelli2023non}%
  \BibitemOpen
  \bibfield  {author} {\bibinfo {author} {\bibfnamefont {A.}~\bibnamefont
  {Dinelli}}, \bibinfo {author} {\bibfnamefont {J.}~\bibnamefont {O'Byrne}},
  \bibinfo {author} {\bibfnamefont {A.}~\bibnamefont {Curatolo}}, \bibinfo
  {author} {\bibfnamefont {Y.}~\bibnamefont {Zhao}}, \bibinfo {author}
  {\bibfnamefont {P.}~\bibnamefont {Sollich}},\ and\ \bibinfo {author}
  {\bibfnamefont {J.}~\bibnamefont {Tailleur}},\ }\bibfield  {title} {\bibinfo
  {title} {Non-reciprocity across scales in active mixtures},\ }\href
  {https://doi.org/10.1038/s41467-023-42713-5} {\bibfield  {journal} {\bibinfo
  {journal} {Nat. Commun.}\ }\textbf {\bibinfo {volume} {14}},\ \bibinfo
  {pages} {7035} (\bibinfo {year} {2023})}\BibitemShut {NoStop}%
\bibitem [{\citenamefont {Ouazan-Reboul}\ \emph {et~al.}(2023)\citenamefont
  {Ouazan-Reboul}, \citenamefont {Agudo-Canalejo},\ and\ \citenamefont
  {Golestanian}}]{ouazan2023self}%
  \BibitemOpen
  \bibfield  {author} {\bibinfo {author} {\bibfnamefont {V.}~\bibnamefont
  {Ouazan-Reboul}}, \bibinfo {author} {\bibfnamefont {J.}~\bibnamefont
  {Agudo-Canalejo}},\ and\ \bibinfo {author} {\bibfnamefont {R.}~\bibnamefont
  {Golestanian}},\ }\bibfield  {title} {\bibinfo {title} {Self-organization of
  primitive metabolic cycles due to non-reciprocal interactions},\ }\href
  {https://doi.org/10.1038/s41467-023-40241-w} {\bibfield  {journal} {\bibinfo
  {journal} {Nat. Commun.}\ }\textbf {\bibinfo {volume} {14}},\ \bibinfo
  {pages} {4496} (\bibinfo {year} {2023})}\BibitemShut {NoStop}%
\bibitem [{\citenamefont {Tucci}\ \emph {et~al.}(2024)\citenamefont {Tucci},
  \citenamefont {Golestanian},\ and\ \citenamefont
  {Saha}}]{tucci2024nonreciprocal}%
  \BibitemOpen
  \bibfield  {author} {\bibinfo {author} {\bibfnamefont {G.}~\bibnamefont
  {Tucci}}, \bibinfo {author} {\bibfnamefont {R.}~\bibnamefont {Golestanian}},\
  and\ \bibinfo {author} {\bibfnamefont {S.}~\bibnamefont {Saha}},\ }\bibfield
  {title} {\bibinfo {title} {Nonreciprocal collective dynamics in a mixture of
  phoretic {Janus} colloids},\ }\href
  {https://doi.org/10.1088/1367-2630/ad50ff} {\bibfield  {journal} {\bibinfo
  {journal} {New J. Phys.}\ }\textbf {\bibinfo {volume} {26}},\ \bibinfo
  {pages} {073006} (\bibinfo {year} {2024})}\BibitemShut {NoStop}%
\bibitem [{\citenamefont {Papenfort}\ and\ \citenamefont
  {Bassler}(2016)}]{Papenfort2016review}%
  \BibitemOpen
  \bibfield  {author} {\bibinfo {author} {\bibfnamefont {K.}~\bibnamefont
  {Papenfort}}\ and\ \bibinfo {author} {\bibfnamefont {B.~L.}\ \bibnamefont
  {Bassler}},\ }\bibfield  {title} {\bibinfo {title} {Quorum sensing
  signal--response systems in gram-negative bacteria},\ }\href
  {https://doi.org/10.1038/nrmicro.2016.89} {\bibfield  {journal} {\bibinfo
  {journal} {Nat. Rev. Microbiol.}\ }\textbf {\bibinfo {volume} {14}},\
  \bibinfo {pages} {576} (\bibinfo {year} {2016})}\BibitemShut {NoStop}%
\bibitem [{\citenamefont {Weiss}\ \emph {et~al.}(2008)\citenamefont {Weiss},
  \citenamefont {Badalamenti}, \citenamefont {Weaver}, \citenamefont {Tascone},
  \citenamefont {Weiss}, \citenamefont {Richard},\ and\ \citenamefont
  {Cirino}}]{Weiss2008}%
  \BibitemOpen
  \bibfield  {author} {\bibinfo {author} {\bibfnamefont {L.~E.}\ \bibnamefont
  {Weiss}}, \bibinfo {author} {\bibfnamefont {J.~P.}\ \bibnamefont
  {Badalamenti}}, \bibinfo {author} {\bibfnamefont {L.~J.}\ \bibnamefont
  {Weaver}}, \bibinfo {author} {\bibfnamefont {A.~R.}\ \bibnamefont {Tascone}},
  \bibinfo {author} {\bibfnamefont {P.~S.}\ \bibnamefont {Weiss}}, \bibinfo
  {author} {\bibfnamefont {T.~L.}\ \bibnamefont {Richard}},\ and\ \bibinfo
  {author} {\bibfnamefont {P.~C.}\ \bibnamefont {Cirino}},\ }\bibfield  {title}
  {\bibinfo {title} {Engineering motility as a phenotypic response to
  {LuxI/R}-dependent quorum sensing in \textit{Escherichia coli}},\ }\href
  {https://doi.org/https://doi.org/10.1002/bit.21862} {\bibfield  {journal}
  {\bibinfo  {journal} {Biotechnol. Bioeng.}\ }\textbf {\bibinfo {volume}
  {100}},\ \bibinfo {pages} {1251} (\bibinfo {year} {2008})}\BibitemShut
  {NoStop}%
\bibitem [{\citenamefont {Liu}\ \emph {et~al.}(2011)\citenamefont {Liu},
  \citenamefont {Fu}, \citenamefont {Liu}, \citenamefont {Ren}, \citenamefont
  {Chau}, \citenamefont {Li}, \citenamefont {Xiang}, \citenamefont {Zeng},
  \citenamefont {Chen}, \citenamefont {Tang}, \citenamefont {Lenz},
  \citenamefont {Cui}, \citenamefont {Huang}, \citenamefont {Hwa},\ and\
  \citenamefont {Huang}}]{Liu2011Science}%
  \BibitemOpen
  \bibfield  {author} {\bibinfo {author} {\bibfnamefont {C.}~\bibnamefont
  {Liu}}, \bibinfo {author} {\bibfnamefont {X.}~\bibnamefont {Fu}}, \bibinfo
  {author} {\bibfnamefont {L.}~\bibnamefont {Liu}}, \bibinfo {author}
  {\bibfnamefont {X.}~\bibnamefont {Ren}}, \bibinfo {author} {\bibfnamefont
  {C.~K.}\ \bibnamefont {Chau}}, \bibinfo {author} {\bibfnamefont
  {S.}~\bibnamefont {Li}}, \bibinfo {author} {\bibfnamefont {L.}~\bibnamefont
  {Xiang}}, \bibinfo {author} {\bibfnamefont {H.}~\bibnamefont {Zeng}},
  \bibinfo {author} {\bibfnamefont {G.}~\bibnamefont {Chen}}, \bibinfo {author}
  {\bibfnamefont {L.-H.}\ \bibnamefont {Tang}}, \bibinfo {author}
  {\bibfnamefont {P.}~\bibnamefont {Lenz}}, \bibinfo {author} {\bibfnamefont
  {X.}~\bibnamefont {Cui}}, \bibinfo {author} {\bibfnamefont {W.}~\bibnamefont
  {Huang}}, \bibinfo {author} {\bibfnamefont {T.}~\bibnamefont {Hwa}},\ and\
  \bibinfo {author} {\bibfnamefont {J.-D.}\ \bibnamefont {Huang}},\ }\bibfield
  {title} {\bibinfo {title} {Sequential establishment of stripe patterns in an
  expanding cell population},\ }\href {https://doi.org/10.1126/science.1209042}
  {\bibfield  {journal} {\bibinfo  {journal} {Science}\ }\textbf {\bibinfo
  {volume} {334}},\ \bibinfo {pages} {238} (\bibinfo {year}
  {2011})}\BibitemShut {NoStop}%
\bibitem [{\citenamefont {Liu}\ \emph {et~al.}(2019)\citenamefont {Liu},
  \citenamefont {Patch}, \citenamefont {Bahar}, \citenamefont {Yllanes},
  \citenamefont {Welch}, \citenamefont {Marchetti}, \citenamefont
  {Thutupalli},\ and\ \citenamefont {Shaevitz}}]{liu2019self}%
  \BibitemOpen
  \bibfield  {author} {\bibinfo {author} {\bibfnamefont {G.}~\bibnamefont
  {Liu}}, \bibinfo {author} {\bibfnamefont {A.}~\bibnamefont {Patch}}, \bibinfo
  {author} {\bibfnamefont {F.}~\bibnamefont {Bahar}}, \bibinfo {author}
  {\bibfnamefont {D.}~\bibnamefont {Yllanes}}, \bibinfo {author} {\bibfnamefont
  {R.~D.}\ \bibnamefont {Welch}}, \bibinfo {author} {\bibfnamefont {M.~C.}\
  \bibnamefont {Marchetti}}, \bibinfo {author} {\bibfnamefont {S.}~\bibnamefont
  {Thutupalli}},\ and\ \bibinfo {author} {\bibfnamefont {J.~W.}\ \bibnamefont
  {Shaevitz}},\ }\bibfield  {title} {\bibinfo {title} {{Self-Driven Phase
  Transitions Drive \textit{Myxococcus xanthus} Fruiting Body Formation}},\
  }\href {https://doi.org/10.1103/PhysRevLett.122.248102} {\bibfield  {journal}
  {\bibinfo  {journal} {Phys. Rev. Lett.}\ }\textbf {\bibinfo {volume} {122}},\
  \bibinfo {pages} {248102} (\bibinfo {year} {2019})}\BibitemShut {NoStop}%
\bibitem [{\citenamefont {Curatolo}\ \emph {et~al.}(2020)\citenamefont
  {Curatolo}, \citenamefont {Zhou}, \citenamefont {Zhao}, \citenamefont {Liu},
  \citenamefont {Daerr}, \citenamefont {Tailleur},\ and\ \citenamefont
  {Huang}}]{curatolo2020cooperative}%
  \BibitemOpen
  \bibfield  {author} {\bibinfo {author} {\bibfnamefont {A.}~\bibnamefont
  {Curatolo}}, \bibinfo {author} {\bibfnamefont {N.}~\bibnamefont {Zhou}},
  \bibinfo {author} {\bibfnamefont {Y.}~\bibnamefont {Zhao}}, \bibinfo {author}
  {\bibfnamefont {C.}~\bibnamefont {Liu}}, \bibinfo {author} {\bibfnamefont
  {A.}~\bibnamefont {Daerr}}, \bibinfo {author} {\bibfnamefont
  {J.}~\bibnamefont {Tailleur}},\ and\ \bibinfo {author} {\bibfnamefont
  {J.}~\bibnamefont {Huang}},\ }\bibfield  {title} {\bibinfo {title}
  {Cooperative pattern formation in multi-component bacterial systems through
  reciprocal motility regulation},\ }\href
  {https://doi.org/10.1038/s41567-020-0964-z} {\bibfield  {journal} {\bibinfo
  {journal} {Nat. Phys.}\ }\textbf {\bibinfo {volume} {16}},\ \bibinfo {pages}
  {1152} (\bibinfo {year} {2020})}\BibitemShut {NoStop}%
\bibitem [{\citenamefont {Fragkopoulos}\ \emph {et~al.}(2021)\citenamefont
  {Fragkopoulos}, \citenamefont {Vachier}, \citenamefont {Frey}, \citenamefont
  {Le~Menn}, \citenamefont {Mazza}, \citenamefont {Wilczek}, \citenamefont
  {Zwicker},\ and\ \citenamefont {B{\"a}umchen}}]{fragkopoulos2021self}%
  \BibitemOpen
  \bibfield  {author} {\bibinfo {author} {\bibfnamefont {A.~A.}\ \bibnamefont
  {Fragkopoulos}}, \bibinfo {author} {\bibfnamefont {J.}~\bibnamefont
  {Vachier}}, \bibinfo {author} {\bibfnamefont {J.}~\bibnamefont {Frey}},
  \bibinfo {author} {\bibfnamefont {F.-M.}\ \bibnamefont {Le~Menn}}, \bibinfo
  {author} {\bibfnamefont {M.~G.}\ \bibnamefont {Mazza}}, \bibinfo {author}
  {\bibfnamefont {M.}~\bibnamefont {Wilczek}}, \bibinfo {author} {\bibfnamefont
  {D.}~\bibnamefont {Zwicker}},\ and\ \bibinfo {author} {\bibfnamefont
  {O.}~\bibnamefont {B{\"a}umchen}},\ }\bibfield  {title} {\bibinfo {title}
  {Self-generated oxygen gradients control collective aggregation of
  photosynthetic microbes},\ }\href {https://doi.org/10.1098/rsif.2021.0553}
  {\bibfield  {journal} {\bibinfo  {journal} {J. R. Soc. Interface.}\ }\textbf
  {\bibinfo {volume} {18}},\ \bibinfo {pages} {20210553} (\bibinfo {year}
  {2021})}\BibitemShut {NoStop}%
\bibitem [{\citenamefont {Courcoubetis}\ \emph {et~al.}(2022)\citenamefont
  {Courcoubetis}, \citenamefont {Gangan}, \citenamefont {Lim}, \citenamefont
  {Guo}, \citenamefont {Haas},\ and\ \citenamefont
  {Boedicker}}]{courcoubetis2022formation}%
  \BibitemOpen
  \bibfield  {author} {\bibinfo {author} {\bibfnamefont {G.}~\bibnamefont
  {Courcoubetis}}, \bibinfo {author} {\bibfnamefont {M.~S.}\ \bibnamefont
  {Gangan}}, \bibinfo {author} {\bibfnamefont {S.}~\bibnamefont {Lim}},
  \bibinfo {author} {\bibfnamefont {X.}~\bibnamefont {Guo}}, \bibinfo {author}
  {\bibfnamefont {S.}~\bibnamefont {Haas}},\ and\ \bibinfo {author}
  {\bibfnamefont {J.~Q.}\ \bibnamefont {Boedicker}},\ }\bibfield  {title}
  {\bibinfo {title} {Formation, collective motion, and merging of macroscopic
  bacterial aggregates},\ }\href {https://doi.org/10.1371/journal.pcbi.1009153}
  {\bibfield  {journal} {\bibinfo  {journal} {PLoS Comput. Biol.}\ }\textbf
  {\bibinfo {volume} {18}},\ \bibinfo {pages} {1} (\bibinfo {year}
  {2022})}\BibitemShut {NoStop}%
\bibitem [{\citenamefont {Anderson}\ and\ \citenamefont
  {Fernandez-Nieves}(2022)}]{anderson2022social}%
  \BibitemOpen
  \bibfield  {author} {\bibinfo {author} {\bibfnamefont {C.}~\bibnamefont
  {Anderson}}\ and\ \bibinfo {author} {\bibfnamefont {A.}~\bibnamefont
  {Fernandez-Nieves}},\ }\bibfield  {title} {\bibinfo {title} {Social
  interactions lead to motility-induced phase separation in fire ants},\ }\href
  {https://doi.org/https://doi.org/10.1038/s41467-022-34181-0} {\bibfield
  {journal} {\bibinfo  {journal} {Nat. Commun.}\ }\textbf {\bibinfo {volume}
  {13}},\ \bibinfo {pages} {6710} (\bibinfo {year} {2022})}\BibitemShut
  {NoStop}%
\bibitem [{\citenamefont {B{\"a}uerle}\ \emph {et~al.}(2018)\citenamefont
  {B{\"a}uerle}, \citenamefont {Fischer}, \citenamefont {Speck},\ and\
  \citenamefont {Bechinger}}]{bauerle2018self}%
  \BibitemOpen
  \bibfield  {author} {\bibinfo {author} {\bibfnamefont {T.}~\bibnamefont
  {B{\"a}uerle}}, \bibinfo {author} {\bibfnamefont {A.}~\bibnamefont
  {Fischer}}, \bibinfo {author} {\bibfnamefont {T.}~\bibnamefont {Speck}},\
  and\ \bibinfo {author} {\bibfnamefont {C.}~\bibnamefont {Bechinger}},\
  }\bibfield  {title} {\bibinfo {title} {Self-organization of active particles
  by quorum sensing rules},\ }\href
  {https://doi.org/10.1038/s41467-018-05675-7} {\bibfield  {journal} {\bibinfo
  {journal} {Nat. Commun.}\ }\textbf {\bibinfo {volume} {9}},\ \bibinfo {pages}
  {1} (\bibinfo {year} {2018})}\BibitemShut {NoStop}%
\bibitem [{\citenamefont {Lavergne}\ \emph {et~al.}(2019)\citenamefont
  {Lavergne}, \citenamefont {Wendehenne}, \citenamefont {B{\"a}uerle},\ and\
  \citenamefont {Bechinger}}]{lavergne2019Science}%
  \BibitemOpen
  \bibfield  {author} {\bibinfo {author} {\bibfnamefont {F.~A.}\ \bibnamefont
  {Lavergne}}, \bibinfo {author} {\bibfnamefont {H.}~\bibnamefont
  {Wendehenne}}, \bibinfo {author} {\bibfnamefont {T.}~\bibnamefont
  {B{\"a}uerle}},\ and\ \bibinfo {author} {\bibfnamefont {C.}~\bibnamefont
  {Bechinger}},\ }\bibfield  {title} {\bibinfo {title} {Group formation and
  cohesion of active particles with visual perception--dependent motility},\
  }\href {https://doi.org/10.1126/science.aau5347} {\bibfield  {journal}
  {\bibinfo  {journal} {Science}\ }\textbf {\bibinfo {volume} {364}},\ \bibinfo
  {pages} {70} (\bibinfo {year} {2019})}\BibitemShut {NoStop}%
\bibitem [{\citenamefont {Chen}\ \emph {et~al.}(2024)\citenamefont {Chen},
  \citenamefont {Lei}, \citenamefont {Xiang}, \citenamefont {Duan},
  \citenamefont {Peng},\ and\ \citenamefont {Zhang}}]{chen2024emergent}%
  \BibitemOpen
  \bibfield  {author} {\bibinfo {author} {\bibfnamefont {J.}~\bibnamefont
  {Chen}}, \bibinfo {author} {\bibfnamefont {X.}~\bibnamefont {Lei}}, \bibinfo
  {author} {\bibfnamefont {Y.}~\bibnamefont {Xiang}}, \bibinfo {author}
  {\bibfnamefont {M.}~\bibnamefont {Duan}}, \bibinfo {author} {\bibfnamefont
  {X.}~\bibnamefont {Peng}},\ and\ \bibinfo {author} {\bibfnamefont
  {H.}~\bibnamefont {Zhang}},\ }\bibfield  {title} {\bibinfo {title} {Emergent
  chirality and hyperuniformity in an active mixture with nonreciprocal
  interactions},\ }\href
  {https://doi.org/https://doi.org/10.1103/PhysRevLett.132.118301} {\bibfield
  {journal} {\bibinfo  {journal} {Phys. Rev. Lett.}\ }\textbf {\bibinfo
  {volume} {132}},\ \bibinfo {pages} {118301} (\bibinfo {year}
  {2024})}\BibitemShut {NoStop}%
\bibitem [{\citenamefont {Cates}\ and\ \citenamefont
  {Tailleur}(2015)}]{cates2015motility}%
  \BibitemOpen
  \bibfield  {author} {\bibinfo {author} {\bibfnamefont {M.~E.}\ \bibnamefont
  {Cates}}\ and\ \bibinfo {author} {\bibfnamefont {J.}~\bibnamefont
  {Tailleur}},\ }\bibfield  {title} {\bibinfo {title} {Motility-induced phase
  separation},\ }\href
  {https://doi.org/https://doi.org/10.1146/annurev-conmatphys-031214-014710}
  {\bibfield  {journal} {\bibinfo  {journal} {Annu. Rev. Condens. Matter
  Phys.}\ }\textbf {\bibinfo {volume} {6}},\ \bibinfo {pages} {219} (\bibinfo
  {year} {2015})}\BibitemShut {NoStop}%
\bibitem [{\citenamefont {Gnan}\ and\ \citenamefont
  {Maggi}(2022)}]{gnan2022critical}%
  \BibitemOpen
  \bibfield  {author} {\bibinfo {author} {\bibfnamefont {N.}~\bibnamefont
  {Gnan}}\ and\ \bibinfo {author} {\bibfnamefont {C.}~\bibnamefont {Maggi}},\
  }\bibfield  {title} {\bibinfo {title} {Critical behavior of quorum-sensing
  active particles},\ }\href {https://doi.org/10.1039/D2SM00654E} {\bibfield
  {journal} {\bibinfo  {journal} {Soft Matter}\ }\textbf {\bibinfo {volume}
  {18}},\ \bibinfo {pages} {7654} (\bibinfo {year} {2022})}\BibitemShut
  {NoStop}%
\bibitem [{\citenamefont {Bertin}\ and\ \citenamefont
  {Solon}(2024)}]{bertin2024biased}%
  \BibitemOpen
  \bibfield  {author} {\bibinfo {author} {\bibfnamefont {E.}~\bibnamefont
  {Bertin}}\ and\ \bibinfo {author} {\bibfnamefont {A.}~\bibnamefont {Solon}},\
  }\bibfield  {title} {\bibinfo {title} {Biased motility-induced phase
  separation: from chemotaxis to traffic jams},\ }\href
  {https://doi.org/10.1088/1742-5468/ad3a5e} {\bibfield  {journal} {\bibinfo
  {journal} {J. Stat. Mech.: Theory Exp.}\ }\textbf {\bibinfo {volume}
  {2024}}\bibinfo  {number} { (5)},\ \bibinfo {pages} {053201}}\BibitemShut
  {NoStop}%
\bibitem [{\citenamefont {Zhou}\ \emph {et~al.}(2024)\citenamefont {Zhou},
  \citenamefont {Li},\ and\ \citenamefont {Marchesoni}}]{zhou2024clustering}%
  \BibitemOpen
\bibfield  {number} {  }\bibfield  {author} {\bibinfo {author} {\bibfnamefont
  {Y.}~\bibnamefont {Zhou}}, \bibinfo {author} {\bibfnamefont {Y.}~\bibnamefont
  {Li}},\ and\ \bibinfo {author} {\bibfnamefont {F.}~\bibnamefont
  {Marchesoni}},\ }\bibfield  {title} {\bibinfo {title} {Clustering of quorum
  sensing colloidal particles},\ }\href
  {https://doi.org/https://doi.org/10.1360/nso/20230081} {\bibfield  {journal}
  {\bibinfo  {journal} {National Science Open}\ }\textbf {\bibinfo {volume}
  {3}},\ \bibinfo {pages} {20230081} (\bibinfo {year} {2024})}\BibitemShut
  {NoStop}%
\bibitem [{\citenamefont {Wittkowski}\ \emph {et~al.}(2014)\citenamefont
  {Wittkowski}, \citenamefont {Tiribocchi}, \citenamefont {Stenhammar},
  \citenamefont {Allen}, \citenamefont {Marenduzzo},\ and\ \citenamefont
  {Cates}}]{wittkowski2014scalar}%
  \BibitemOpen
  \bibfield  {author} {\bibinfo {author} {\bibfnamefont {R.}~\bibnamefont
  {Wittkowski}}, \bibinfo {author} {\bibfnamefont {A.}~\bibnamefont
  {Tiribocchi}}, \bibinfo {author} {\bibfnamefont {J.}~\bibnamefont
  {Stenhammar}}, \bibinfo {author} {\bibfnamefont {R.~J.}\ \bibnamefont
  {Allen}}, \bibinfo {author} {\bibfnamefont {D.}~\bibnamefont {Marenduzzo}},\
  and\ \bibinfo {author} {\bibfnamefont {M.~E.}\ \bibnamefont {Cates}},\
  }\bibfield  {title} {\bibinfo {title} {Scalar $\varphi^4$ field theory for
  active-particle phase separation},\ }\href
  {https://doi.org/10.1038/ncomms5351} {\bibfield  {journal} {\bibinfo
  {journal} {Nat. Commun.}\ }\textbf {\bibinfo {volume} {5}},\ \bibinfo {pages}
  {4351} (\bibinfo {year} {2014})}\BibitemShut {NoStop}%
\bibitem [{\citenamefont {Dinelli}\ \emph {et~al.}(2024)\citenamefont
  {Dinelli}, \citenamefont {O'Byrne},\ and\ \citenamefont
  {Tailleur}}]{dinelli2024fluctuating}%
  \BibitemOpen
  \bibfield  {author} {\bibinfo {author} {\bibfnamefont {A.}~\bibnamefont
  {Dinelli}}, \bibinfo {author} {\bibfnamefont {J.}~\bibnamefont {O'Byrne}},\
  and\ \bibinfo {author} {\bibfnamefont {J.}~\bibnamefont {Tailleur}},\
  }\bibfield  {title} {\bibinfo {title} {Fluctuating hydrodynamics of active
  particles interacting via taxis and quorum sensing: static and dynamics},\
  }\href {https://doi.org/10.1088/1751-8121/ad72bc} {\bibfield  {journal}
  {\bibinfo  {journal} {J. Phys. A}\ }\textbf {\bibinfo {volume} {57}},\
  \bibinfo {pages} {395002} (\bibinfo {year} {2024})}\BibitemShut {NoStop}%
\bibitem [{\citenamefont {Frohoff-H{\"u}lsmann}\ and\ \citenamefont
  {Thiele}(2023)}]{frohoff2023nonreciprocal}%
  \BibitemOpen
  \bibfield  {author} {\bibinfo {author} {\bibfnamefont {T.}~\bibnamefont
  {Frohoff-H{\"u}lsmann}}\ and\ \bibinfo {author} {\bibfnamefont
  {U.}~\bibnamefont {Thiele}},\ }\bibfield  {title} {\bibinfo {title}
  {{Nonreciprocal Cahn-Hilliard model emerges as a universal amplitude
  equation}},\ }\href
  {https://doi.org/https://doi.org/10.1103/PhysRevLett.131.107201} {\bibfield
  {journal} {\bibinfo  {journal} {Phys. Rev. Lett.}\ }\textbf {\bibinfo
  {volume} {131}},\ \bibinfo {pages} {107201} (\bibinfo {year}
  {2023})}\BibitemShut {NoStop}%
\bibitem [{\citenamefont {Frohoff-H{\"u}lsmann}\ \emph
  {et~al.}(2021)\citenamefont {Frohoff-H{\"u}lsmann}, \citenamefont {Wrembel},\
  and\ \citenamefont {Thiele}}]{frohoff2021suppression}%
  \BibitemOpen
  \bibfield  {author} {\bibinfo {author} {\bibfnamefont {T.}~\bibnamefont
  {Frohoff-H{\"u}lsmann}}, \bibinfo {author} {\bibfnamefont {J.}~\bibnamefont
  {Wrembel}},\ and\ \bibinfo {author} {\bibfnamefont {U.}~\bibnamefont
  {Thiele}},\ }\bibfield  {title} {\bibinfo {title} {{Suppression of coarsening
  and emergence of oscillatory behavior in a Cahn-Hilliard model with
  nonvariational coupling}},\ }\href
  {https://doi.org/https://doi.org/10.1103/PhysRevE.103.042602} {\bibfield
  {journal} {\bibinfo  {journal} {Phys. Rev. E}\ }\textbf {\bibinfo {volume}
  {103}},\ \bibinfo {pages} {042602} (\bibinfo {year} {2021})}\BibitemShut
  {NoStop}%
\bibitem [{\citenamefont {Brauns}\ and\ \citenamefont
  {Marchetti}(2024)}]{brauns2024nonreciprocal}%
  \BibitemOpen
  \bibfield  {author} {\bibinfo {author} {\bibfnamefont {F.}~\bibnamefont
  {Brauns}}\ and\ \bibinfo {author} {\bibfnamefont {M.~C.}\ \bibnamefont
  {Marchetti}},\ }\bibfield  {title} {\bibinfo {title} {Nonreciprocal pattern
  formation of conserved fields},\ }\href
  {https://doi.org/https://doi.org/10.1103/PhysRevX.14.021014} {\bibfield
  {journal} {\bibinfo  {journal} {Phys. Rev. X}\ }\textbf {\bibinfo {volume}
  {14}},\ \bibinfo {pages} {021014} (\bibinfo {year} {2024})}\BibitemShut
  {NoStop}%
\bibitem [{\citenamefont {Rana}\ and\ \citenamefont
  {Golestanian}(2024{\natexlab{a}})}]{rana2023defect}%
  \BibitemOpen
  \bibfield  {author} {\bibinfo {author} {\bibfnamefont {N.}~\bibnamefont
  {Rana}}\ and\ \bibinfo {author} {\bibfnamefont {R.}~\bibnamefont
  {Golestanian}},\ }\bibfield  {title} {\bibinfo {title} {{Defect Solutions of
  the Nonreciprocal Cahn-Hilliard Model: Spirals and Targets}},\ }\href
  {https://doi.org/10.1103/PhysRevLett.133.078301} {\bibfield  {journal}
  {\bibinfo  {journal} {Phys. Rev. Lett.}\ }\textbf {\bibinfo {volume} {133}},\
  \bibinfo {pages} {078301} (\bibinfo {year} {2024}{\natexlab{a}})}\BibitemShut
  {NoStop}%
\bibitem [{\citenamefont {Rana}\ and\ \citenamefont
  {Golestanian}(2024{\natexlab{b}})}]{rana2024defect}%
  \BibitemOpen
  \bibfield  {author} {\bibinfo {author} {\bibfnamefont {N.}~\bibnamefont
  {Rana}}\ and\ \bibinfo {author} {\bibfnamefont {R.}~\bibnamefont
  {Golestanian}},\ }\bibfield  {title} {\bibinfo {title} {Defect interactions
  in the nonreciprocal {Cahn-Hilliard} model},\ }\href
  {https://arxiv.org/abs/2407.16547} {\bibfield  {journal} {\bibinfo  {journal}
  {arXiv preprint arXiv2407.16547}\ } (\bibinfo {year}
  {2024}{\natexlab{b}})}\BibitemShut {NoStop}%
\bibitem [{\citenamefont {Solon}\ \emph
  {et~al.}(2018{\natexlab{a}})\citenamefont {Solon}, \citenamefont
  {Stenhammar}, \citenamefont {Cates}, \citenamefont {Kafri},\ and\
  \citenamefont {Tailleur}}]{solon2018generalized1}%
  \BibitemOpen
  \bibfield  {author} {\bibinfo {author} {\bibfnamefont {A.~P.}\ \bibnamefont
  {Solon}}, \bibinfo {author} {\bibfnamefont {J.}~\bibnamefont {Stenhammar}},
  \bibinfo {author} {\bibfnamefont {M.~E.}\ \bibnamefont {Cates}}, \bibinfo
  {author} {\bibfnamefont {Y.}~\bibnamefont {Kafri}},\ and\ \bibinfo {author}
  {\bibfnamefont {J.}~\bibnamefont {Tailleur}},\ }\bibfield  {title} {\bibinfo
  {title} {Generalized thermodynamics of phase equilibria in scalar active
  matter},\ }\href {https://doi.org/https://doi.org/10.1103/PhysRevE.97.020602}
  {\bibfield  {journal} {\bibinfo  {journal} {Phys. Rev. E}\ }\textbf {\bibinfo
  {volume} {97}},\ \bibinfo {pages} {020602} (\bibinfo {year}
  {2018}{\natexlab{a}})}\BibitemShut {NoStop}%
\bibitem [{\citenamefont {Solon}\ \emph
  {et~al.}(2018{\natexlab{b}})\citenamefont {Solon}, \citenamefont
  {Stenhammar}, \citenamefont {Cates}, \citenamefont {Kafri},\ and\
  \citenamefont {Tailleur}}]{solon2018generalized2}%
  \BibitemOpen
  \bibfield  {author} {\bibinfo {author} {\bibfnamefont {A.~P.}\ \bibnamefont
  {Solon}}, \bibinfo {author} {\bibfnamefont {J.}~\bibnamefont {Stenhammar}},
  \bibinfo {author} {\bibfnamefont {M.~E.}\ \bibnamefont {Cates}}, \bibinfo
  {author} {\bibfnamefont {Y.}~\bibnamefont {Kafri}},\ and\ \bibinfo {author}
  {\bibfnamefont {J.}~\bibnamefont {Tailleur}},\ }\bibfield  {title} {\bibinfo
  {title} {Generalized thermodynamics of motility-induced phase separation:
  phase equilibria, laplace pressure, and change of ensembles},\ }\href
  {https://doi.org/10.1088/1367-2630/aaccdd} {\bibfield  {journal} {\bibinfo
  {journal} {New J. Phys.}\ }\textbf {\bibinfo {volume} {20}},\ \bibinfo
  {pages} {075001} (\bibinfo {year} {2018}{\natexlab{b}})}\BibitemShut
  {NoStop}%
\bibitem [{\citenamefont {Greve}\ \emph {et~al.}(2024)\citenamefont {Greve},
  \citenamefont {Lovato}, \citenamefont {Frohoff-H{\"u}lsmann},\ and\
  \citenamefont {Thiele}}]{greve2024maxwell}%
  \BibitemOpen
  \bibfield  {author} {\bibinfo {author} {\bibfnamefont {D.}~\bibnamefont
  {Greve}}, \bibinfo {author} {\bibfnamefont {G.}~\bibnamefont {Lovato}},
  \bibinfo {author} {\bibfnamefont {T.}~\bibnamefont {Frohoff-H{\"u}lsmann}},\
  and\ \bibinfo {author} {\bibfnamefont {U.}~\bibnamefont {Thiele}},\
  }\bibfield  {title} {\bibinfo {title} {{Maxwell construction for a
  nonreciprocal Cahn-Hilliard model}},\ }\href
  {https://doi.org/10.48550/arXiv.2402.08634} {\bibfield  {journal} {\bibinfo
  {journal} {arXiv preprint arXiv:2402.08634}\ } (\bibinfo {year}
  {2024})}\BibitemShut {NoStop}%
\bibitem [{\citenamefont {Saha}(2024)}]{saha2024phase}%
  \BibitemOpen
  \bibfield  {author} {\bibinfo {author} {\bibfnamefont {S.}~\bibnamefont
  {Saha}},\ }\bibfield  {title} {\bibinfo {title} {Phase coexistence in the
  nonreciprocal {Cahn-Hilliard} model},\ }\href
  {https://doi.org/10.48550/arXiv.2402.10057} {\bibfield  {journal} {\bibinfo
  {journal} {arXiv preprint arXiv:2402.10057}\ } (\bibinfo {year}
  {2024})}\BibitemShut {NoStop}%
\bibitem [{\citenamefont {Tsyganov}\ \emph {et~al.}(2003)\citenamefont
  {Tsyganov}, \citenamefont {Brindley}, \citenamefont {Holden},\ and\
  \citenamefont {Biktashev}}]{tsyganov2003quasisoliton}%
  \BibitemOpen
  \bibfield  {author} {\bibinfo {author} {\bibfnamefont {M.}~\bibnamefont
  {Tsyganov}}, \bibinfo {author} {\bibfnamefont {J.}~\bibnamefont {Brindley}},
  \bibinfo {author} {\bibfnamefont {A.}~\bibnamefont {Holden}},\ and\ \bibinfo
  {author} {\bibfnamefont {V.}~\bibnamefont {Biktashev}},\ }\bibfield  {title}
  {\bibinfo {title} {Quasisoliton interaction of pursuit-evasion waves in a
  predator-prey system},\ }\href
  {https://doi.org/https://doi.org/10.1103/PhysRevLett.91.218102} {\bibfield
  {journal} {\bibinfo  {journal} {Phys. Rev. Lett.}\ }\textbf {\bibinfo
  {volume} {91}},\ \bibinfo {pages} {218102} (\bibinfo {year}
  {2003})}\BibitemShut {NoStop}%
\bibitem [{\citenamefont {Yamanaka}\ and\ \citenamefont
  {Ohta}(2014)}]{yamanaka2014collision}%
  \BibitemOpen
  \bibfield  {author} {\bibinfo {author} {\bibfnamefont {S.}~\bibnamefont
  {Yamanaka}}\ and\ \bibinfo {author} {\bibfnamefont {T.}~\bibnamefont
  {Ohta}},\ }\bibfield  {title} {\bibinfo {title} {Collision dynamics of
  traveling bands in systems of deformable self-propelled particles},\ }\href
  {https://doi.org/https://doi.org/10.1103/PhysRevE.90.042927} {\bibfield
  {journal} {\bibinfo  {journal} {Phys. Rev. E}\ }\textbf {\bibinfo {volume}
  {90}},\ \bibinfo {pages} {042927} (\bibinfo {year} {2014})}\BibitemShut
  {NoStop}%
\bibitem [{\citenamefont {Takatori}\ and\ \citenamefont
  {Brady}(2015)}]{takatori2015towards}%
  \BibitemOpen
  \bibfield  {author} {\bibinfo {author} {\bibfnamefont {S.~C.}\ \bibnamefont
  {Takatori}}\ and\ \bibinfo {author} {\bibfnamefont {J.~F.}\ \bibnamefont
  {Brady}},\ }\bibfield  {title} {\bibinfo {title} {Towards a thermodynamics of
  active matter},\ }\href
  {https://doi.org/https://doi.org/10.1103/PhysRevE.91.032117} {\bibfield
  {journal} {\bibinfo  {journal} {Phys. Rev. E}\ }\textbf {\bibinfo {volume}
  {91}},\ \bibinfo {pages} {032117} (\bibinfo {year} {2015})}\BibitemShut
  {NoStop}%
\bibitem [{\citenamefont {Paliwal}\ \emph {et~al.}(2018)\citenamefont
  {Paliwal}, \citenamefont {Rodenburg}, \citenamefont {van Roij},\ and\
  \citenamefont {Dijkstra}}]{paliwal2018chemical}%
  \BibitemOpen
  \bibfield  {author} {\bibinfo {author} {\bibfnamefont {S.}~\bibnamefont
  {Paliwal}}, \bibinfo {author} {\bibfnamefont {J.}~\bibnamefont {Rodenburg}},
  \bibinfo {author} {\bibfnamefont {R.}~\bibnamefont {van Roij}},\ and\
  \bibinfo {author} {\bibfnamefont {M.}~\bibnamefont {Dijkstra}},\ }\bibfield
  {title} {\bibinfo {title} {Chemical potential in active systems: predicting
  phase equilibrium from bulk equations of state?},\ }\href
  {https://doi.org/10.1088/1367-2630/aa9b4d} {\bibfield  {journal} {\bibinfo
  {journal} {New J. Phys.}\ }\textbf {\bibinfo {volume} {20}},\ \bibinfo
  {pages} {015003} (\bibinfo {year} {2018})}\BibitemShut {NoStop}%
\bibitem [{\citenamefont {Hermann}\ \emph {et~al.}(2021)\citenamefont
  {Hermann}, \citenamefont {de~las Heras},\ and\ \citenamefont
  {Schmidt}}]{hermann2021phase}%
  \BibitemOpen
  \bibfield  {author} {\bibinfo {author} {\bibfnamefont {S.}~\bibnamefont
  {Hermann}}, \bibinfo {author} {\bibfnamefont {D.}~\bibnamefont {de~las
  Heras}},\ and\ \bibinfo {author} {\bibfnamefont {M.}~\bibnamefont
  {Schmidt}},\ }\bibfield  {title} {\bibinfo {title} {Phase separation of
  active {Brownian} particles in two dimensions: anything for a quiet life},\
  }\href {https://doi.org/https://doi.org/10.1080/00268976.2021.1902585}
  {\bibfield  {journal} {\bibinfo  {journal} {Mol. Phys.}\ }\textbf {\bibinfo
  {volume} {119}},\ \bibinfo {pages} {e1902585} (\bibinfo {year}
  {2021})}\BibitemShut {NoStop}%
\bibitem [{\citenamefont {Speck}(2021)}]{speck2021coexistence}%
  \BibitemOpen
  \bibfield  {author} {\bibinfo {author} {\bibfnamefont {T.}~\bibnamefont
  {Speck}},\ }\bibfield  {title} {\bibinfo {title} {Coexistence of active
  {Brownian} disks: {Van der Waals} theory and analytical results},\ }\href
  {https://doi.org/https://doi.org/10.1103/PhysRevE.103.012607} {\bibfield
  {journal} {\bibinfo  {journal} {Phys. Rev. E}\ }\textbf {\bibinfo {volume}
  {103}},\ \bibinfo {pages} {012607} (\bibinfo {year} {2021})}\BibitemShut
  {NoStop}%
\bibitem [{\citenamefont {Omar}\ \emph {et~al.}(2023)\citenamefont {Omar},
  \citenamefont {Row}, \citenamefont {Mallory},\ and\ \citenamefont
  {Brady}}]{omar2023mechanical}%
  \BibitemOpen
  \bibfield  {author} {\bibinfo {author} {\bibfnamefont {A.~K.}\ \bibnamefont
  {Omar}}, \bibinfo {author} {\bibfnamefont {H.}~\bibnamefont {Row}}, \bibinfo
  {author} {\bibfnamefont {S.~A.}\ \bibnamefont {Mallory}},\ and\ \bibinfo
  {author} {\bibfnamefont {J.~F.}\ \bibnamefont {Brady}},\ }\bibfield  {title}
  {\bibinfo {title} {Mechanical theory of nonequilibrium coexistence and
  motility-induced phase separation},\ }\href
  {https://doi.org/https://doi.org/10.1073/pnas.2219900120} {\bibfield
  {journal} {\bibinfo  {journal} {Proc. Natl. Acad. Sci. U.S.A.}\ }\textbf
  {\bibinfo {volume} {120}},\ \bibinfo {pages} {e2219900120} (\bibinfo {year}
  {2023})}\BibitemShut {NoStop}%
\bibitem [{\citenamefont {Cates}(2022)}]{cates2022active}%
  \BibitemOpen
  \bibfield  {author} {\bibinfo {author} {\bibfnamefont {M.~E.}\ \bibnamefont
  {Cates}},\ }\bibfield  {title} {\bibinfo {title} {{Active Field Theories}},\
  }in\ \href {https://doi.org/10.1093/oso/9780192858313.003.0006} {\emph
  {\bibinfo {booktitle} {{Active Matter and Nonequilibrium Statistical Physics:
  Lecture Notes of the Les Houches Summer School: Volume 112, September
  2018}}}}\ (\bibinfo  {publisher} {Oxford University Press},\ \bibinfo {year}
  {2022})\BibitemShut {NoStop}%
\bibitem [{\citenamefont {K{\"u}rsten}\ and\ \citenamefont
  {Ihle}(2020)}]{kursten2020dry}%
  \BibitemOpen
  \bibfield  {author} {\bibinfo {author} {\bibfnamefont {R.}~\bibnamefont
  {K{\"u}rsten}}\ and\ \bibinfo {author} {\bibfnamefont {T.}~\bibnamefont
  {Ihle}},\ }\bibfield  {title} {\bibinfo {title} {Dry active matter exhibits a
  self-organized cross sea phase},\ }\href
  {https://doi.org/10.1103/PhysRevLett.125.188003} {\bibfield  {journal}
  {\bibinfo  {journal} {Phys. Rev. Lett.}\ }\textbf {\bibinfo {volume} {125}},\
  \bibinfo {pages} {188003} (\bibinfo {year} {2020})}\BibitemShut {NoStop}%
\bibitem [{\citenamefont {Chat{\'e}}(2020)}]{chate2020dry}%
  \BibitemOpen
  \bibfield  {author} {\bibinfo {author} {\bibfnamefont {H.}~\bibnamefont
  {Chat{\'e}}},\ }\bibfield  {title} {\bibinfo {title} {Dry aligning dilute
  active matter},\ }\href
  {https://doi.org/10.1146/annurev-conmatphys-031119-050752} {\bibfield
  {journal} {\bibinfo  {journal} {Annu. Rev. Condens. Matter Phys.}\ }\textbf
  {\bibinfo {volume} {11}},\ \bibinfo {pages} {189} (\bibinfo {year}
  {2020})}\BibitemShut {NoStop}%
\bibitem [{\citenamefont {Kondo}\ and\ \citenamefont
  {Miura}(2010)}]{kondo2010reaction}%
  \BibitemOpen
  \bibfield  {author} {\bibinfo {author} {\bibfnamefont {S.}~\bibnamefont
  {Kondo}}\ and\ \bibinfo {author} {\bibfnamefont {T.}~\bibnamefont {Miura}},\
  }\bibfield  {title} {\bibinfo {title} {Reaction-diffusion model as a
  framework for understanding biological pattern formation},\ }\href
  {https://doi.org/10.1126/science.117904} {\bibfield  {journal} {\bibinfo
  {journal} {Science}\ }\textbf {\bibinfo {volume} {329}},\ \bibinfo {pages}
  {1616} (\bibinfo {year} {2010})}\BibitemShut {NoStop}%
\bibitem [{\citenamefont {Reichenbach}\ \emph {et~al.}(2007)\citenamefont
  {Reichenbach}, \citenamefont {Mobilia},\ and\ \citenamefont
  {Frey}}]{reichenbach2007noise}%
  \BibitemOpen
  \bibfield  {author} {\bibinfo {author} {\bibfnamefont {T.}~\bibnamefont
  {Reichenbach}}, \bibinfo {author} {\bibfnamefont {M.}~\bibnamefont
  {Mobilia}},\ and\ \bibinfo {author} {\bibfnamefont {E.}~\bibnamefont
  {Frey}},\ }\bibfield  {title} {\bibinfo {title} {Noise and correlations in a
  spatial population model with cyclic competition},\ }\href
  {https://doi.org/https://doi.org/10.1103/PhysRevLett.99.238105} {\bibfield
  {journal} {\bibinfo  {journal} {Phys. Rev. Lett.}\ }\textbf {\bibinfo
  {volume} {99}},\ \bibinfo {pages} {238105} (\bibinfo {year}
  {2007})}\BibitemShut {NoStop}%
\bibitem [{\citenamefont {Vanag}\ and\ \citenamefont
  {Epstein}(2009)}]{vanag2009cross}%
  \BibitemOpen
  \bibfield  {author} {\bibinfo {author} {\bibfnamefont {V.~K.}\ \bibnamefont
  {Vanag}}\ and\ \bibinfo {author} {\bibfnamefont {I.~R.}\ \bibnamefont
  {Epstein}},\ }\bibfield  {title} {\bibinfo {title} {Cross-diffusion and
  pattern formation in reaction--diffusion systems},\ }\href
  {https://doi.org/10.1039/B813825G} {\bibfield  {journal} {\bibinfo  {journal}
  {Phys. Chem. Chem. Phys.}\ }\textbf {\bibinfo {volume} {11}},\ \bibinfo
  {pages} {897} (\bibinfo {year} {2009})}\BibitemShut {NoStop}%
\bibitem [{\citenamefont {Luo}\ and\ \citenamefont
  {Zwicker}(2023)}]{luo2023influence}%
  \BibitemOpen
  \bibfield  {author} {\bibinfo {author} {\bibfnamefont {C.}~\bibnamefont
  {Luo}}\ and\ \bibinfo {author} {\bibfnamefont {D.}~\bibnamefont {Zwicker}},\
  }\bibfield  {title} {\bibinfo {title} {Influence of physical interactions on
  spatiotemporal patterns},\ }\href
  {https://doi.org/https://doi.org/10.1103/PhysRevE.108.034206} {\bibfield
  {journal} {\bibinfo  {journal} {Phys. Rev. E}\ }\textbf {\bibinfo {volume}
  {108}},\ \bibinfo {pages} {034206} (\bibinfo {year} {2023})}\BibitemShut
  {NoStop}%
\bibitem [{\citenamefont {Aranson}\ and\ \citenamefont
  {Kramer}(2002)}]{aranson2002world}%
  \BibitemOpen
  \bibfield  {author} {\bibinfo {author} {\bibfnamefont {I.~S.}\ \bibnamefont
  {Aranson}}\ and\ \bibinfo {author} {\bibfnamefont {L.}~\bibnamefont
  {Kramer}},\ }\bibfield  {title} {\bibinfo {title} {The world of the complex
  {Ginzburg-Landau} equation},\ }\href
  {https://doi.org/https://doi.org/10.1103/RevModPhys.74.99} {\bibfield
  {journal} {\bibinfo  {journal} {Rev. Mod. Phys.}\ }\textbf {\bibinfo {volume}
  {74}},\ \bibinfo {pages} {99} (\bibinfo {year} {2002})}\BibitemShut {NoStop}%
\bibitem [{\citenamefont {Cross}\ and\ \citenamefont
  {Hohenberg}(1993)}]{cross1993pattern}%
  \BibitemOpen
  \bibfield  {author} {\bibinfo {author} {\bibfnamefont {M.~C.}\ \bibnamefont
  {Cross}}\ and\ \bibinfo {author} {\bibfnamefont {P.~C.}\ \bibnamefont
  {Hohenberg}},\ }\bibfield  {title} {\bibinfo {title} {Pattern formation
  outside of equilibrium},\ }\href
  {https://doi.org/https://doi.org/10.1103/RevModPhys.65.851} {\bibfield
  {journal} {\bibinfo  {journal} {Rev. Mod. Phys.}\ }\textbf {\bibinfo {volume}
  {65}},\ \bibinfo {pages} {851} (\bibinfo {year} {1993})}\BibitemShut
  {NoStop}%
\bibitem [{\citenamefont {Parteli}\ \emph {et~al.}(2011)\citenamefont
  {Parteli}, \citenamefont {Andrade~Jr},\ and\ \citenamefont
  {Herrmann}}]{parteli2011transverse}%
  \BibitemOpen
  \bibfield  {author} {\bibinfo {author} {\bibfnamefont {E.~J.}\ \bibnamefont
  {Parteli}}, \bibinfo {author} {\bibfnamefont {J.~S.}\ \bibnamefont
  {Andrade~Jr}},\ and\ \bibinfo {author} {\bibfnamefont {H.~J.}\ \bibnamefont
  {Herrmann}},\ }\bibfield  {title} {\bibinfo {title} {Transverse instability
  of dunes},\ }\href
  {https://doi.org/https://doi.org/10.1103/PhysRevLett.107.188001} {\bibfield
  {journal} {\bibinfo  {journal} {Phys. Rev. Lett.}\ }\textbf {\bibinfo
  {volume} {107}},\ \bibinfo {pages} {188001} (\bibinfo {year}
  {2011})}\BibitemShut {NoStop}%
\bibitem [{\citenamefont {Alert}\ \emph {et~al.}(2022)\citenamefont {Alert},
  \citenamefont {Mart{\'\i}nez-Calvo},\ and\ \citenamefont
  {Datta}}]{alert2022cellular}%
  \BibitemOpen
  \bibfield  {author} {\bibinfo {author} {\bibfnamefont {R.}~\bibnamefont
  {Alert}}, \bibinfo {author} {\bibfnamefont {A.}~\bibnamefont
  {Mart{\'\i}nez-Calvo}},\ and\ \bibinfo {author} {\bibfnamefont {S.~S.}\
  \bibnamefont {Datta}},\ }\bibfield  {title} {\bibinfo {title} {Cellular
  sensing governs the stability of chemotactic fronts},\ }\href
  {https://doi.org/https://doi.org/10.1103/PhysRevLett.128.148101} {\bibfield
  {journal} {\bibinfo  {journal} {Phys. Rev. Lett.}\ }\textbf {\bibinfo
  {volume} {128}},\ \bibinfo {pages} {148101} (\bibinfo {year}
  {2022})}\BibitemShut {NoStop}%
\bibitem [{\citenamefont {Allen}\ and\ \citenamefont
  {Waclaw}(2018)}]{allen2019bacterial}%
  \BibitemOpen
  \bibfield  {author} {\bibinfo {author} {\bibfnamefont {R.~J.}\ \bibnamefont
  {Allen}}\ and\ \bibinfo {author} {\bibfnamefont {B.}~\bibnamefont {Waclaw}},\
  }\bibfield  {title} {\bibinfo {title} {Bacterial growth: a statistical
  physicist's guide},\ }\href {https://doi.org/10.1088/1361-6633/aae546}
  {\bibfield  {journal} {\bibinfo  {journal} {Rep. Prog. Phys.}\ }\textbf
  {\bibinfo {volume} {82}},\ \bibinfo {pages} {016601} (\bibinfo {year}
  {2018})}\BibitemShut {NoStop}%
\bibitem [{\citenamefont {Hendrey}\ \emph {et~al.}(2000)\citenamefont
  {Hendrey}, \citenamefont {Nam}, \citenamefont {Guzdar},\ and\ \citenamefont
  {Ott}}]{hendrey2000target}%
  \BibitemOpen
  \bibfield  {author} {\bibinfo {author} {\bibfnamefont {M.}~\bibnamefont
  {Hendrey}}, \bibinfo {author} {\bibfnamefont {K.}~\bibnamefont {Nam}},
  \bibinfo {author} {\bibfnamefont {P.}~\bibnamefont {Guzdar}},\ and\ \bibinfo
  {author} {\bibfnamefont {E.}~\bibnamefont {Ott}},\ }\bibfield  {title}
  {\bibinfo {title} {Target waves in the complex {Ginzburg-Landau} equation},\
  }\href {https://doi.org/https://doi.org/10.1103/PhysRevE.62.7627} {\bibfield
  {journal} {\bibinfo  {journal} {Phys. Rev. E}\ }\textbf {\bibinfo {volume}
  {62}},\ \bibinfo {pages} {7627} (\bibinfo {year} {2000})}\BibitemShut
  {NoStop}%
\bibitem [{\citenamefont {Jiang}\ \emph {et~al.}(2009)\citenamefont {Jiang},
  \citenamefont {Zhou}, \citenamefont {Perc}, \citenamefont {Huang},\ and\
  \citenamefont {Wang}}]{jiang2009emergence}%
  \BibitemOpen
  \bibfield  {author} {\bibinfo {author} {\bibfnamefont {L.-L.}\ \bibnamefont
  {Jiang}}, \bibinfo {author} {\bibfnamefont {T.}~\bibnamefont {Zhou}},
  \bibinfo {author} {\bibfnamefont {M.}~\bibnamefont {Perc}}, \bibinfo {author}
  {\bibfnamefont {X.}~\bibnamefont {Huang}},\ and\ \bibinfo {author}
  {\bibfnamefont {B.-H.}\ \bibnamefont {Wang}},\ }\bibfield  {title} {\bibinfo
  {title} {Emergence of target waves in paced populations of cyclically
  competing species},\ }\href {https://doi.org/10.1088/1367-2630/11/10/103001}
  {\bibfield  {journal} {\bibinfo  {journal} {New J. Phys.}\ }\textbf {\bibinfo
  {volume} {11}},\ \bibinfo {pages} {103001} (\bibinfo {year}
  {2009})}\BibitemShut {NoStop}%
\bibitem [{\citenamefont {Lifshitz}\ and\ \citenamefont
  {Slyozov}(1961)}]{Lifshitz1961JPCS}%
  \BibitemOpen
  \bibfield  {author} {\bibinfo {author} {\bibfnamefont {I.}~\bibnamefont
  {Lifshitz}}\ and\ \bibinfo {author} {\bibfnamefont {V.}~\bibnamefont
  {Slyozov}},\ }\bibfield  {title} {\bibinfo {title} {The kinetics of
  precipitation from supersaturated solid solutions},\ }\href
  {https://doi.org/https://doi.org/10.1016/0022-3697(61)90054-3} {\bibfield
  {journal} {\bibinfo  {journal} {J. Phys. Chem. Solids}\ }\textbf {\bibinfo
  {volume} {19}},\ \bibinfo {pages} {35} (\bibinfo {year} {1961})}\BibitemShut
  {NoStop}%
\bibitem [{\citenamefont {Bray}(2002)}]{BrayAdvPhys2002}%
  \BibitemOpen
  \bibfield  {author} {\bibinfo {author} {\bibfnamefont {A.~J.}\ \bibnamefont
  {Bray}},\ }\bibfield  {title} {\bibinfo {title} {Theory of phase-ordering
  kinetics},\ }\href {https://doi.org/10.1080/00018730110117433} {\bibfield
  {journal} {\bibinfo  {journal} {Adv. Phys.}\ }\textbf {\bibinfo {volume}
  {51}},\ \bibinfo {pages} {481} (\bibinfo {year} {2002})}\BibitemShut
  {NoStop}%
\bibitem [{\citenamefont {Stenhammar}\ \emph {et~al.}(2014)\citenamefont
  {Stenhammar}, \citenamefont {Marenduzzo}, \citenamefont {Allen},\ and\
  \citenamefont {Cates}}]{Stenhammar2014SoftMatt}%
  \BibitemOpen
  \bibfield  {author} {\bibinfo {author} {\bibfnamefont {J.}~\bibnamefont
  {Stenhammar}}, \bibinfo {author} {\bibfnamefont {D.}~\bibnamefont
  {Marenduzzo}}, \bibinfo {author} {\bibfnamefont {R.~J.}\ \bibnamefont
  {Allen}},\ and\ \bibinfo {author} {\bibfnamefont {M.~E.}\ \bibnamefont
  {Cates}},\ }\bibfield  {title} {\bibinfo {title} {{Phase behaviour of active
  {Brownian} particles: the role of dimensionality}},\ }\href
  {https://doi.org/10.1039/C3SM52813H} {\bibfield  {journal} {\bibinfo
  {journal} {Soft Matter}\ }\textbf {\bibinfo {volume} {10}},\ \bibinfo {pages}
  {1489} (\bibinfo {year} {2014})}\BibitemShut {NoStop}%
\bibitem [{\citenamefont {Shi}\ \emph {et~al.}(2020)\citenamefont {Shi},
  \citenamefont {Fausti}, \citenamefont {Chat{\'e}}, \citenamefont {Nardini},\
  and\ \citenamefont {Solon}}]{shi2020self}%
  \BibitemOpen
  \bibfield  {author} {\bibinfo {author} {\bibfnamefont {X.-q.}\ \bibnamefont
  {Shi}}, \bibinfo {author} {\bibfnamefont {G.}~\bibnamefont {Fausti}},
  \bibinfo {author} {\bibfnamefont {H.}~\bibnamefont {Chat{\'e}}}, \bibinfo
  {author} {\bibfnamefont {C.}~\bibnamefont {Nardini}},\ and\ \bibinfo {author}
  {\bibfnamefont {A.}~\bibnamefont {Solon}},\ }\bibfield  {title} {\bibinfo
  {title} {Self-organized critical coexistence phase in repulsive active
  particles},\ }\href {https://doi.org/10.1103/PhysRevLett.125.168001}
  {\bibfield  {journal} {\bibinfo  {journal} {Phys. Rev. Lett.}\ }\textbf
  {\bibinfo {volume} {125}},\ \bibinfo {pages} {168001} (\bibinfo {year}
  {2020})}\BibitemShut {NoStop}%
\bibitem [{\citenamefont {Caporusso}\ \emph {et~al.}(2023)\citenamefont
  {Caporusso}, \citenamefont {Cugliandolo}, \citenamefont {Digregorio},
  \citenamefont {Gonnella}, \citenamefont {Levis},\ and\ \citenamefont
  {Suma}}]{caporusso2023dynamics}%
  \BibitemOpen
  \bibfield  {author} {\bibinfo {author} {\bibfnamefont {C.~B.}\ \bibnamefont
  {Caporusso}}, \bibinfo {author} {\bibfnamefont {L.~F.}\ \bibnamefont
  {Cugliandolo}}, \bibinfo {author} {\bibfnamefont {P.}~\bibnamefont
  {Digregorio}}, \bibinfo {author} {\bibfnamefont {G.}~\bibnamefont
  {Gonnella}}, \bibinfo {author} {\bibfnamefont {D.}~\bibnamefont {Levis}},\
  and\ \bibinfo {author} {\bibfnamefont {A.}~\bibnamefont {Suma}},\ }\bibfield
  {title} {\bibinfo {title} {{Dynamics of Motility-Induced clusters: coarsening
  beyond Ostwald ripening}},\ }\href
  {https://doi.org/https://doi.org/10.1103/PhysRevLett.131.068201} {\bibfield
  {journal} {\bibinfo  {journal} {Phys. Rev. Lett.}\ }\textbf {\bibinfo
  {volume} {131}},\ \bibinfo {pages} {068201} (\bibinfo {year}
  {2023})}\BibitemShut {NoStop}%
\bibitem [{\citenamefont {Duan}\ \emph {et~al.}(2024)\citenamefont {Duan},
  \citenamefont {Golestanian},\ and\ \citenamefont {Mahault}}]{TBP}%
  \BibitemOpen
  \bibfield  {author} {\bibinfo {author} {\bibfnamefont {Y.}~\bibnamefont
  {Duan}}, \bibinfo {author} {\bibfnamefont {R.}~\bibnamefont {Golestanian}},\
  and\ \bibinfo {author} {\bibfnamefont {B.}~\bibnamefont {Mahault}},\
  }\href@noop {} {\emph {\bibinfo {title} {To be published}}}\ (\bibinfo {year}
  {2024})\BibitemShut {NoStop}%
\bibitem [{\citenamefont {Toner}\ and\ \citenamefont
  {Tu}(1995)}]{toner1995long}%
  \BibitemOpen
  \bibfield  {author} {\bibinfo {author} {\bibfnamefont {J.}~\bibnamefont
  {Toner}}\ and\ \bibinfo {author} {\bibfnamefont {Y.}~\bibnamefont {Tu}},\
  }\bibfield  {title} {\bibinfo {title} {Long-range order in a two-dimensional
  dynamical {XY} model: how birds fly together},\ }\href
  {https://doi.org/https://doi.org/10.1103/PhysRevLett.75.4326} {\bibfield
  {journal} {\bibinfo  {journal} {Phys. Rev. Lett.}\ }\textbf {\bibinfo
  {volume} {75}},\ \bibinfo {pages} {4326} (\bibinfo {year}
  {1995})}\BibitemShut {NoStop}%
\bibitem [{\citenamefont {Mahault}\ \emph {et~al.}(2019)\citenamefont
  {Mahault}, \citenamefont {Ginelli},\ and\ \citenamefont
  {Chat{\'e}}}]{mahault2019quantitative}%
  \BibitemOpen
  \bibfield  {author} {\bibinfo {author} {\bibfnamefont {B.}~\bibnamefont
  {Mahault}}, \bibinfo {author} {\bibfnamefont {F.}~\bibnamefont {Ginelli}},\
  and\ \bibinfo {author} {\bibfnamefont {H.}~\bibnamefont {Chat{\'e}}},\
  }\bibfield  {title} {\bibinfo {title} {{Quantitative Assessment of the Toner
  and Tu Theory of Polar Flocks}},\ }\href
  {https://doi.org/https://doi.org/10.1103/PhysRevLett.123.218001} {\bibfield
  {journal} {\bibinfo  {journal} {Phys. Rev. Lett.}\ }\textbf {\bibinfo
  {volume} {123}},\ \bibinfo {pages} {218001} (\bibinfo {year}
  {2019})}\BibitemShut {NoStop}%
\bibitem [{\citenamefont {Duan}\ \emph {et~al.}(2021)\citenamefont {Duan},
  \citenamefont {Mahault}, \citenamefont {Ma}, \citenamefont {Shi},\ and\
  \citenamefont {Chat{\'e}}}]{duan2021breakdown}%
  \BibitemOpen
  \bibfield  {author} {\bibinfo {author} {\bibfnamefont {Y.}~\bibnamefont
  {Duan}}, \bibinfo {author} {\bibfnamefont {B.}~\bibnamefont {Mahault}},
  \bibinfo {author} {\bibfnamefont {Y.-q.}\ \bibnamefont {Ma}}, \bibinfo
  {author} {\bibfnamefont {X.-q.}\ \bibnamefont {Shi}},\ and\ \bibinfo {author}
  {\bibfnamefont {H.}~\bibnamefont {Chat{\'e}}},\ }\bibfield  {title} {\bibinfo
  {title} {{Breakdown of Ergodicity and Self-Averaging in Polar Flocks with
  Quenched Disorder}},\ }\href
  {https://doi.org/https://doi.org/10.1103/PhysRevLett.126.178001} {\bibfield
  {journal} {\bibinfo  {journal} {Phys. Rev. Lett.}\ }\textbf {\bibinfo
  {volume} {126}},\ \bibinfo {pages} {178001} (\bibinfo {year}
  {2021})}\BibitemShut {NoStop}%
\bibitem [{\citenamefont {Besse}\ \emph {et~al.}(2022)\citenamefont {Besse},
  \citenamefont {Chat{\'e}},\ and\ \citenamefont
  {Solon}}]{besse2022metastability}%
  \BibitemOpen
  \bibfield  {author} {\bibinfo {author} {\bibfnamefont {M.}~\bibnamefont
  {Besse}}, \bibinfo {author} {\bibfnamefont {H.}~\bibnamefont {Chat{\'e}}},\
  and\ \bibinfo {author} {\bibfnamefont {A.}~\bibnamefont {Solon}},\ }\bibfield
   {title} {\bibinfo {title} {Metastability of constant-density flocks},\
  }\href {https://doi.org/https://doi.org/10.1103/PhysRevLett.129.268003}
  {\bibfield  {journal} {\bibinfo  {journal} {Phys. Rev. Lett.}\ }\textbf
  {\bibinfo {volume} {129}},\ \bibinfo {pages} {268003} (\bibinfo {year}
  {2022})}\BibitemShut {NoStop}%
\bibitem [{\citenamefont {Benvegnen}\ \emph {et~al.}(2023)\citenamefont
  {Benvegnen}, \citenamefont {Granek}, \citenamefont {Ro}, \citenamefont
  {Yaacoby}, \citenamefont {Chat{\'e}}, \citenamefont {Kafri}, \citenamefont
  {Mukamel}, \citenamefont {Solon},\ and\ \citenamefont
  {Tailleur}}]{benvegnen2023metastability}%
  \BibitemOpen
  \bibfield  {author} {\bibinfo {author} {\bibfnamefont {B.}~\bibnamefont
  {Benvegnen}}, \bibinfo {author} {\bibfnamefont {O.}~\bibnamefont {Granek}},
  \bibinfo {author} {\bibfnamefont {S.}~\bibnamefont {Ro}}, \bibinfo {author}
  {\bibfnamefont {R.}~\bibnamefont {Yaacoby}}, \bibinfo {author} {\bibfnamefont
  {H.}~\bibnamefont {Chat{\'e}}}, \bibinfo {author} {\bibfnamefont
  {Y.}~\bibnamefont {Kafri}}, \bibinfo {author} {\bibfnamefont
  {D.}~\bibnamefont {Mukamel}}, \bibinfo {author} {\bibfnamefont
  {A.}~\bibnamefont {Solon}},\ and\ \bibinfo {author} {\bibfnamefont
  {J.}~\bibnamefont {Tailleur}},\ }\bibfield  {title} {\bibinfo {title}
  {Metastability of discrete-symmetry flocks},\ }\href
  {https://doi.org/https://doi.org/10.1103/PhysRevLett.131.218301} {\bibfield
  {journal} {\bibinfo  {journal} {Phys. Rev. Lett.}\ }\textbf {\bibinfo
  {volume} {131}},\ \bibinfo {pages} {218301} (\bibinfo {year}
  {2023})}\BibitemShut {NoStop}%
\bibitem [{\citenamefont {Codina}\ \emph {et~al.}(2022)\citenamefont {Codina},
  \citenamefont {Mahault}, \citenamefont {Chat\'e}, \citenamefont {Dobnikar},
  \citenamefont {Pagonabarraga},\ and\ \citenamefont {Shi}}]{codina2022small}%
  \BibitemOpen
  \bibfield  {author} {\bibinfo {author} {\bibfnamefont {J.}~\bibnamefont
  {Codina}}, \bibinfo {author} {\bibfnamefont {B.}~\bibnamefont {Mahault}},
  \bibinfo {author} {\bibfnamefont {H.}~\bibnamefont {Chat\'e}}, \bibinfo
  {author} {\bibfnamefont {J.}~\bibnamefont {Dobnikar}}, \bibinfo {author}
  {\bibfnamefont {I.}~\bibnamefont {Pagonabarraga}},\ and\ \bibinfo {author}
  {\bibfnamefont {X.-q.}\ \bibnamefont {Shi}},\ }\bibfield  {title} {\bibinfo
  {title} {{Small Obstacle in a Large Polar Flock}},\ }\href
  {https://doi.org/10.1103/PhysRevLett.128.218001} {\bibfield  {journal}
  {\bibinfo  {journal} {Phys. Rev. Lett.}\ }\textbf {\bibinfo {volume} {128}},\
  \bibinfo {pages} {218001} (\bibinfo {year} {2022})}\BibitemShut {NoStop}%
\bibitem [{\citenamefont {Saha}\ and\ \citenamefont
  {Golestanian}(2022)}]{saha2022effervescent}%
  \BibitemOpen
  \bibfield  {author} {\bibinfo {author} {\bibfnamefont {S.}~\bibnamefont
  {Saha}}\ and\ \bibinfo {author} {\bibfnamefont {R.}~\bibnamefont
  {Golestanian}},\ }\bibfield  {title} {\bibinfo {title} {Effervescent waves in
  a binary mixture with nonreciprocal couplings},\ }\href
  {https://doi.org/10.48550/arXiv.2208.14985} {\bibfield  {journal} {\bibinfo
  {journal} {arXiv preprint arXiv:2208.14985}\ } (\bibinfo {year}
  {2022})}\BibitemShut {NoStop}%
\bibitem [{\citenamefont {Besse}\ \emph {et~al.}(2023)\citenamefont {Besse},
  \citenamefont {Fausti}, \citenamefont {Cates}, \citenamefont {Delamotte},\
  and\ \citenamefont {Nardini}}]{besse2023interface}%
  \BibitemOpen
  \bibfield  {author} {\bibinfo {author} {\bibfnamefont {M.}~\bibnamefont
  {Besse}}, \bibinfo {author} {\bibfnamefont {G.}~\bibnamefont {Fausti}},
  \bibinfo {author} {\bibfnamefont {M.~E.}\ \bibnamefont {Cates}}, \bibinfo
  {author} {\bibfnamefont {B.}~\bibnamefont {Delamotte}},\ and\ \bibinfo
  {author} {\bibfnamefont {C.}~\bibnamefont {Nardini}},\ }\bibfield  {title}
  {\bibinfo {title} {Interface roughening in nonequilibrium phase-separated
  systems},\ }\href
  {https://doi.org/https://doi.org/10.1103/PhysRevLett.130.187102} {\bibfield
  {journal} {\bibinfo  {journal} {Phys. Rev. Lett.}\ }\textbf {\bibinfo
  {volume} {130}},\ \bibinfo {pages} {187102} (\bibinfo {year}
  {2023})}\BibitemShut {NoStop}%
\bibitem [{\citenamefont {Osat}\ \emph {et~al.}(2024)\citenamefont {Osat},
  \citenamefont {Metson}, \citenamefont {Kardar},\ and\ \citenamefont
  {Golestanian}}]{osat2024escaping}%
  \BibitemOpen
  \bibfield  {author} {\bibinfo {author} {\bibfnamefont {S.}~\bibnamefont
  {Osat}}, \bibinfo {author} {\bibfnamefont {J.}~\bibnamefont {Metson}},
  \bibinfo {author} {\bibfnamefont {M.}~\bibnamefont {Kardar}},\ and\ \bibinfo
  {author} {\bibfnamefont {R.}~\bibnamefont {Golestanian}},\ }\bibfield
  {title} {\bibinfo {title} {Escaping kinetic traps using nonreciprocal
  interactions},\ }\href
  {https://doi.org/https://doi.org/10.1103/PhysRevLett.133.028301} {\bibfield
  {journal} {\bibinfo  {journal} {Phys. Rev. Lett.}\ }\textbf {\bibinfo
  {volume} {133}},\ \bibinfo {pages} {028301} (\bibinfo {year}
  {2024})}\BibitemShut {NoStop}%
\bibitem [{\citenamefont {Kourbane-Houssene}\ \emph {et~al.}(2018)\citenamefont
  {Kourbane-Houssene}, \citenamefont {Erignoux}, \citenamefont {Bodineau},\
  and\ \citenamefont {Tailleur}}]{kourbane2018exact}%
  \BibitemOpen
  \bibfield  {author} {\bibinfo {author} {\bibfnamefont {M.}~\bibnamefont
  {Kourbane-Houssene}}, \bibinfo {author} {\bibfnamefont {C.}~\bibnamefont
  {Erignoux}}, \bibinfo {author} {\bibfnamefont {T.}~\bibnamefont {Bodineau}},\
  and\ \bibinfo {author} {\bibfnamefont {J.}~\bibnamefont {Tailleur}},\
  }\bibfield  {title} {\bibinfo {title} {Exact hydrodynamic description of
  active lattice gases},\ }\href
  {https://doi.org/https://doi.org/10.1103/PhysRevLett.120.268003} {\bibfield
  {journal} {\bibinfo  {journal} {Phys. Rev. Lett.}\ }\textbf {\bibinfo
  {volume} {120}},\ \bibinfo {pages} {268003} (\bibinfo {year}
  {2018})}\BibitemShut {NoStop}%
\bibitem [{\citenamefont {K\"ursten}\ and\ \citenamefont
  {Ihle}(2021)}]{kursten2021quantitative}%
  \BibitemOpen
  \bibfield  {author} {\bibinfo {author} {\bibfnamefont {R.}~\bibnamefont
  {K\"ursten}}\ and\ \bibinfo {author} {\bibfnamefont {T.}~\bibnamefont
  {Ihle}},\ }\bibfield  {title} {\bibinfo {title} {Quantitative kinetic theory
  of flocking with three-particle closure},\ }\href
  {https://doi.org/10.1103/PhysRevE.104.034604} {\bibfield  {journal} {\bibinfo
   {journal} {Phys. Rev. E}\ }\textbf {\bibinfo {volume} {104}},\ \bibinfo
  {pages} {034604} (\bibinfo {year} {2021})}\BibitemShut {NoStop}%
\bibitem [{\citenamefont {Martin}\ \emph {et~al.}(2023)\citenamefont {Martin},
  \citenamefont {Seara}, \citenamefont {Avni}, \citenamefont {Fruchart},\ and\
  \citenamefont {Vitelli}}]{martin2023exact}%
  \BibitemOpen
  \bibfield  {author} {\bibinfo {author} {\bibfnamefont {D.}~\bibnamefont
  {Martin}}, \bibinfo {author} {\bibfnamefont {D.}~\bibnamefont {Seara}},
  \bibinfo {author} {\bibfnamefont {Y.}~\bibnamefont {Avni}}, \bibinfo {author}
  {\bibfnamefont {M.}~\bibnamefont {Fruchart}},\ and\ \bibinfo {author}
  {\bibfnamefont {V.}~\bibnamefont {Vitelli}},\ }\bibfield  {title} {\bibinfo
  {title} {An exact model for the transition to collective motion in
  nonreciprocal active matter},\ }\href
  {https://doi.org/10.48550/arXiv.2307.08251} {\bibfield  {journal} {\bibinfo
  {journal} {arXiv preprint arXiv:2307.08251}\ } (\bibinfo {year}
  {2023})}\BibitemShut {NoStop}%
\bibitem [{\citenamefont {Mason}\ \emph {et~al.}(2024)\citenamefont {Mason},
  \citenamefont {Jack},\ and\ \citenamefont {Bruna}}]{mason2024dynamical}%
  \BibitemOpen
  \bibfield  {author} {\bibinfo {author} {\bibfnamefont {J.}~\bibnamefont
  {Mason}}, \bibinfo {author} {\bibfnamefont {R.~L.}\ \bibnamefont {Jack}},\
  and\ \bibinfo {author} {\bibfnamefont {M.}~\bibnamefont {Bruna}},\ }\bibfield
   {title} {\bibinfo {title} {Dynamical patterns in active-passive particle
  mixtures with non-reciprocal interactions: Exact hydrodynamic analysis},\
  }\href {https://doi.org/10.48550/arXiv.2408.03932} {\bibfield  {journal}
  {\bibinfo  {journal} {arXiv preprint arXiv:2408.03932}\ } (\bibinfo {year}
  {2024})}\BibitemShut {NoStop}%
\bibitem [{\citenamefont {Marenda}\ \emph {et~al.}(2016)\citenamefont
  {Marenda}, \citenamefont {Zanardo}, \citenamefont {Trovato}, \citenamefont
  {Seno},\ and\ \citenamefont {Squartini}}]{marenda2016modeling}%
  \BibitemOpen
  \bibfield  {author} {\bibinfo {author} {\bibfnamefont {M.}~\bibnamefont
  {Marenda}}, \bibinfo {author} {\bibfnamefont {M.}~\bibnamefont {Zanardo}},
  \bibinfo {author} {\bibfnamefont {A.}~\bibnamefont {Trovato}}, \bibinfo
  {author} {\bibfnamefont {F.}~\bibnamefont {Seno}},\ and\ \bibinfo {author}
  {\bibfnamefont {A.}~\bibnamefont {Squartini}},\ }\bibfield  {title} {\bibinfo
  {title} {Modeling quorum sensing trade-offs between bacterial cell density
  and system extension from open boundaries},\ }\href
  {https://www.nature.com/articles/srep39142} {\bibfield  {journal} {\bibinfo
  {journal} {Sci. Rep.}\ }\textbf {\bibinfo {volume} {6}},\ \bibinfo {pages}
  {39142} (\bibinfo {year} {2016})}\BibitemShut {NoStop}%
\end{thebibliography}%

\end{document}